\documentclass{jpp}
\usepackage{graphicx}
\usepackage{epstopdf, epsfig}
\usepackage{eurosym}
\usepackage{graphicx}
\usepackage{caption,subcaption}
\usepackage{float}
\usepackage[normalem]{ulem}
\usepackage{color}
\usepackage{soul}

\usepackage{bbold}
\usepackage{amsmath,amssymb}
\usepackage{braket}
\usepackage[english]{babel}
\usepackage{ulem}
\usepackage{cancel}

\newcommand{\beq}{\begin{equation}}
\newcommand{\eeq}{\end{equation}}

\def\mD{ \mathcal{D} }

\def\mE{ \mathcal{E} }

\def\mJ{ \mathcal{J} }

\def\mW{ \mathcal{W} }

\newcommand{\btu}{\tilde{\bf u}}

\newcommand{\tu}{\tilde{ u}}
\newcommand{\tb}{\tilde{ b}}

\newcommand{\tS}{\tilde{S}}

\newcommand{\bu}{{\bf u}}

\newcommand{\bk}{{\bf k}}

\newcommand{\bx}{{\bf x}}
\newcommand{\br}{{\bf r}}

\newcommand{\grad}{\nabla  }

\newcommand{\ba}{{\bf a}}
\newcommand{\bb}{{\bf b}}

\newcommand{\bS}{{\bf S}}

\newcommand{\lb}{\label}
\newcommand{\ol}{\widetilde}
\newcommand{\bdot}{{\mbox{\boldmath $\cdot$}}}

\newcommand{\btau}{{\mbox{\boldmath $\tau$}}}

\def\be{\begin{equation}}
\def\ee{\end{equation}}

\newcommand{\fltr}{\widetilde}
\newcommand{\OL}{\overline}
\newcommand{\fb}{\fltr\bb}
\newcommand{\fu}{\fltr\bu}
\newcommand{\fS}{\fltr\bS}

\newcommand{\SOmega}{\fltr\Omega}

\def\NEW#1{{\textcolor{black}{#1}}}            % question

\shorttitle{Local fluxes in MHD turbulence}
\shortauthor{A. Alexakis and S. Chibbaro}

\title{ Local fluxes in MHD Turbulence}

\author{ Alexandros Alexakis \aff{1}
\corresp{\email{alexakis@phys.ens.fr}}\& Sergio Chibbaro  \aff{2}
   }

\affiliation{Laboratoire de Physique de l’Ecole Normale Supérieure, ENS, Université PSL, CNRS, Sorbonne Université, Université Paris-Diderot, Sorbonne Paris Cité, Paris, France, \aff{2} Universit\'e Paris-Saclay, CNRS, LISN, 91400 Orsay, France }

\begin{document}

\maketitle

\begin{abstract}
Using highly resolved direct numerical simulations we examine the statistical properties of the local energy flux rate $\Pi_\ell(x)$ towards small scales for three isotropic turbulent magnetohydrodynamic flows, which differ in strength and structure of the magnetic field. 
We analyse the cascade process both in the kinetic and magnetic energy, disentangling the different flux contributions to the overall energy dynamics.
The results show that the probability distribution of the local energy flux develops long tails related to extreme events, similar to the hydrodynamic case.    %
The different terms of the energy flux display different properties and show sensitivity on the type of the flow examined. 
We further examine the joint pdf between the local energy flux and the gradients of the involved fields. The results point out a correlation with the magnetic field gradients, showing however a dispersion much stronger than what is observed in hydrodynamic flows. 
Finally, it is also shown that the local energy flux shows \NEW{some} dependence on the local amplitude of the magnetic field.
The present results have implications for subgrid scale models that we discuss.

\end{abstract}

%%%%%%%%%%%%%%%%%%%%%%%%%%%%%%%%%%%%%%%%%%%%%%%%%%%%%%%%%%%%%%%%%%%%%%%
%%%%%%%%%%%%%%%%%%%%%%%%%%%%%%%%%%%%%%%%%%%%%%%%%%%%%%%%%%%%%%%%%%%%%%%
\section{Introduction    }    %%%%%%%%%%%%%%%%%%%%%%%%%%%%%%%%%%%%%%%%%
\label{sec:intro}            
%%%%%%%%%%%%%%%%%%%%%%%%%%%%%%%%%%%%%%%%%
%%%%%%%%%%%%%%%%%%%%%%%%%%%%%%%%%%%%%%%%%%%%%%%%%%%%%%%%%%%%%%%%%%%%%%%
%%%%%%%%%%%%%%%%%%%%%%%%%%%%%%%%%%%%%%%%%%%%%%%%%%%%%%%%%%%%%%%%%%%%%%%
%\begin{keywords}
%turbulence,absolute equilibrium
%\end{keywords}

%Turbulence prevails ionized mater in the universe. As a result
%are made of or host gas in a ionized state.
Most stellar objects from stars to galaxies are \NEW{made of  hot gas} in a ionized state. %incompressible 
The magneto-hydrodynamic (MHD) equations give the simplest description of the dynamics involved \citep{goldstein1995magnetohydrodynamic,battaner1996astrophysical,biskamp2003magnetohydrodynamic,verma2004statistical,mckee2007theory,bruno2013solar,galtier2016introduction} and give a general framework relevant for dynamo and many plasma phenomena \citep{landau2013electrodynamics,brandenburg2005astrophysical,davidson2002introduction}.  
At large Reynolds numbers (small dissipation parameters) the MHD turbulent dynamics 
become turbulent sharing many features with hydrodynamic turbulence.
%and is based upon the fundamental concepts of fluid turbulence \citep{}.
The key idea is the existence of {\em cascade} processes,  meaning that  inviscid invariants are 
%related to a quasi-equilibrium process able to 
transferred  across scales so that they are efficiently dissipated at the smallest scales \citep{Mon_75,Fri_95,alexakis2018cascades}.
This idea sketched by Richardson, was at the basis of the phenomenological statistical theory by Kolmogorov \citep{kolmogorov1941local,kraichnan1971inertial} and is the cornerstone of our understanding of turbulence.
\NEW{Different theories have been proposed that attempt to describe quantitatively the behavior of MHD turbulence \citep{iroshnikov1964turbulence,kraichnan1965inertial}.  
Yet, MHD turbulence displays an even  more complex behaviour than Hydrodynamic one with diverse regimes and recently new theories have been proposed
\citep{goldreich1995toward,boldyrev2006spectrum,beresnyak2011spectral}.
A variety of methods have been applied in MHD to test and constrain existing theories. Two point correlation functions have led to some constraints on third order structure functions \citep{politano1998karman,galtier2009exact}.
Scale by scale analysis in Fourier space that has been extensively 
used in hydrodynamics \citep{chen2003joint,domaradzki2009locality,teaca2011locality,alexakis2007turbulent,verma2004statistical,verma2019energy} has been applied with some success in MHD as well
\citep{verma2021variable,dar2001energy,teaca2009energy,alexakis2005shell,mininni2005shell}.
%MHD thus still needs new concepts and a more thorough physical understanding in particular to develop accurate models.
}

%\NEW{  }

New insights on the cascade physics have been obtained analysing the scale-by-scale budgets of energy \NEW{of filtered fields}, including the fluctuations of the flux \citep{Eyink:2006p1379,dubrulle2019beyond}.
Notably, this kind of complex tools have been used to get insights on the physical mechanisms in 2D and 3D fluid turbulence %,Chen:2006p1741
\citep{piomelli1991subgrid,eyink2009localness,eyink2005locality,liu1994properties,chen2006kelvin} and  more recently, scale-by-scale \NEW{filtering} analysis has been started to be applied also to MHD turbulence to highlight and understand some specific processes \citep{eyink2013flux,galtier2018origin,bian2019decoupled}.
%%%
\NEW{ The interest of such tools is witnessed by its use in many other problems such as two-fluid plasma models \citep{camporeale2018coherent},  hybrid-kinetic models \citep{cerri2020space},  gyro-kinetics \citep{teaca2021sub}, and  full-kinetic systems \citep{eyink2018cascades,yang2022pressure}.}
%%%
More generally, this kind of approach seems promising to go beyond standard statistical information and access to energetic processes, which are usually the most important for applications, and are nowadays commonly used in a variety of cases \citep{danaila2001turbulent,casciola2003scale,sorriso2007observation,cimarelli2013paths,valori2020weak,innocenti2021direct}. Our work follows this path in order to get similar understanding for MHD turbulence. 

%On the other hand, 
Such an understanding of the cascade process is paramount for astrophysical and industrial applications for the following reason. 
While the numerical solution of the fundamental equations is in principle possible, the vast range of scales excited in such objects prohibits any direct numerical calculation that resolves
all the scales down to the dissipation ones. 
%%%
However, if all scales are not properly resolved, the numerical approach is tantamount to apply a coarse-graining to the initial problem discarding a part of the information, related to the small scales not resolved.
As a result all simulations of astrophysical flows require some modeling of smaller unresolved scales that are responsible for the energy dissipation.
Such modeling will thus have to compensate the energy transfers between resolved and unresolved scales so that 
it correctly captures the  energy dissipation of the flow. 
%%%
The construction of such models requires a deep and quantitative understanding of how turbulence transfers energy to small scales. 

This practice to compute only large scales and to put forward approximate models at small scales is referred to as Large-eddy simulations (LES), and has a long history in fluid turbulence in particular for  practical applications \citep{leonard1975energy,germano1992turbulence,meneveau2000scale,sagaut2001large,lesieur2005large,Pope_turbulent}.
In hydrodynamic turbulence the development and testing of such models is thus well advanced. 
A particular class of hydrodynamic models quantify the dissipation energy by a an eddy-viscosity term whose coefficient depends on the gradients of the resolved flow \citep{smagorinsky1963general,germano1991dynamic,germano1992turbulence}, and 
many studies have been devoted in quantifying this dependence \citep{vreman1994realizability,vreman1997large,borue1998local,meneveau2000scale}.
%In the last years,  new analysis of this kind of modelling has been made possible by the use of scale-by-scale analysis \citep{linkmann2018multi,biferale2019self,buzzicotti2018effect,alexakis2020local}

In MHD, LES is less developed and tested than in fluid turbulence.
At the same time the complexity of the  MHD is increased as more channels exist for the energy to be transferred to the smaller scales and since there are two fields involved there are more possibilities for the dependence of the eddy viscosity term. 
%Moreover, comparison with experiments is practically impossible, since the physical relevant problems are mainly astrophysical plasmas. 
\NEW{ A convenient} way to assess subgrid models is %therefore 
through direct numerical simulations, which are however more difficult than in hydrodynamics and only recently DNS \NEW{with a sufficient resolution allowing for an inertial range to be present have been made available.}
Yet, Smagorinski-like models have been adapted through formal analogy in MHD since the 90' \citep{theobald1994subgrid,muller2002dynamic,verma2004large,bian2021scaling}. 
These models still represent the only available strategy for subgrid modelling \citep{miesch2015large}, and 
they have been recently generalised even to compressible \citep{chernyshov2014subgrid,vlaykov2016nonlinear,grete2017comparative} 
and %MHD, 
relativistic MHD \citep{vigano2019extension,carrasco2020gradient}.
%\NEW{and to gyro-kinetic simulations of plasma turbulence in tokamaks \cite{banon2014applications}.}

Although DNS has been used to partially assess the validity of such closures \citep{agullo2001large,miesch2015large,kessar2016effect}, a thorough analysis is still needed, given the importance of the problem.
Notably, in the last years,  new analysis of this kind of modelling has been made possible by the use of scale-by-scale analysis \citep{linkmann2018multi,biferale2019self,buzzicotti2018effect,alexakis2020local}.
In these works, it has been pointed out that a deeper understanding of the cascade process, including fluctuations is valuable also to improve modelling.

The purpose of this work is to shed some light on this direction by examining different MHD turbulence flows and the statistical properties of the local energy flux towards the smaller scales, bringing to MHD this original approach already applied to hydrodynamic turbulence.
In particular, we analyse the scale-by-scale budgets of kinetic and magnetic energy of highly resolved DNS. 
That allows to provide new information on the cascade process, including fluctuations, and thus
to go beyond usual investigation about the validity  of subgrid models.
\NEW{As a final introductory comment, it is worth emphasising that our goal is not to put forward and/or test some specific model but rather to get some insights on the modelling via DNS. }

%The rest of the paper is structured as follows. In the next section we present the theoretical formulation of the energy fluxes for MHD turbulence. In section \ref{sec:results} we present the numerical simulations and in section \ref{sec:pdfs} the statistical  analysis of the energy flux. We conclude in the last section.

%%%%%%%%%%%%%%%%%%%%%%%%%%%%%%%%%%%%%%%%%%%%%%%%%%%%%%%%%%%%%%%%%%%%%%%
%%%%%%%%%%%%%%%%%%%%%%%%%%%%%%%%%%%%%%%%%%%%%%%%%%%%%%%%%%%%%%%%%%%%%%%
%\section{Hydro}               %%%%%%%%%%%%%%%%%%%%%%%%%%%%%%%%%%%%%%%%%
%\label{sec:intro}             %%%%%%%%%%%%%%%%%%%%%%%%%%%%%%%%%%%%%%%%%
%%%%%%%%%%%%%%%%%%%%%%%%%%%%%%%%%%%%%%%%%%%%%%%%%%%%%%%%%%%%%%%%%%%%%%%
%%%%%%%%%%%%%%%%%%%%%%%%%%%%%%%%%%%%%%%%%%%%%%%%%%%%%%%%%%%%%%%%%%%%%%%

%% FIG 1
%\begin{figure*}
%\centering
%\includegraphics[width=0.47\textwidth]{RS_Q_gauss_HD0_total.pdf}
%\includegraphics[width=0.47\textwidth]{RS_Q_sharp_HD0_total.pdf}
%\caption{ Hydro RESULTS: Gaussian vs Sharp filter. Asymmetry is more clear for the Gaussian filter. }
%\label{fig4}
%\end{figure*}

%%%%%%%%%%%%%%%%%%%%%%%%%%%%%%%%%%%%%%%%%%%%%%%%%%%%%%%%%%%%%%%%%%%%%%%
%%%%%%%%%%%%%%%%%%%%%%%%%%%%%%%%%%%%%%%%%%%%%%%%%%%%%%%%%%%%%%%%%%%%%%%
\section{Theoretical formulation}   %%%%%%%%%%%%%%%%%%%%%%%%%%%%%%%%%%%
\label{sec:theory}      
\subsection{Main Equations}

We begin by considering 
the magneto-hydrodynamic equations in the incompressible non-relativistic limit, given by \citep{biskamp2003magnetohydrodynamic}
%\be
%\nabla\cdot {\bf v}=0
%  \label{mass}
%\ee
%
\begin{eqnarray}
%\rho \ ({\partial {\bf v}\over\partial t}+{\bf v}\cdot\nabla{\bf v}) &=&
%   -\nabla P +  {\bf J}\times{\bf b}+
%{\mu}\nabla^2\ {\bf v}=\\
{\partial {\bf u}\over\partial t}+{\bf u}\cdot\nabla{\bf u} &=&
   -\nabla p +  {\bf b} \cdot \nabla {\bf b} +
{\nu}\nabla^2 {\bf u} 
\label{momentum} \\
%\end{eqnarray}
%
%
%\begin{eqnarray}
%{\partial {\bf b}\over\partial t}&=&\nabla\times{\bf v}\times{\bf b} 
%  - {\eta} \nabla^2 {\bf b} = \\
  {\partial {\bf b}\over\partial t}+{\bf u}\cdot \nabla{\bf b}&=& {\bf b} \cdot \nabla{\bf u}
  + {\eta} \nabla^2 {\bf b} ,
  \label{faraday} \\
  \nabla\cdot {\bf v}=0,   
%  \label{mass}
& &
%\\
\nabla\cdot{\bf b} = 0,
\label{divb}
\end{eqnarray}
%
%\be
%\nabla\cdot{\bf b} = 0\label{divb}
%\ee
where the modified pressure $p$ is given by $p=P/\rho+b^2/2$, ${\bf b}={\bf B}\sqrt{4\pi \rho}$, and %where:~~
$\rho$ is the mass density,
${\bf u}({\bf x}, t) $ is the flow velocity,
$P({\bf x}, t)$ is the thermal pressure,
${\bf B}({\bf x}, t) $ is the magnetic induction field
$\nu$ is the viscosity and
$\eta$ the magnetic diffusivity.
%We have thus splitted the Lorentz force in its isotropic, included in the modified pressure, and anisotropic part.
%$\zeta_{ij} ({\bf x}, t) = e_{ij} - {2\over 3}\nabla\cdot {\bf v} \ \delta_{ij}$ is the viscous
%stress tensor, $e_{ij} = (\partial_j v_i + \partial_i v_j)$ is the strain tensor
%and $\gamma$ is the adiabatic ratio.
%%%%%%%%%%%%%%%%%%%%%%%%%%%%%%%%%%%
%%%%%%%%%%%%%%%%%%%%%%%%%%%%%%%%%%%%%%%%%%%%%%%%%%%%%%%%%%%%%%%%%%%%%%%
%%%%%%%%%%%%%%%%%%%%%%%%%%%%%%%%%%%%%%%%%%%%%%%%%%%%%%%%%%%%%%%%%%%%%%%
\subsection{Scale-By-scale Analysis}
\NEW{A considerable amount of work in analysing the transfer of energy among scales has been performed in Fourier space with significant results
%that allows for significant analytical treatment
\citep{domaradzki2009locality,alexakis2007turbulent,teaca2011locality,
verma2004statistical,verma2019energy,verma2021variable,teaca2009energy,alexakis2005shell,mininni2005shell}. 
This kind of approach while useful in understanding and quantifying the scale-by-scale energy budget of turbulence, does not allow to simply link the related energy transfers with local properties of the flow and thus limits their applicability to subgrid scale models.} %~\citep{verma2021variable}.}

\NEW{ 
To analyse the local in space scale-by-scale dynamics,
we use the local filtering approach \citep{germano1992turbulence}}.
We introduce a filter such that 
\be \fltr{\ba}_\ell(\bx) = \int dr^3\, G_\ell(\br) \ba(\bx+\br) \lb{filter} \ee
where $G({\bf r})$ is a smooth filtering function, 
%non-negative, 
spatially localized and such that $\int dr^3 \ G({\bf r})=1$, and $\int dr^3 \ \vert {\bf r} \vert ^2 G({\bf r}) \approx 1$. The function $G_\ell$ is rescaled with $\ell$ as $G_\ell ({\bf r}) = \ell^{-3}G({\bf  r}/\ell)$. 
Applying this filter to the MHD equations one obtains \citep{aluie2017coarse}:

\be \partial_{t} \ol{\bu}_\ell +  (\ol{\bu}_\ell\cdot\nabla)\ol{\bu}_\ell = -\grad\ol{p}_{\ell} 
+ (\fb_{\ell}\bdot\grad)\fb_{\ell} -\grad \cdot (\btau_\ell^{uu}-\btau_\ell^{bb})
+\nu\nabla^{2}\ol{\bu}_\ell.
\lb{u-eq-ell} \ee

\be \partial_{t} \ol{\bb}_\ell  +  
(\ol{\bu}_\ell\cdot\grad)\ol{\bb}_\ell -(\ol{\bb}_\ell\cdot\grad)\ol{\bu}_\ell 
= - \nabla \cdot (\btau^{ub}_\ell-\btau^{bu}_\ell)
+ \eta\nabla^{2}\ol{\bb}_\ell
\lb{b-eq-ell} \ee

\be \grad\cdot\ol{\bb}_\ell=\grad\cdot\ol{\bu}_\ell=0.
\lb{div-free-ell} \ee
Here  $\btau^{uu}_\ell,\btau^{bb}_\ell,\btau^{ub}_\ell$ and $\btau^{bu}_\ell,$ are the 
{\it subscale stress tensors} which describe the force exerted 
on scales larger than $\ell$ by fluctuations at scales smaller than $\ell$. 
%It is the sum of two contributions, $\btau_\ell^{ub} 
%=\btau_\ell^u-\btau_\ell^b$ where
They are defined as
\be 
\btau^{uu}_\ell = \ol{(\bu\bu)}_\ell -\fu_{\ell}\fu_{\ell}, \qquad 
%\lb{R-stress} 
%\ee
%is the \emph {subscale Reynolds stress} and 
%\be 
\btau^{bb}_\ell =\ol{(\bb\bb)}_\ell - \fb_{\ell}\fb_{\ell} 
\lb{M-stress} 
\ee 
are the {\it subscale Reynolds stress} and  the {\it subscale Maxwell stress}
respectively and finally
\be 
\btau^{ub}_\ell = \ol{(\bu\bb)}_\ell - \fu_{\ell}\fb_{\ell}, \qquad 
\btau^{bu}_\ell = \ol{(\bb\bu)}_\ell - \fb_{\ell}\fu_{\ell} 
\lb{cross-stress} 
\ee 
are cross-field tensors for which one is the transpose of the other $\btau^{bu}_\ell=[\btau^{ub}_\ell]^T$.

Using this notation we can write an equation for the large scale energy densities
$\tilde{\mE}^u=\frac{1}{2}|\fu|^2$ and 
$\tilde{\mE}^b=\frac{1}{2}|\fb|^2$ as
%\be 
%\btau^{b}_\ell =[\ol{(\bb\bb)}_\ell - \fb_{\ell}\fb_{\ell}] 
%\lb{M-stress} 
%\ee 

\begin{eqnarray}
\partial_t \tilde{\mE}_\ell^u + \nabla \cdot {\mJ}^u &=& -\Pi_\ell^{uu} - \Pi_\ell^{bb} + \mW_{L} -\mathcal{D}_u \\
\partial_t \tilde{\mE}^b_\ell + \nabla \cdot {\mJ}^b &=& -\Pi_\ell^{bu} -\Pi_\ell^{ub} - \mW_{L} -\mathcal{D}_b
\end{eqnarray}
where the currents $\mJ^u,\mJ^b$ are given by 
\begin{eqnarray}
\mJ^u_j &=& \left(\frac{1}{2}|\fu|^2 +\OL{p} \right) \ol{u}_j  + (\tau^{uu}_{\ell,ij} - \tau^{bb}_{\ell,ij})\ol{u}_i   - \nu  \frac{1}{2}\partial_j|\fu|^{2} 
%+ (\fu\cdot\fb)\ol{b}_j 
\\
\mJ^b_j &=& \left(\frac{1}{2}   |\fb|^2       \right) \ol{u}_j  + (\tau^{ub}_{\ell,ij} - \tau^{bu}_{\ell,ij})\ol{b}_i - (\fu\cdot\fb)\ol{b}_j   -\eta (\fb \times \nabla \times \fb ) 
\end{eqnarray}
and express the transport of energy in space. The rate of work done on the flow \NEW{originating from} the Lorentz  force is given by
\beq
 \mW_{L} = \ol{u}_i\ol{b}_j\partial_j\ol{b}_i.
\eeq 
\NEW{This term is responsible for the transfer of kinetic energy to magnetic energy in the large scales.}
The viscous and Ohmic energy dissipation rates are given by 
\beq
 \mD_u  = \nu (\partial_{j}\ol{u}_{i})(\partial_{j}\ol{u}_{i})\quad\mathrm{and}\quad 
 \mD_b  = \eta (\nabla \times \fb)\cdot (\nabla \times \fb)
\eeq  
and express the rate energy is dissipated by viscous and Ohmic effects respectively. 
\NEW{These terms can be shown to be negligible if the filter scale is large and the viscous/resistive coefficients are small
\citep{aluie2017coarse}.}
Finally the four fluxes in scale space $\Pi_\ell^{uu},\Pi_\ell^{bb},\Pi_\ell^{ub},\Pi_\ell^{bu},$
explicitly given by:
\begin{eqnarray}
\Pi_\ell^{uu} =-\tau_{\ell,ij}^{ub}\partial_{i}\ol{u}_{j}
              =-[ \ol{(u_iu_j)}_\ell - \tu_{\ell,i}\tu_{\ell,j}]\partial_{i}\ol{u}_{j} \label{flux1} \\ 
\Pi_\ell^{bb} =+\tau_{\ell,ij}^{bb}\partial_{i}\ol{u}_{j}
              =+[ \ol{(b_ib_j)}_\ell - \tb_{\ell,i}\tb_{\ell,j}]\partial_{i}\ol{u}_{j} \label{flux2a} \\
\Pi_\ell^{ub} =-\tau_{\ell,ij}^{ub}\partial_{i}\ol{b}_{j}
              =-[ \ol{(u_ib_j)}_\ell - \tu_{\ell,i}\tb_{\ell,j}]\partial_{i}\ol{b}_{j} \label{flux3} \\
\Pi_\ell^{bu} =+\tau_{\ell,ij}^{bu}\partial_{i}\ol{b}_{j}
              =+[ \ol{(b_iu_j)}_\ell - \tb_{\ell,i}\tu_{\ell,j}]\partial_{i}\ol{b}_{j} \label{flux4}
\label{flux2}
\end{eqnarray}
express the rate of gain (if $\Pi_\ell<0$) or loss (if $\Pi_\ell>0$) of energy of the large scales to the small filtered-out scales. 
Their sum gives the total energy flux
\beq \Pi_\ell(\bx) = \Pi_\ell^{uu}(\bx)+\Pi_\ell^{bb}(\bx)+\Pi_\ell^{ub}(\bx)+\Pi_\ell^{bu}(\bx)
\eeq 

\NEW{ 
The fluxes are chosen so that they are invariant under
Galilean transformations $\bu \to \bu+{\bf U}_0$ and also under $\bb \to \bb+{\bf B}_0$ for any flow realization $\bf u,b$, 
where $\bf U_0,B_0$ are constant in space vector fields.
\NEW{The importance of Galilean invariance has been noted in the 
past \citep{speziale1985galilean,aluie2009localness,aluie2010scale}.}
We note however that while the defined fluxes are invariant under $\bb \to \bb+{\bf B}_0$ the MHD equations are not. Therefore, the introduction of a constant ${\bf B}_0$ in the dynamics of the MHD equations will alter the statistical behavior of the fields $\bf u,b$ and as a result of the fluxes as well. }

\NEW{ 
The fluxes in (\ref{flux1})-(\ref{flux4})  
comprise the main object of the present work. }

\subsection{Filters}

Although we formulate our filtering procedure in the physical space, since we will be working in periodic domains it is useful to define the filters through their Fourier transforms 
\be \hat{G}_q (\bk) = \int G_\ell({\bf x}) e^{i\bf k\cdot x}  d {\bx} \ee
where $q=1/\ell$ \NEW{is the wavenumber corresponding to the filter length $\ell$, not to be confused with the Fourier wavenumber $k$.
}
For the first filter we consider a Gaussian kernel
\be \hat{G}_q(\bk)=\exp\left[-\frac{k^2}{2q^2}\right]. 
\label{gauss}\ee 
For an infinite domain this filter corresponds to the Gaussian filter in real space $G_\ell(r) =\exp(-\frac{1}{2}r^2/\ell^2) /(2\pi \ell^{2})^{3/2}$. 
We note that this filtering is not a projection and in general
$\ol{({\btu_\ell})_\ell}\ne \btu_\ell $. 
The second filter we are going to consider is a sharp spectral filter such that 
\be
\ol{\bu}_\ell(\bx,t)=\sum_{\vert \bk \vert < q}\hat{\bu}(\bk,t) e^{i\bk \cdot \bx } .
\label{sharp}
\ee
This filtering is  a projector 
$\ol{({\btu_\ell})_\ell}= \btu_\ell $
and is based on a Galerkin truncation for all wavenumbers larger than the given cutoff $q=1/\ell$. 
\NEW{With regard to the representation of energy fluxes,}
this filter has been shown in the past not to be optimal as it leads to a wider fluctuations of the local energy flux \citep{buzzicotti2018effect,alexakis2020local}, something  
that as we will show also holds in MHD.

\section{Sub-scale stress modeling}     %%%%%%%%%%%%%%%%%%%%%%%%%%%%%%%%%%%%
%%%%%%%%%%%%%%%%%%%%%%%%%%%%%%%%%%%%%%%%%%%%%%%%%%%%%%%%%%%%%%%%%%%%%%%
%%%%%%%%%%%%%%%%%%%%%%%%%%%%%%%%%%%%%%%%%%%%%%%%%%%%%%%%%%%%%%%%%%%%%%%
 In this section, we provide some information about typical modelling strategy. \NEW{Our work focuses on fundamental issues, that allows us to build a precise framework and to point to possible applications.}
 
In a simulation for which only the large smoothed out fields $\fu,\fb$ fields are followed dynamically the effect of the small unresolved scales on the resolved scales need to be captured by modeling the sub-scale stress tensors $\btau_\ell^{uu},\btau_\ell^{ub},\btau_\ell^{bu},\btau_\ell^{bb}$. In hydrodynamics, the simplest perhaps such model 
is formed by assuming that $\btau^{uu}$ takes the form
\beq
\btau^{uu}_{\ell,ij} = \nu^e_{\ell,ijkl} \nabla_k \fu_l^{}
\eeq 
where the eddy viscosity tensor ${\bf \nu}^e$ is a function of space and time and needs to be prescribed from the filtered field $\fu$ in order for the filtered equations to be closed. 
It needs to satisfy $\langle \nabla_i \fu_j^{}\nu^e_{\ell,ijkl} \nabla_k \fu_l^{}\rangle=\epsilon> 0$ and acts thus on average as a sink of energy.  Furthermore, 
%if $\nu^e$ depends on space 
the divergence of $\nabla \cdot \btau^{uu}_\ell$ is not necessarily zero so its
projection to divergence free vector fields needs to be considered by adding 
$p'=\nabla^{-2} \nabla_i \nabla_j \tau^{uu}_{\ell,ij} $ to the 
pressure.  
Since the system is Galilean invariant $\nu^e$ cannot depend on the values of $\fu$ but only on its gradients $\nabla \fu, \nabla\nabla \fu,\dots$, thus in the simplest case $\nu_\ell^e$ is just a function of $\nabla \fu$. If we further assume that isotropy is present at the small scales   $\nu^e[\nabla \fu]$ will only depend on the invariants (under rotations and reflections) of the strain tensor $\nabla \fu$. In the now classical Smagorinsky approach 
the sub-scale Reynolds stress tensor is modeled as \citep{smagorinsky1963general}
\beq
\btau^{uu}_{\ell,ij} -\frac{1}{3}\btau^{uu}_{\ell,kk} = \nu_\ell^e \ol{S}_{ij} =  \frac{\nu^e}{2} \left( \partial_i \tu_j + \partial_j \tu_i \right) 
\eeq 
where $\fS$ is the symmetric part of the filtered strain tensor. 
The $\nu^e$ is a scalar defined as~\citep{smagorinsky1963general}
\beq
\nu_\ell^e = c \ell^{2} |\fS| = c \ell^{2} \sqrt{ \tS_{ij}\tS_{ij}}
\label{Smago}
\eeq 
where $\ell$ is the filtering lengthscale and $c$ 
\NEW{is the Smagorinsky constant,}
an order one non-dimensional coefficient . Eq. (\ref{Smago}) gives the only combination of $\ell$ and $|\fS|$ with dimensions of viscosity. We note that in general $\nu_\ell^e$ depends on space and needs to be taken in to account for the pressure so that the divergence free condition for $\fu$ is satisfied.

This model implies that the flux of energy to the small scales is given by 
\beq 
\Pi_\ell(\bx)= c \ell^{2} |\fS|^3 .
\label{SmagoRel}
\eeq 
After Smagorinsky's work other models that also take into account the anti-symmetric part of the stress tensor
\beq
\SOmega_{ij} =  \frac{1}{2} \left( \partial_i \tu_j - \partial_j \tu_i \right)
\eeq 
have been developed. Approximating the sub-scale stress with its extreme local expression, the nonlinear Clark model is obtained \citep{meneveau2000scale}:
\begin{equation}
\tau_\ell^{uu}\approx 
\dfrac{1}{3}C_2 \ell^2 \left(\ol{\bf S}_\ell^2+
\ol{\bf \Omega}_\ell^2+\ol{\bf \Omega}_\ell\ol{\bf S}_\ell-\ol{\bf S}_\ell\ol{\bf \Omega}_\ell \right )~,
\label{clark}
\end{equation}
where  both strain and vorticity participate in the dynamics \citep{misra1997vortex,borue1998local,Ten_90}
and leads to modeled local energy flux:
\begin{equation}
    \Pi_\ell\approx \dfrac{1}{3}C_2 \ell^2 [-\mathrm{Tr}(\ol{\bf S}_\ell^3)+3\mathrm{Tr}(\ol{\bf S}_\ell \ol{\bf \Omega}_\ell^2)]~.
\end{equation}
\NEW{The development and assessment of such models has been often guided by  {\it a priori} analysis~ \citep{piomelli1991subgrid,meneveau1994statistics,borue1998local,meneveau2000scale}, that is studying the filtered DNS field properties rather than resorting to actual LES.
Following this approach}, in %our previous work
\cite{alexakis2020local} we have demonstrated that there is indeed a strong correlation between $ \Pi_\ell$ and $|\fS|$ verifying the Smagorinsky relation 
(\ref{SmagoRel}) for the mean value of $\Pi_\ell$ although strong fluctuations in $ \Pi_\ell$ were 
also measured making $\Pi_\ell$ not a strictly positive quantity.

In MHD a similar type of modeling is considerably more difficult. % for the reasons we describe bellow: 
%\begin{enumerate}
% \item[(i)] 
  First of all
 there are three sub-scale turbulent stresses that need to be modeled $\btau_\ell^{uu},\btau_\ell^{ub},\btau_\ell^{bb}$. 
 %From these three tensors $\btau_\ell^{uu}$ should reduce to the hydrodynamic one in the absence of magnetic fields. 
 %The tensor $\btau_\ell^{ub}$ can depend on the gradients of both magnetic and velocity field while the last one should depend only on magnetic field gradients. 
 In the past literature these dependencies are modeled based on the symmetric and anti-symmetric parts of the stress tensors of the two fields: 
 \beq
 \ol{S}_{ul,ij}= \frac{1}{2}\left(  \partial_i \ol{u}_j + \partial_j \ol{u}_i  \right), \quad
 \ol{\Omega}_{ul,ij}=\frac{1}{2}\left(  \partial_i \ol{u}_j - \partial_j \ol{u}_i  \right), \quad
 \eeq
 \beq
 \ol{S}_{bl,ij}= \frac{1}{2}\left(  \partial_i \ol{b}_j + \partial_j \ol{b}_i  \right), \quad
 \ol{\Omega}_{bl,ij}=\frac{1}{2}\left(  \partial_i \ol{b}_j - \partial_j \ol{b}_i  \right). 
 \eeq 
 An in depth discussion of different models used can be found in \cite{muller2002dynamic,miesch2015large}. 
 %We note that since there are two stress tensors $\nabla \fu, \nabla \fb  $ the isotropic eddy-viscosities can depend not only on their invariant (under rotations) properties but also on their relative orientation
 %, and their relative orientation with respect to $\fb$.
 %
 %\item[(ii)]  
 %Although 
 Furthermore, although
 the system is still Galilean invariant, it is not invariant under transformation $\bf b \to b+B_0$ where $\bf B_0$ is a constant in space magnetic field. \NEW{The statistics of the sub-filter scale fields  can depend on the local value of $\fb$  and as a result
 so will the relevant eddie-viscosities that attempt to model their effect.} 
 %We note that while $\btau_\ell^{uu},\btau_\ell^{ub},\btau_\ell^{bb}$ remain unchanged under the transformation $\bf b \to b+B_0$ the dynamics of the small scale fields ${\bf b}'={\bf b}-\fb$, ${\bf u}'={\bf u}-\fu$ will depend on $\bf B_0$ making the statistics  of $\btau_\ell^{uu},\btau_\ell^{ub},\btau_\ell^{bb}$ depend on $\bf B_0$. 
 %
% \item[(iii)] 
Finally,
in the presence of helicity the small scales are known to transfer energy to the large scales in a mean way by the alpha dynamo mechanism that modeling needs to take into account.
%in which case tha magnetic field tensors are modeled so that  
%\beq 
%\nabla \cdot (\btau_\ell^{ub}-\btau_\ell^{bu}) = 
%\nabla \times (\alpha \fb) 
%\eeq 
% 
% \item[(iv)] 
% Finally,
% since three tensors that represent the different sub-scale stresses, there are  three eddy viscosity/diffusivity   coefficients to be used to close the relationships. 
 %Two of them are clearly understood. One represent the relation between $\btau_\ell^{uu}$ and $\fS_{u\ell}$; The second relates  $\btau_\ell^{bb}$ to $\mathbf{\Omega}_{b\ell}$. The first is therefore an eddy-viscosity coefficient responsible for kinetic-energy dissipation, while the second is an eddy-resistivity linked to Ohm dissipation. These coefficients are therefore related to large-scale energy, and should be positive-defined. The other two couple instead the velocity with the magnetic fields, and are therefore connected with helicity $h=\fu \cdot \fb$. While in flows with small helicity these two terms might be considered as negligible \citep{miesch2015large}, in general that should not be true.

%\end{enumerate}
Given the large number of the properties of the smoothed fields that the modeling of the sub-scale turbulent stresses can depend on, it is important to try to limit the possibilities identifying the parameters that play the most important role. This is what we are attempting to do in the following sections using direct numerical simulations for which the turbulent stresses can be directly measured.
%and their relations with the field amplitude and  gradients can be tested.
\NEW{ Thus, we need to stress that we do not test a posteriori any proposed model. Instead we are trying a priori to find relations that hold between the local energy flux and the gradients of the flow so that
they can be used in the construction of new models.}

%%%%%%%%%%%%%%%%%%%%%%%%%%%%%%%%%%%%%%%%%%%%%%%%%%%%%%%%%%%%%%%%%%%%%%%
%%%%%%%%%%%%%%%%%%%%%%%%%%%%%%%%%%%%%%%%%%%%%%%%%%%%%%%%%%%%%%%%%%%%%%%%%%%%%%%%%%%%%%%%%%%%%%%%%%%%%%%%%%%%%%%%%%%%%%%%%%%%%%%%%%%%%%%%%%%%%%
%%%%%%%%%%%%%%%%%%%%%%%%%%%%%%%%%%%%%%%%%%%%%%%%%%%%%%%%%%%%%%%%%%%%%%%
\section{Numerical Simulations}     %%%%%%%%%%%%%%%%%%%%%%%%%%%%%%%%%%%
\label{sec:results}                  %%%%%%%%%%%%%%%%%%%%%%%%%%%%%%%%%%%
%%%%%%%%%%%%%%%%%%%%%%%%%%%%%%%%%%%%%%%%%%%%%%%%%%%%%%%%%%%%%%%%%%%%%%%
%%%%%%%%%%%%%%%%%%%%%%%%%%%%%%%%%%%%%%%%%%%%%%%%%%%%%%%%%%%%%%%%%%%%%%%
\subsection{Numerical Set-up}
To investigate the local energy fluxes described before we use the results from direct numerical simulations DNS \NEW{using the pseudo-spectral code {\sc ghost} \citep{mininni2011hybrid} }
evolving the MHD equations (\ref{momentum})-(\ref{divb}) in a cubic triple periodic domain of side $L=2\pi$ \NEW{so that $|\bf k|=1$ give the smallest non-zero wavenumber }. 
The forcing is limited only to wave numbers with $|{\bf k}|\le k_f=2$ in all cases and was random and delta correlated in time so that the mean energy injection rate
$\epsilon= \left\langle {\bf u \cdot f_u}\right\rangle 
         + \left\langle {\bf b \cdot f_b}\right\rangle=1$ 
is independent of the flow state. All runs have the same energy injection rate and unit Prandtl number with $\nu=\mu=0.0001$.
\NEW{
The involved Reynolds number based on the input parameters $\epsilon,L$ and $\nu$ is given by $Re=\epsilon^{1/3}k_f^{4/3}/\nu$ that is thus fixed to $Re=3968$.}
The resolution  was fixed for all runs at $N=1024$ grid points in each direction \NEW{ that was sufficient for the flow to be well resolved so that an exponential decrease of the energy spectrum is observed at large $k$.
%M{\color{red}(the mesh size $\Delta x$ is of the order of the Kolmogorov scale)} 
No artificial dissipation was used.}

Four different cases were considered. In the first, no magnetic forcing was introduced and the magnetic field was set identically to zero $\bf b=0$ so that the flow reduced to a hydrodynamic run, repeating essentially the results of \cite{alexakis2020local}. This run is used as a reference to identify the differences of MHD from hydrodynamic runs and referred in the subsequent figures as `Hydro'.  The remaining three runs were designed to have different levels of magnetic energy as in \cite{alexakis2013large}. % that can alter the statistical properties of the flow.
The second run was thus a dynamo run. No magnetic forcing was used but an initial
small magnetic field was introduced that was amplified by the turbulent motions until a steady state was reached where the magnetic energy fluctuated around a mean value. We note that the forcing was mirror symmetric so no helicity is injected in the system. A result of the absence of helicity and the randomness of the flow is that magnetic energy is concentrated in the small scales. This run is referred to in the figures as `dynamo'.
In the third run a magnetic forcing was also used of equal magnitude as the mechanical flow. The resulting flow at steady state has magnetic energy at equipartition with the kinetic energy at all scales. The helicity and magnetic helicity injection rate for this flow is also set to zero.
Results from this run are referred to as `moderate'. 
Finally the fourth flow has also both mechanically and magnetically forced but in this case the magnetic helicity was weak but not zero. This has lead to the formation of a slowly growing large scale helical magnetic field. As a result this flow has magnetic energy that exceeds the kinetic energy of the flow.  
This run is referred to in the figures as `strong'.

\NEW{It is worth noting that all the statistics we present in the following are obtained through time-averaging over a large amount of samples during the steady state to assure statistical convergence. 
Whenever possible, several field realizations have been used also to average over space, thanks to spatial homogeneity. 
}
%% FIG 1
\begin{figure}
\centering
\includegraphics[width=0.4\textwidth]{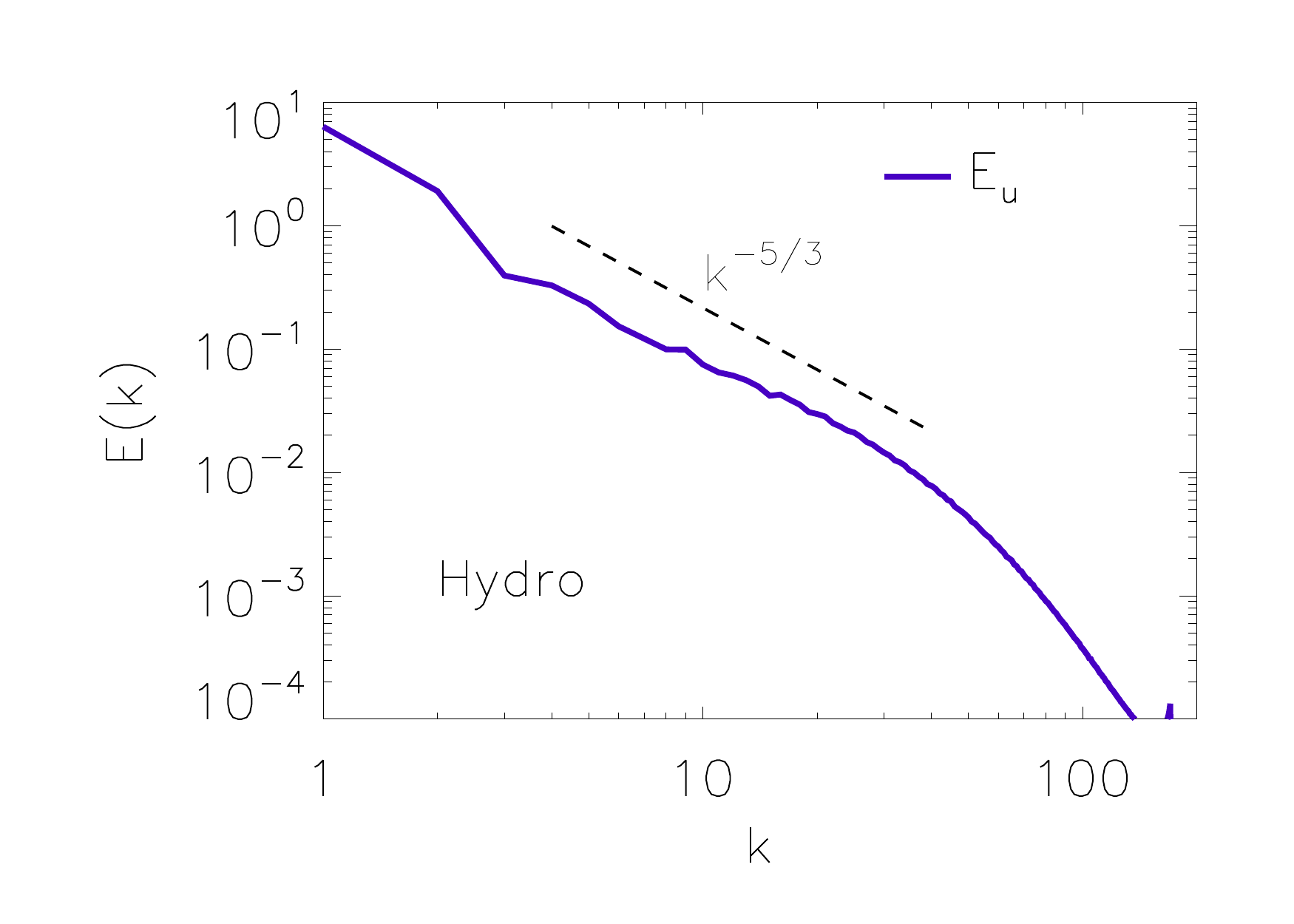}
\includegraphics[width=0.4\textwidth]{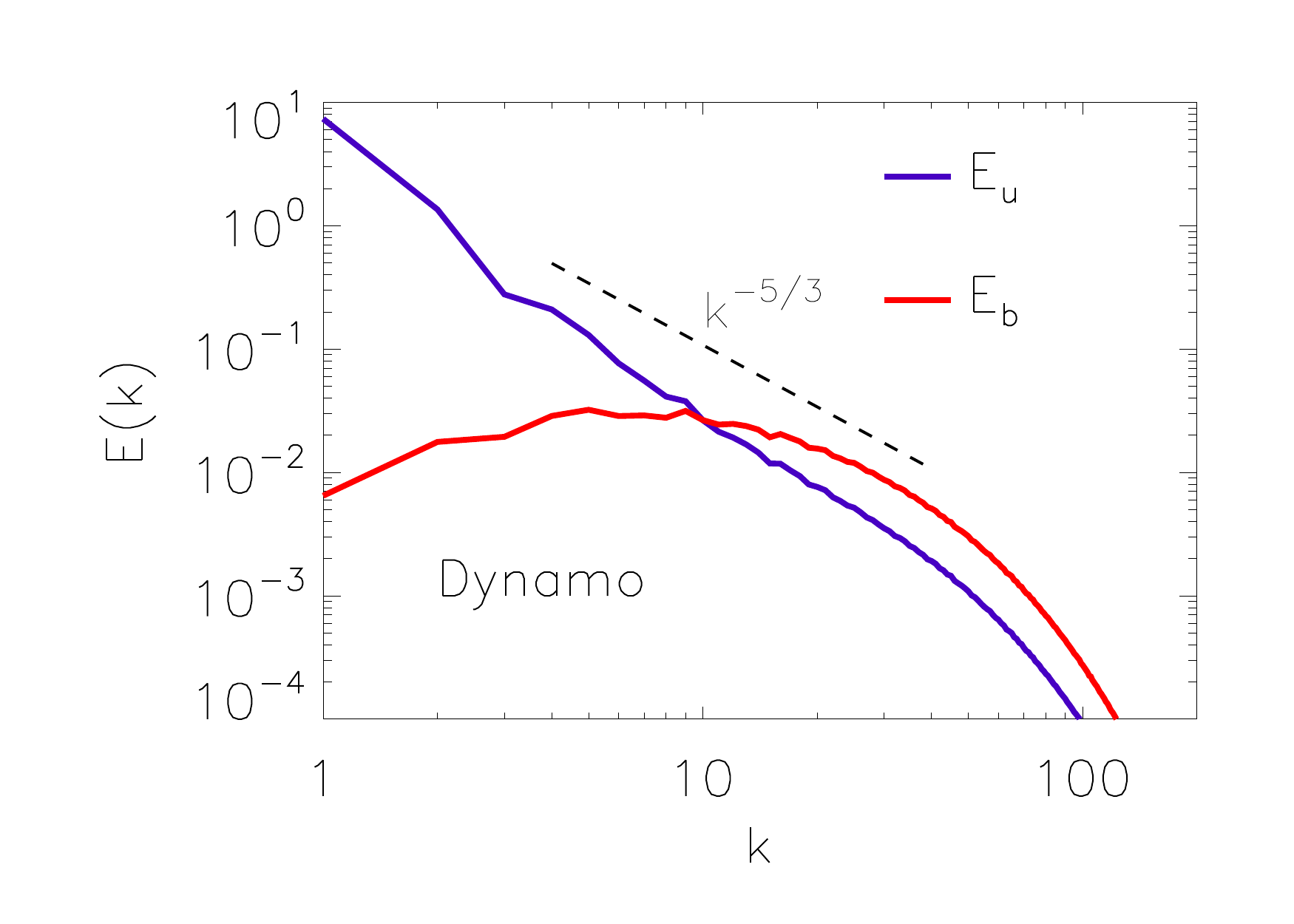}
\includegraphics[width=0.4\textwidth]{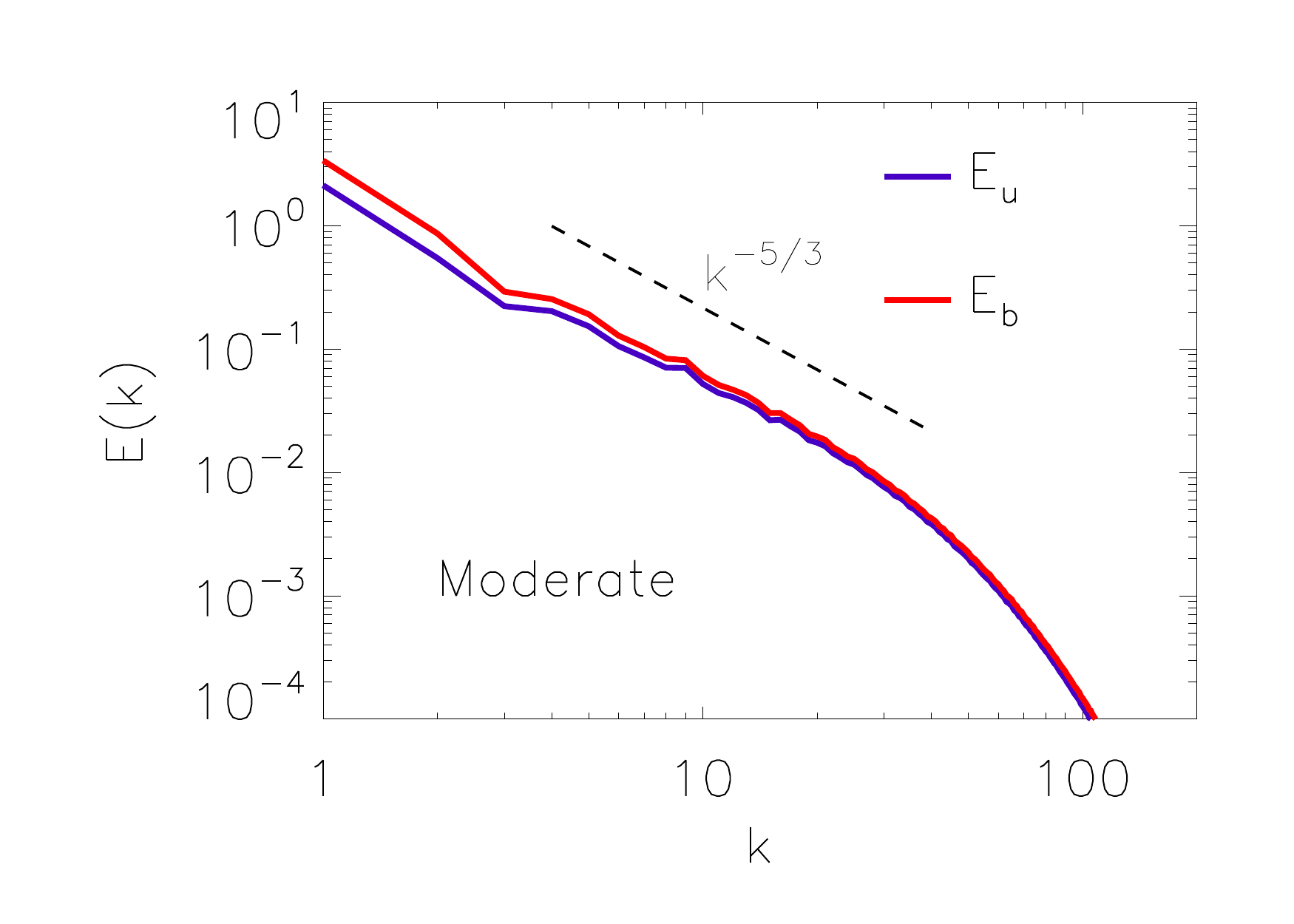}
\includegraphics[width=0.4\textwidth]{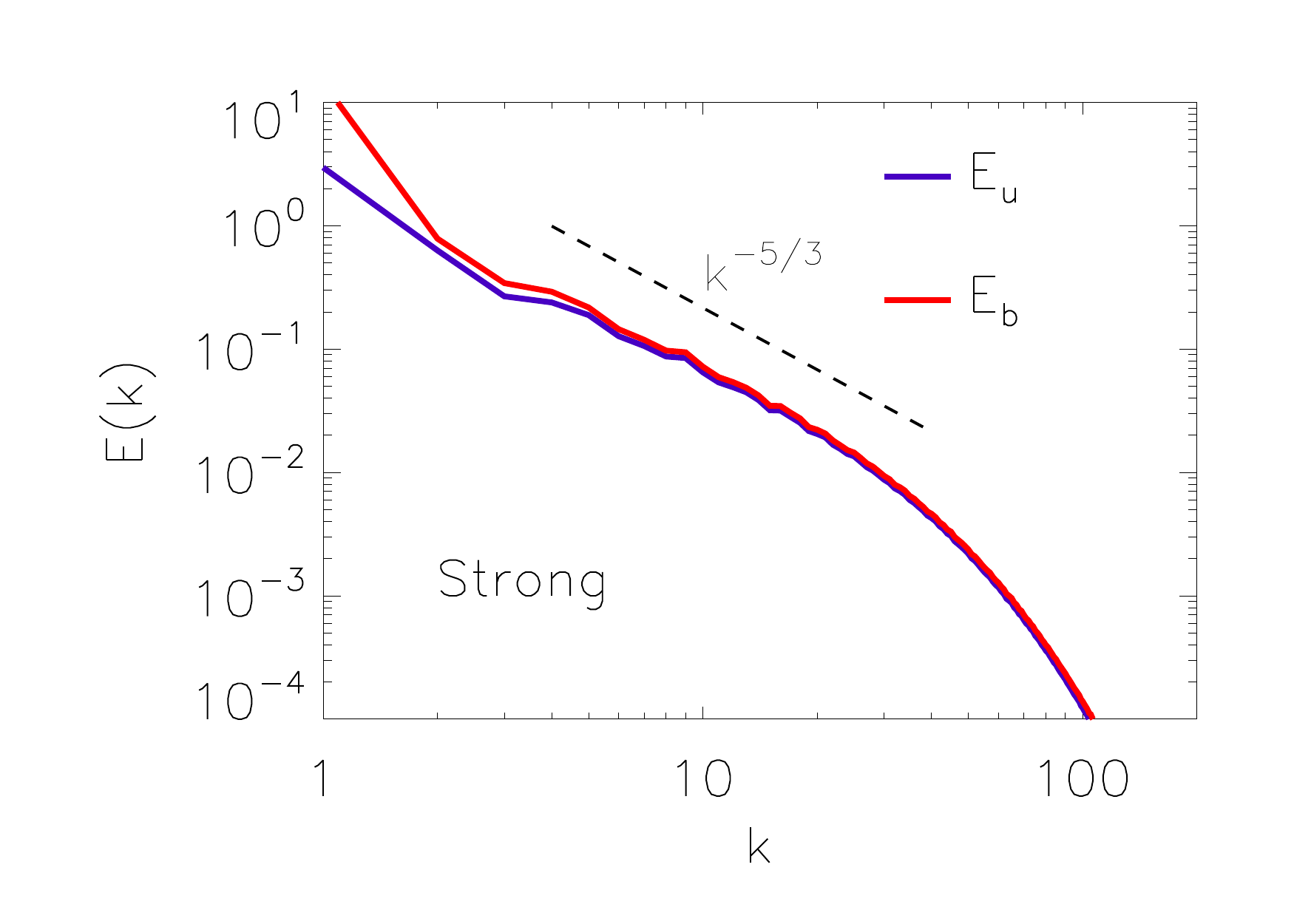}
\caption{Energy Spectra for the 4 different cases examined. %\NOTE{Axis labels in the first figure are different (not power)}
}
\label{fig:Spectra}
\end{figure}
%
%All the statistical observables considered in the present work have been obtained through time-averaging. 
\subsection{Energy Spectra and Magnetic field}
Figure \ref{fig:Spectra} shows the resulting \NEW{time-averaged} energy spectra for the four runs at the steady state. \NEW{For the dynamo run in particular analysis is made at the state such that magnetic energy does not grow anymore. } All runs show energy spectra compatible with a $k^{-5/3}$ power-law exponent. The magnetically forced runs `moderate' and `strong' have equipartition magnetic energy across wave numbers except for $k=1$ in the 'strong' case for which magnetic energy is larger.  
For the dynamo run the magnetic energy is weaker than kinetic at small wave numbers but the reverse is true for the large wavenumbers. This behavior is typical for randomly forced non-helical small scale dynamos \citep{schekochihin2004simulations,moll2011universality,brandenburg2012current}. 
Dynamo flows with steady non-helical forcing, like Taylor-Green for example \citep{ponty2008linear}, produce spectra closer to the `moderate' run.
This figure points out also that a quasi-inertial range 
is roughly displayed over about a decade between $k=4$ and $k=64$, as standard with present resolution.
%The spectrum shape indicates that in all configurations studied, including that with non-zero cross helicity and large-scale magnetic field, effects related to Alf\'en waves are globally negligible.

%% FIG 2
\begin{figure}
\centering
\includegraphics[width=0.3\textwidth]{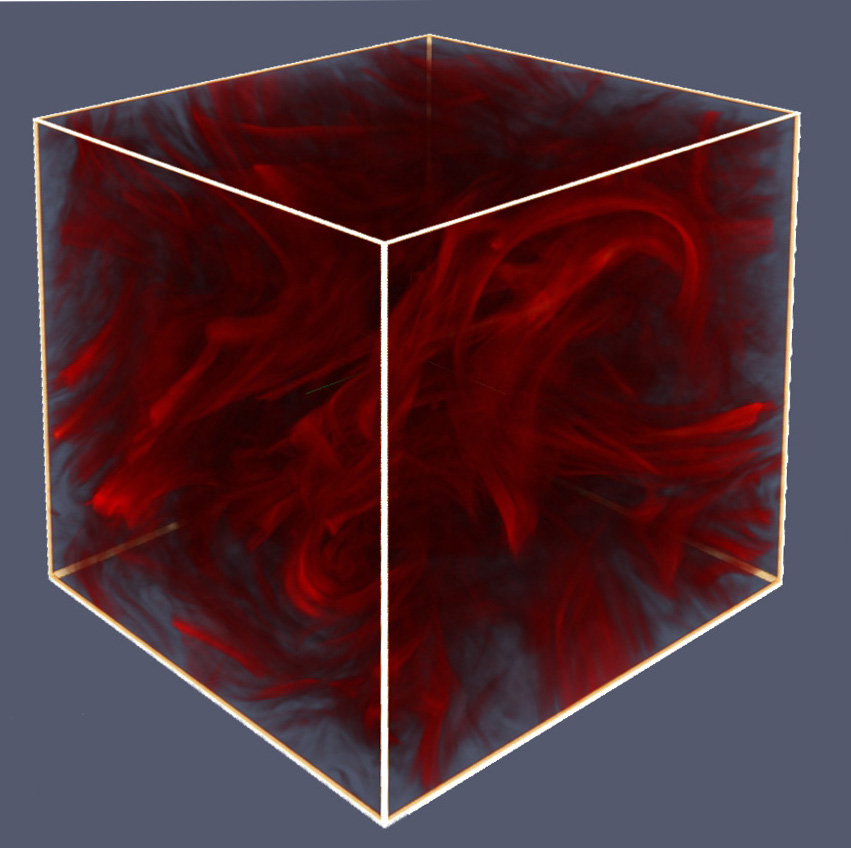}
\includegraphics[width=0.3\textwidth]{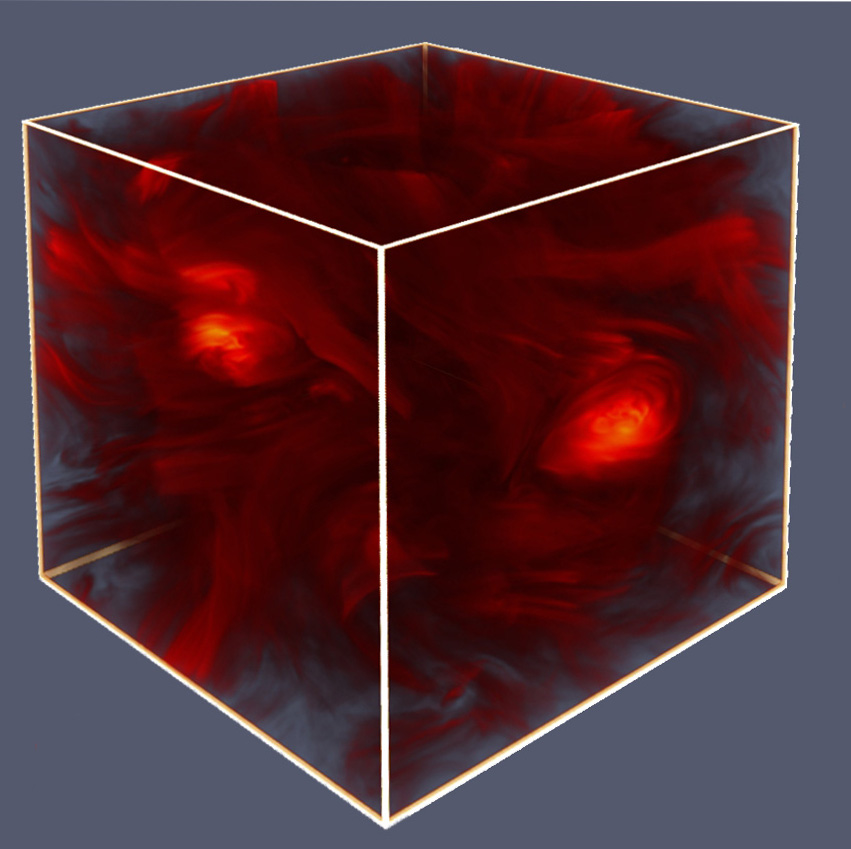}\\
\includegraphics[width=0.3\textwidth]{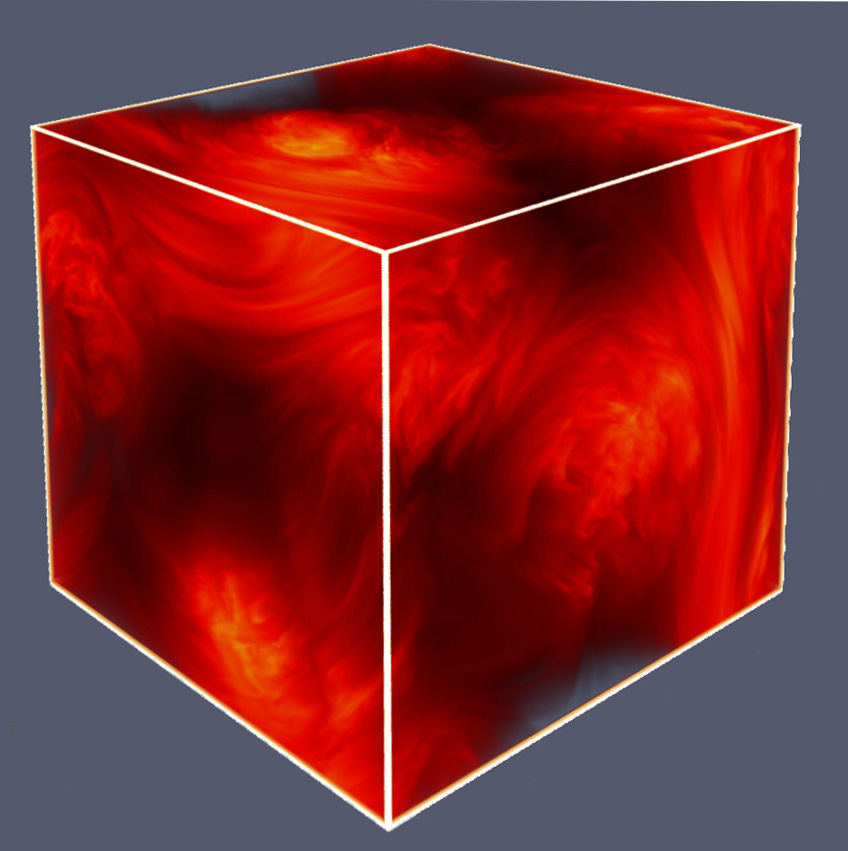}
\includegraphics[width=0.3\textwidth]{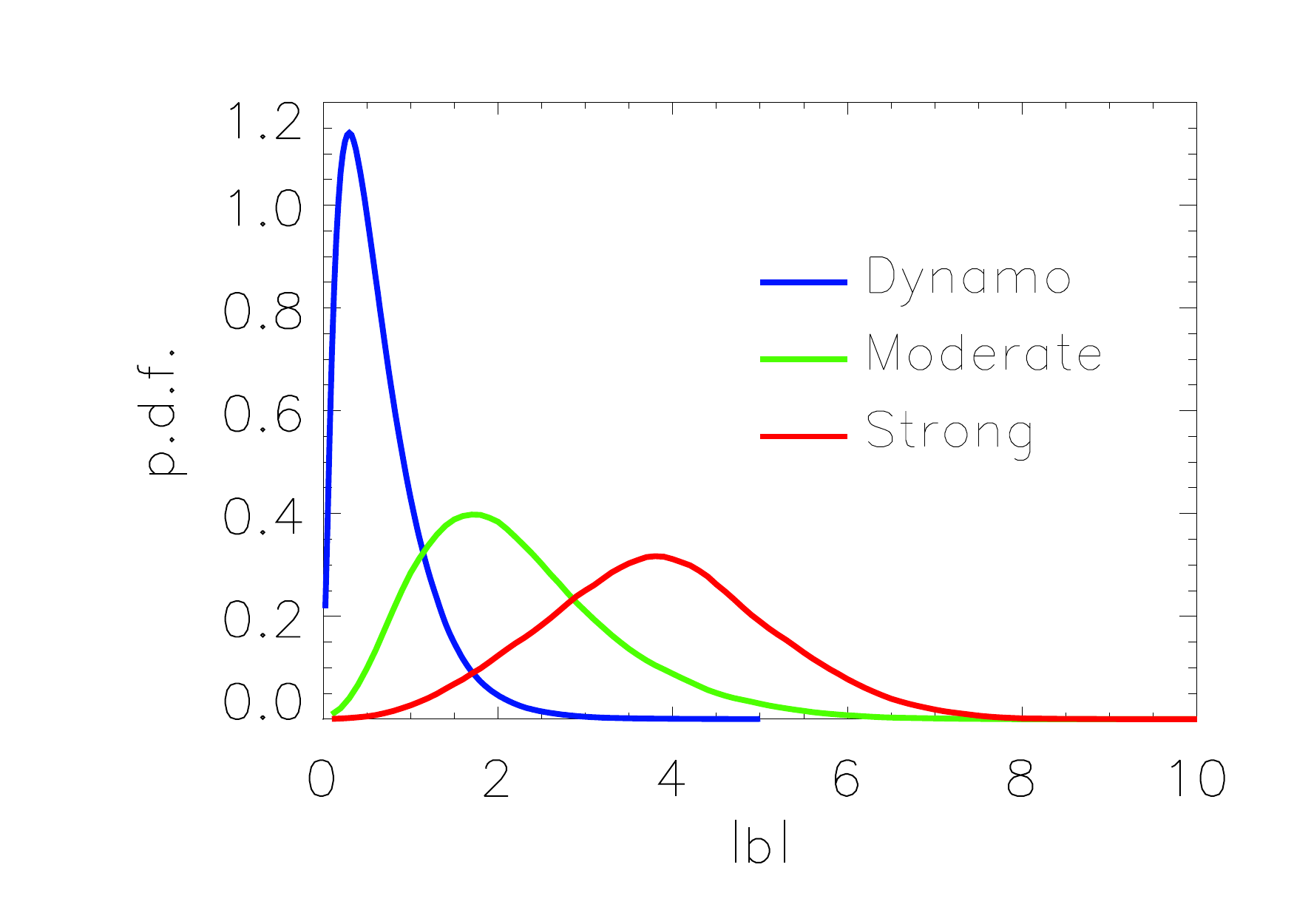}
\caption{ \NEW{Top panels: Visualisations of the
magnetic energy density for
the dynamo (top left)
the moderate case (top right) and
the strong case (bottom left).
The bottom panel shows 
the PDF of $|{\bf b}|$ for the 3 different MHD flows examined.}
% HDO is the Hydro run, 
% MHD0 is a Dynamo Run with $E_b \ll E_u$,
% MHD1 $E_b\simeq E_u$, (mechanically and magnetically forced)
% MHD2 $E_b\gg E_u$   (mechanically and magnetically forced with small helicity injection). 
}
\label{fig2}
\end{figure}
In figure \ref{fig2}, it is possible to get some qualitative insights on the different cases studied. Notably, the visualization of the magnetic energy permits to observe the structure of the large scales, through its space distribution. 
The dynamo case displays very little dispersion of the magnetic field amplitude, and energy is concentrated in flux tubes. 
%no large-scale magnetic field.
The moderate case shows a  larger distribution and some regions of intense field. 
Finally the strong case shows a much more widely distributed energy, with some regions characterised by large values of the magnetic field, and in particular large-scale structures are visible. 
%Furthermore, the current-sheets typical of MHD turbulence are visible.
%
To give a quantitative picture, we also show in figure \ref{fig2} the probability distribution of the magnetic amplitude $|\bf b|$ for the three MHD cases examined.
In the dynamo case the maximum is at small $|{\bf b}|$, with a peaked distribution. The width of the distribution increases in the other two cases, with a broad distribution for the strong case.
In this last case, the tails appear substantial.
%%%%%%%%%%%%%%%%%%%%%%%%%%%%%%%%%%%%%%%%%%%%%%%%%%%%%%%%%%%%%%%%%%%%%%%
%%%%%%%%%%%%%%%%%%%%%%%%%%%%%%%%%%%%%%%%%%%%%%%%%%%%%%%%%%%%%%%%%%%%%%%
\section{Statistics of the local energy flux}  %%%%%%%%%%%%%%%%%%%%%%%%
\label{sec:loc-flux}             %%%%%%%%%%%%%%%%%%%%%%%%%%%%%%%%%%%%%%%%%
%%%%%%%%%%%%%%%%%%%%%%%%%%%%%%%%%%%%%%%%%%%%%%%%%%%%%%%%%%%%%%%%%%%%%%%
%%%%%%%%%%%%%%%%%%%%%%%%%%%%%%%%%%%%%%%%%%%%%%%%%%%%%%%%%%%%%%%%%%%%%%%
%% FIG 3
\begin{figure}
\centering
\includegraphics[width=0.47\textwidth]{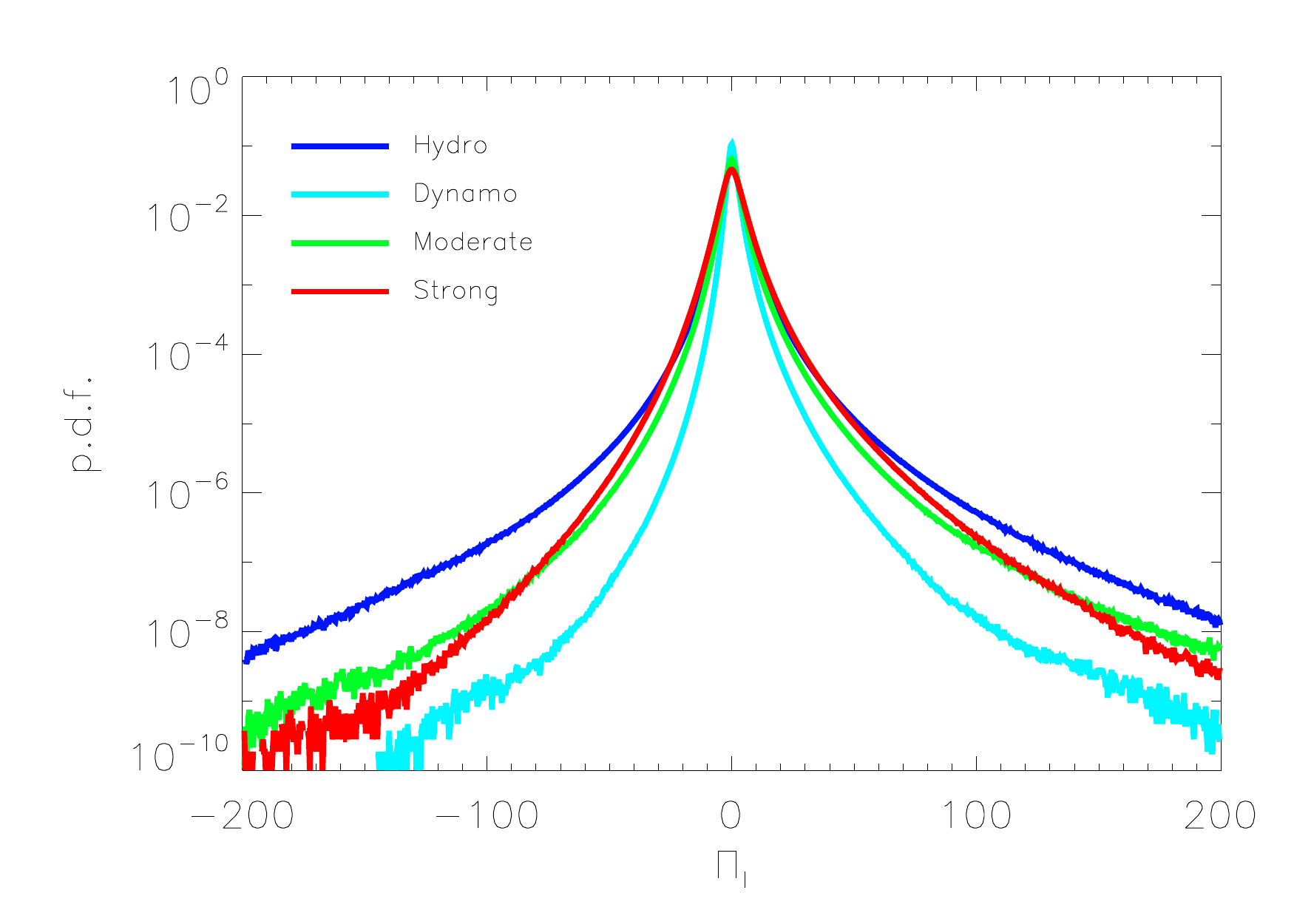}
\includegraphics[width=0.47\textwidth]{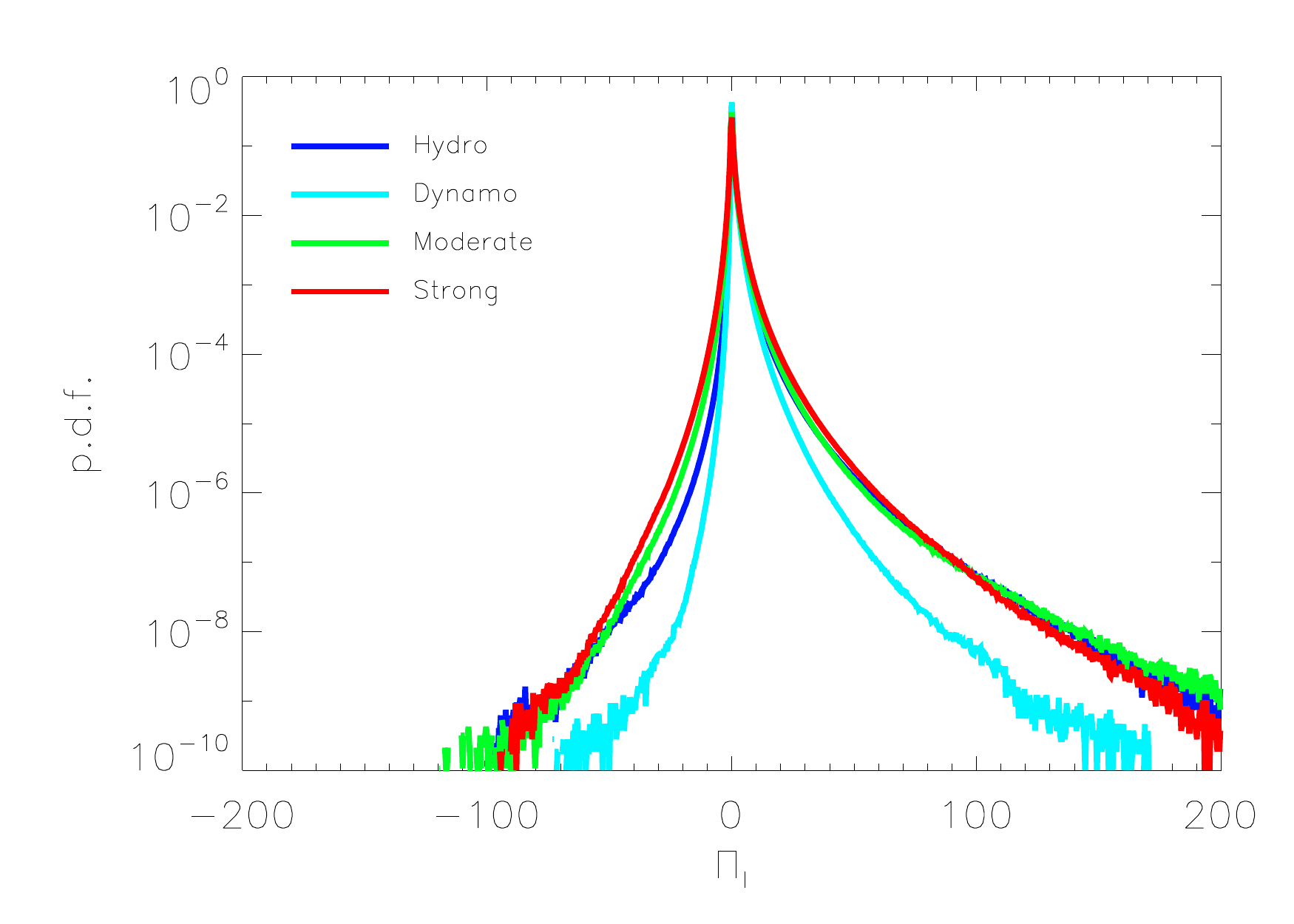}
\caption{Pdf of the total nonlinear flux $\Pi_\ell(x)$ at $q=32$ for a sharp spectral filter on the  left and a Gauss filter on the right, for the 4 different runs.
%, 
%  HDO is the Hydro run, 
% MHD0 is a Dynamo Run with $E_b \ll E_u$,
 %MHD1 $E_b\simeq E_u$, (mechanically and magnetically forced)
 %MHD2 $E_b\gg E_u$   (mechanically and magnetically forced with small helicity injection).
  }
\label{fig3}
\end{figure}
%% FIG 4
\begin{figure}
\centering
\includegraphics[width=0.48\textwidth]{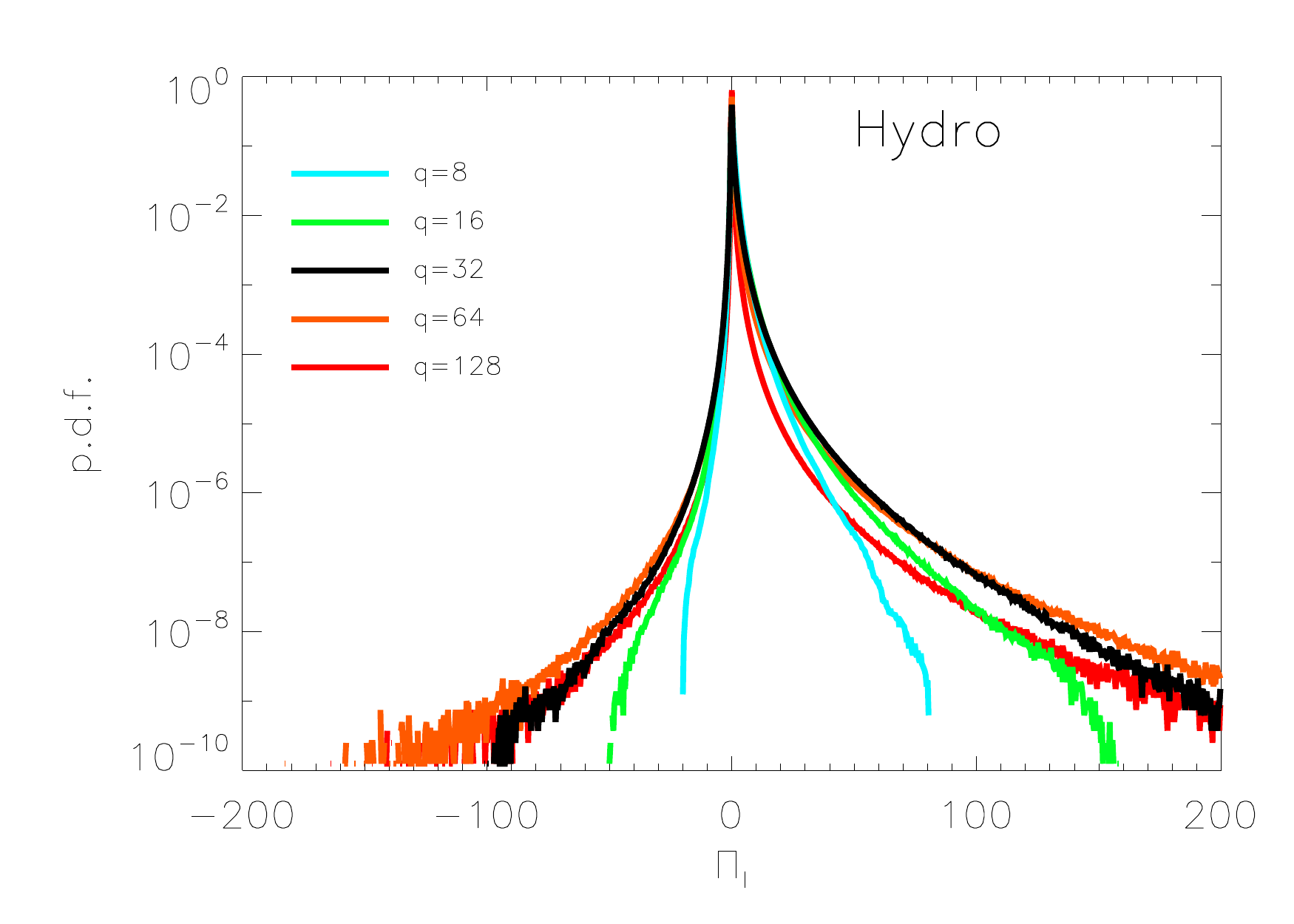}
\includegraphics[width=0.48\textwidth]{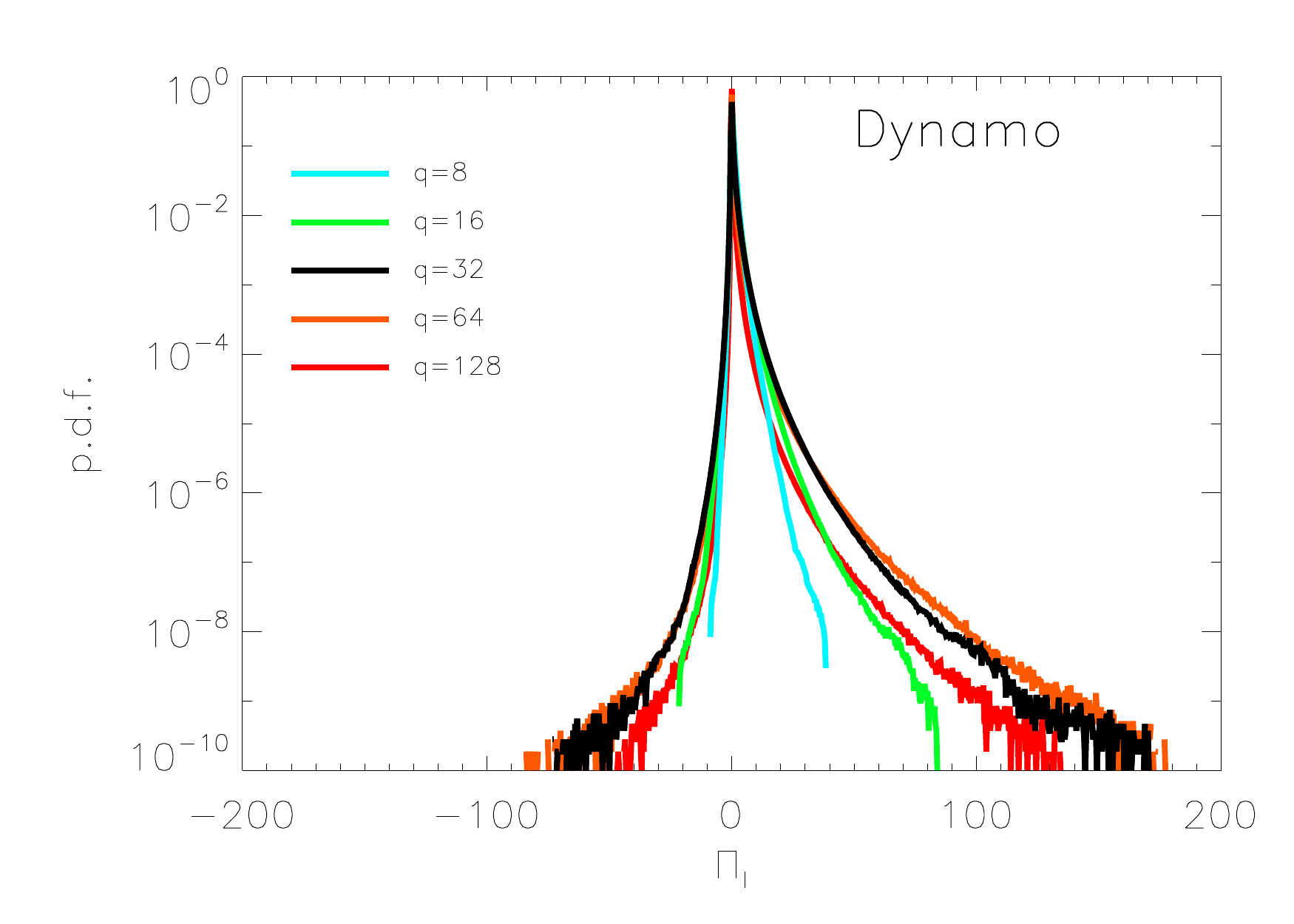}
\includegraphics[width=0.48\textwidth]{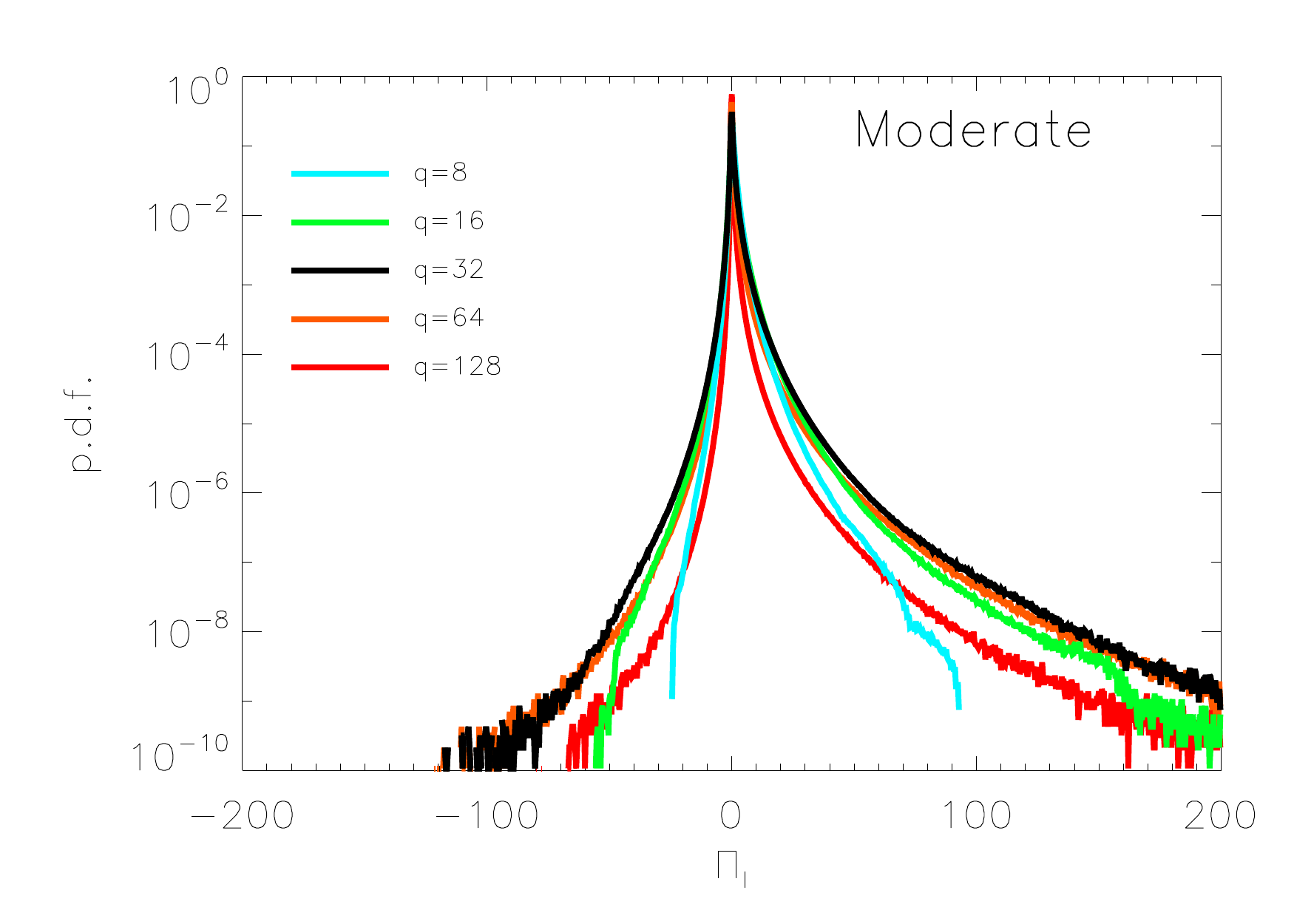}
\includegraphics[width=0.48\textwidth]{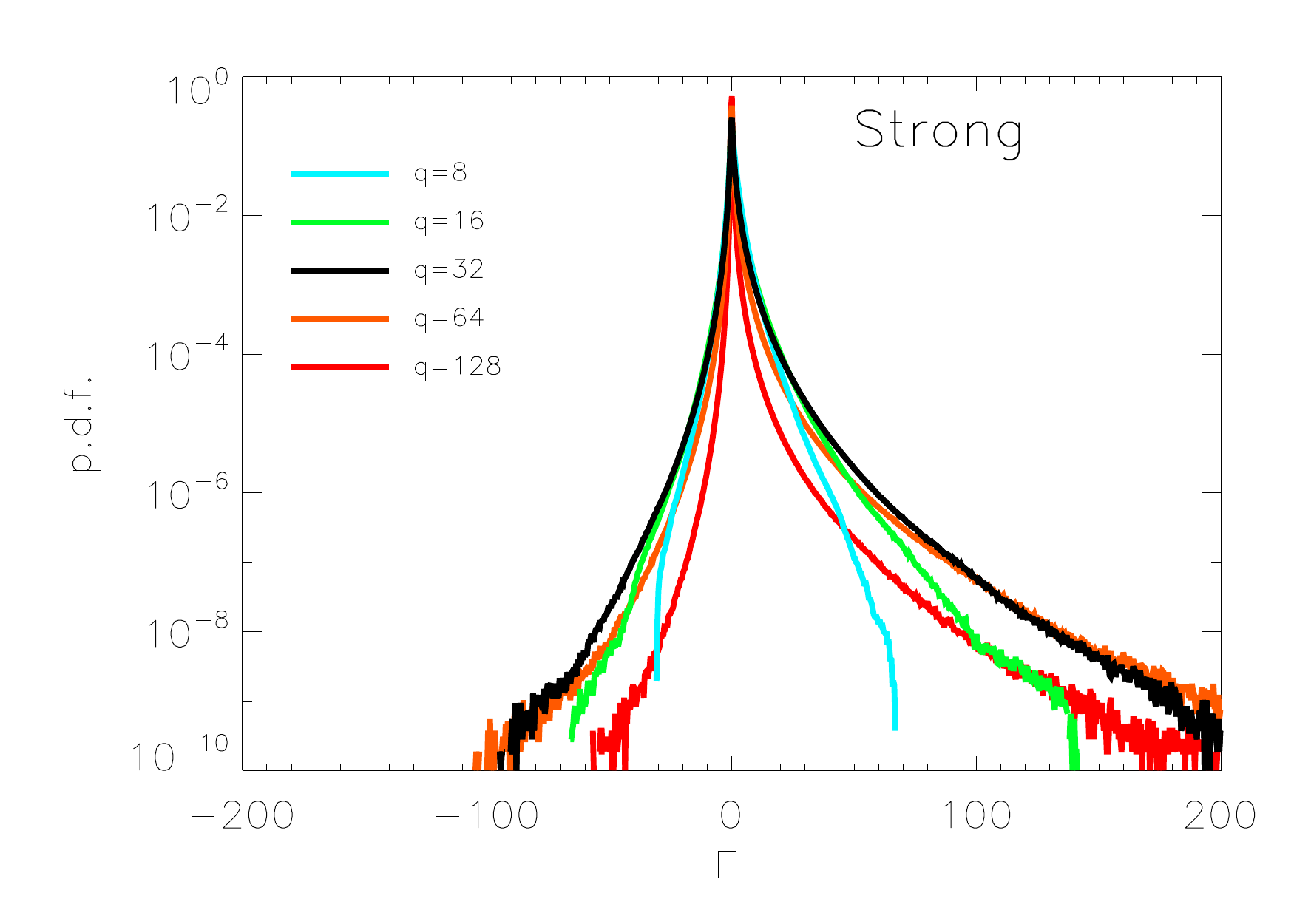}
\caption{ Pdf of the Energy flux for different $q$ for the three MHD cases. In the first panel, the hydro case is also plotted for comparison.}
\label{fig4}
\end{figure}
The local energy fluxes 
$\Pi_\ell^{uu}(x)$, 
$\Pi_\ell^{ub}(x)$,
$\Pi_\ell^{bu}(x)$, 
$\Pi_\ell^{bb}(x)$  and $\Pi_\ell(x)$ were calculated for the four different runs at their steady state. 
%\NEW{ For each run several field realisations at different time intervals were used to calculate the probability density function (p.d.f) for the values of $\Pi_\ell(x)$ in space and time.}
Figure \ref{fig3} shows this pdf for the four different runs using a sharp-spectral filter (left) and Gaussian filter (right) for a filter length $q=32$, 
which corresponds to a value at around the end of the inertial range.

%Figure \ref{fig3} points out that, 
As in hydrodynamics, in MHD the sharp-spectral filter leads to a wider and more symmetric pdf of $\Pi_\ell$. Interestingly, the pure hydro curve presents the strongest tails, while with the Gaussian filter the profiles were found always similar. 
\NEW{Considering the representation of the fluxes, the sharp-filter is thus found to enhance fluctuations displaying many negative events, making it \NEW{more} difficult to use directly in LES. The reason for this behavior is the fact that the sharp filter is not localised in space and not positive-definite. For this reason, it appears less useful for the analysis of the energy fluxes and it is abandoned in the following.
In this sense, the present analysis confirms previous results obtained in the pure hydrodynamic case~ \citep{buzzicotti2018effect,alexakis2020local}.
}

The Gaussian filter leads for all four cases to a skewed pdf with non-Gaussian (stretched exponential) tails. The tails for all cases extend to values hundred times more than the mean value $\langle \Pi_\ell \rangle=1$, implying that, although rare, these events can play a significant role in the dynamics.
The most notable result when comparing the four different cases is that the dynamo run has weaker tails than the other flows.  It turns out hence that the dynamo dynamics suppresses efficiently extreme energy flux fluctuations.  
Small differences are experienced between the moderate and strong cases, indicating that fluid fluctuations are not affected by the large helical magnetic field, which is present only in the strong case.
It is possible to observe just a slight increase in the probability of negative events with respect to positive ones. 
Still, not much difference is found neither comparing the moderate/strong cases to the hydro case.
%Although this case appears a little less intermittent, all the curves are similar.

The statistical behaviour of the energy flux $\Pi_\ell$ is further studied in Figure \ref{fig4}. 
Here we present the probability density function of the flux $\Pi_\ell$ at different scales $\ell=1/q$,
%\emph{i.e.} wavenumbers $q$, 
for the 4 different cases. 
The hydro results are given for comparison.
The first remark is that as said above, the dynamo case is quite different from the others, with more shrunk distributions, though the distribution remains tailed and skewed. All the cases displays similar changes with going towards smaller scales. The profiles are much less wide in the large-scale range $q=8,16$, notably for $q=8$. Then, the distributions are basically indistinguishable for $q=32,64$, which are more or less in the far inertial range, and finally the tails are again reduced for $q=128$, which is in the dissipation range. As expected, intermittency is maximum at the end of the inertial range. Slight differences between the moderate and the strong case can be detected, yet not significant. %With respect to the simpler Hydro case, it would seem that the magnetic term has a small impact at very small scales, with a depression of the tails. 

%% FIG 5
\begin{figure}
\centering
\includegraphics[width=0.48\textwidth]{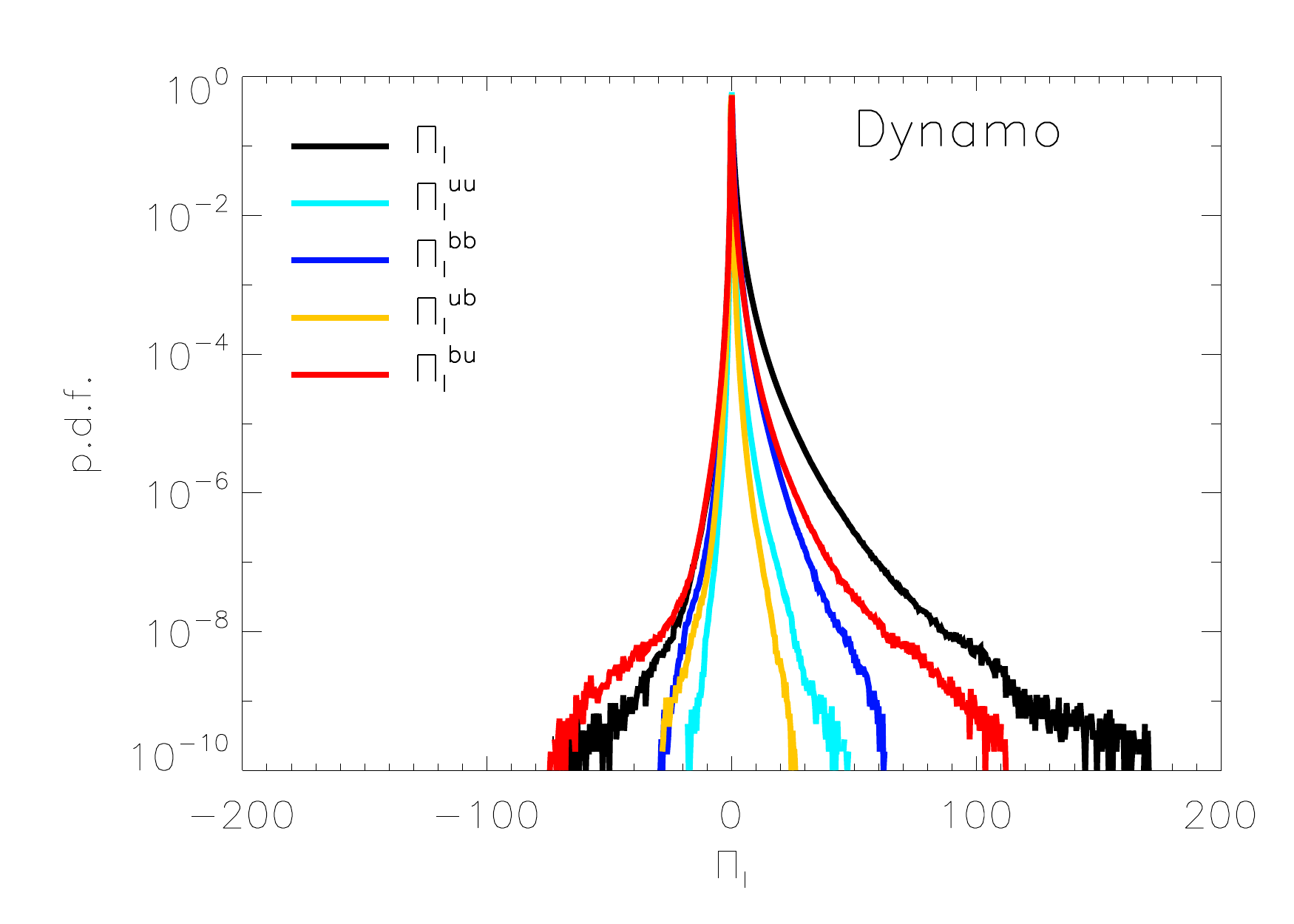}
\includegraphics[width=0.48\textwidth]{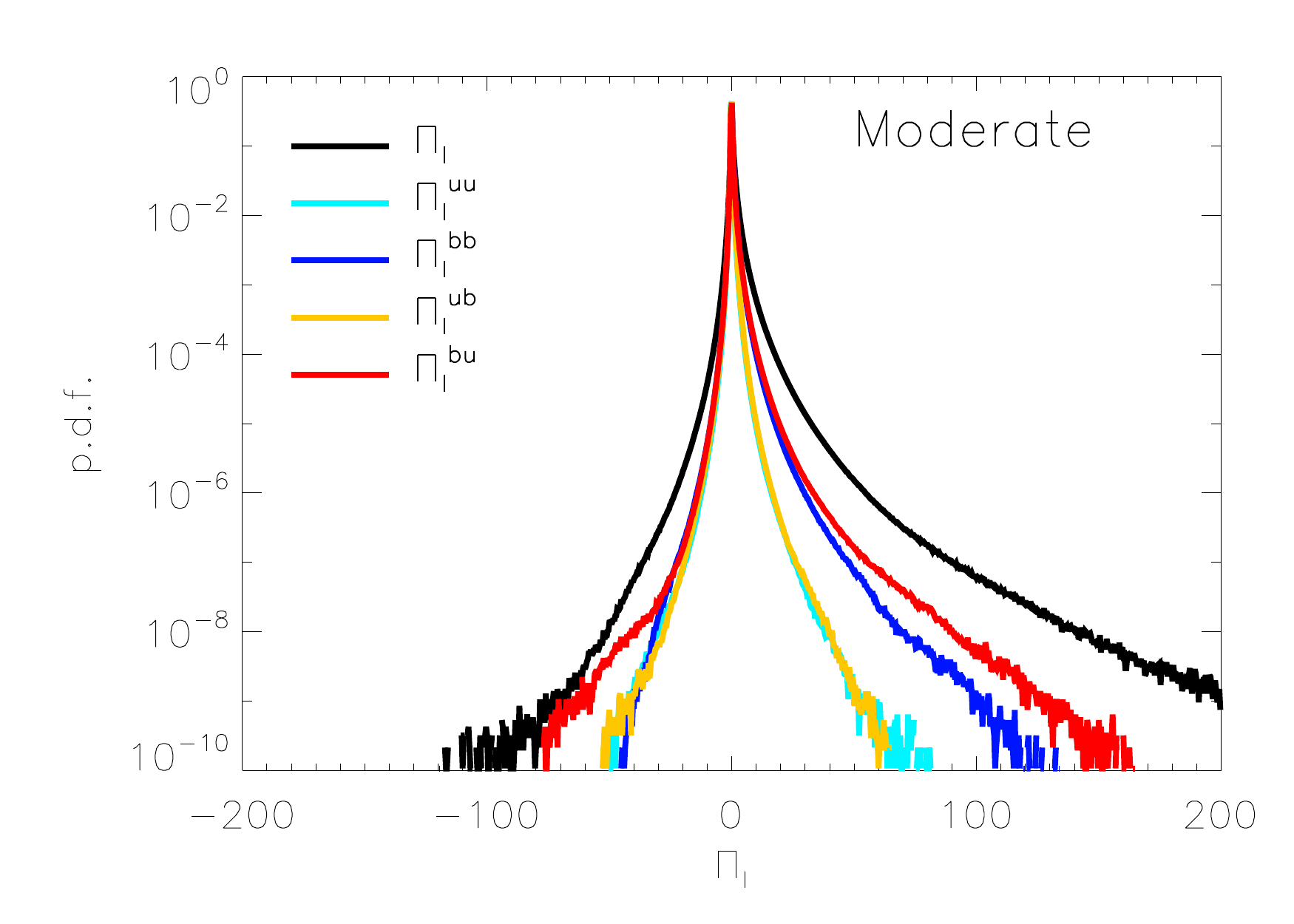}
\includegraphics[width=0.48\textwidth]{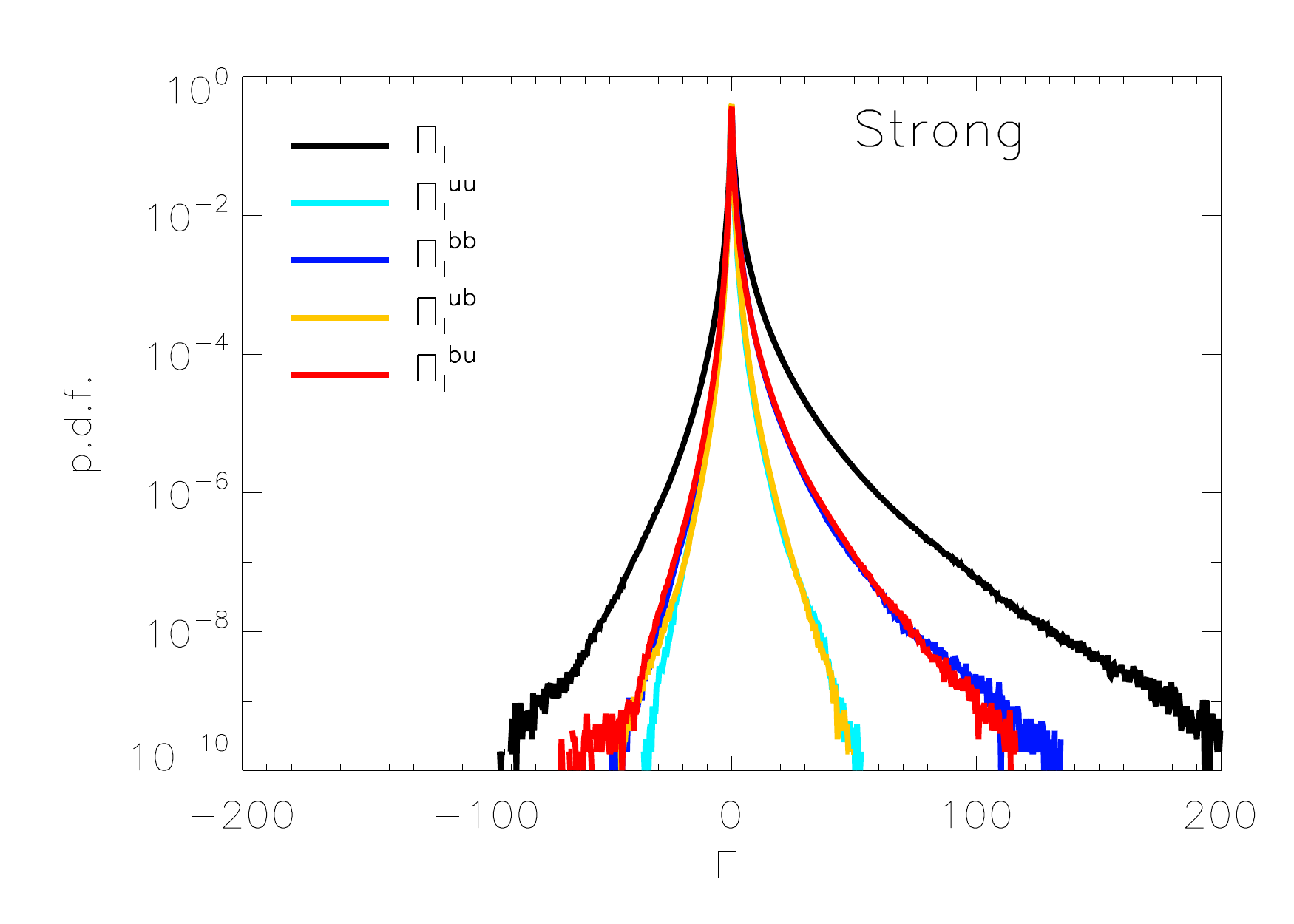}
\caption{The Different components of the Energy flux \NEW{for $q=32$}. The total flux is also plotted for information. The three MHD cases are considered.   }
\label{fig5}
\end{figure}
Now, we turn our attention to the four different components of the total energy flux  $\Pi_\ell(x)$. In Figure \ref{fig5}. We compare all the terms 
$\Pi_\ell^{uu}(x)$, 
$\Pi_\ell^{ub}(x)$,
$\Pi_\ell^{bu}(x)$, 
$\Pi_\ell^{bb}(x)$
composing $\Pi_\ell$ as given by Eqs. (\ref{flux1})-(\ref{flux4}). %The total  energy flux $T_\ell=\Pi_\ell^u+\Pi_\ell^b$ is also plotted.
\NEW{The fluxes show different behavior in the three different cases examined.}
%We note that in the dynamo case there is a mean transfer of kinetic energy to magnetic energy both by the $\Pi^{bb}_\ell$ flux and
%the $\mW_L$ term in the large scales (that we do not examine here).
%This alters the transfer balance between the different cases.

The first striking observation is that in all cases the pure hydrodynamic flux, 
%due to the work of the sub-scale stresses against large-scale strain, is the smallest one. 
%\NEW{({\it ie} the flux caused by the work of the sub-scale velocity stresses against large-scale velocity strain), 
is the smallest one.
Most notably, it shows the least important negative fluctuations. 
It was not possible to infer that from looking only at the total flux $\Pi_\ell$, which is found in Fig. \ref{fig2} to be similar in the hydro and all the MHD cases.
In the moderate and strong cases, the hydrodynamic flux $\Pi_\ell^{uu}$ is basically the same as the cross magnetic-hydro flux $\Pi_\ell^{ub}$, while the two are different for the dynamo case.
As a consequence, in all cases the terms found to be dominant are 
 $ \Pi^{bu}_\ell$ and $ \Pi^{bb}_\ell$, that is the blue and red lines.
 Therefore, the coupling magnetic term contributing to the kinetic energy flux dominates the pure hydro term. Since the total flux in the hydro case is much similar, except for the dynamo case, it appears that the cascade readjusts to distribute much of the flux in this term, although the total kinetic flux remains unchanged.
% NEW
%The dominant term in the magnetic flux is related to large-scale magnetic field gradients parallel to the velocity \NEW{$\Pi^{bu}_\ell$}. Instead, the term in which the gradients are parallel to the magnetic field is always the smallest one. 
It is interesting to remark also that in the strong case, where both the velocity and the magnetic fields are fully developed, there is a nice symmetry between the two fluxes, with $\Pi_\ell^{uu}\approx \Pi_\ell^{ub}$, and $\Pi_\ell^{bu}\approx \Pi_\ell^{bb}$. 
That points out to two very similar cascade processes \NEW{ and could be related to the equipartition of the magnetic and kinetic energy cascade recently conjectured in \cite{bian2019decoupled}.}
\NEW{While for the moderate case this symmetry appears also holding apart from the extreme events, the dynamo case breaks it}  
%\NEW{and it is stronger at the Ohmic scales} \NOTE{I don't understand well the point about the Ohmic scale}
underlying different physics of the cascade process.

Finally, it is useful to note that all the terms, and hence the total flux, are production terms, that is the flux is in average positive towards smaller scales, even though there may be a significant local negative flux.
\NEW{We note also that in the dynamo case there is a mean transfer of kinetic energy to magnetic energy both by the $\Pi^{bb}_\ell$ flux and
the $\mW_L$ term in the large scales (that we do not examine here).
This may alter the transfer balance between the different cases.}
Concerning the negative fluxes, it is important to remark that, unlike the positive tails, the negative parts of the pdfs are basically the same for all components. This is most notably true for the strong case, but also in the other two cases the differences are limited to the region of extremely rare events, where statistical errors may be substantial.
That shows that a change in large scale amplitude of the magnetic field can affect the smaller scale forward cascade process. 
%% FIG X
%\begin{figure}
%\centering
%\includegraphics[width=0.48\textwidth]{RS_UB_gauss_uuu_MHD1.pdf}
%\includegraphics[width=0.48\textwidth]{RS_UB_gauss_bub_MHD1.pdf}
%\caption{ MHD RESULTS: Pdfs for different $q$  for `uuu' left and 
%`bub' right for the MHD1 run. 'uuu' is self-simiular! 'bub' not! }
%\label{fig4}
%\end{figure}

%%%%%%%%%%%%%%%%%%%%%%%%%%%%%%%%%%%%%%%%%%%%%%%%%%%%%%%%%%%%%%%%%%%%%%%
%%%%%%%%%%%%%%%%%%%%%%%%%%%%%%%%%%%%%%%%%%%%%%%%%%%%%%%%%%%%%%%%%%%%%%%
%%%%%%%%%%%%%%%%%%%%%%%%%%%%%%%%%%%%%%%%%%%%%%%%%%%%%%%%%%%%%%%%%%%%%%%
\section{Joint pdfs}         %%%%%%%%%%%%%%%%%%%%%%%%%%%%%%%%%%%%%%%%%%
\label{sec:pdfs}             %%%%%%%%%%%%%%%%%%%%%%%%%%%%%%%%%%%%%%%%%%
%%%%%%%%%%%%%%%%%%%%%%%%%%%%%%%%%%%%%%%%%%%%%%%%%%%%%%%%%%%%%%%%%%%%%%%
%%%%%%%%%%%%%%%%%%%%%%%%%%%%%%%%%%%%%%%%%%%%%%%%%%%%%%%%%%%%%%%%%%%%%%%
\subsection{Field Gradients}                              %%%%%%%%%%%%%
\label{fieldgrad}                                         %%%%%%%%%%%%%
%%%%%%%%%%%%%%%%%%%%%%%%%%%%%%%%%%%%%%%%%%%%%%%%%%%%%%%%%%%%%%%%%%%%%%%

%% FIG 6
\begin{figure}
\begin{center}
\begin{subfigure}[b]{0.5\textwidth}
\centering
\includegraphics[width=0.45\textwidth]{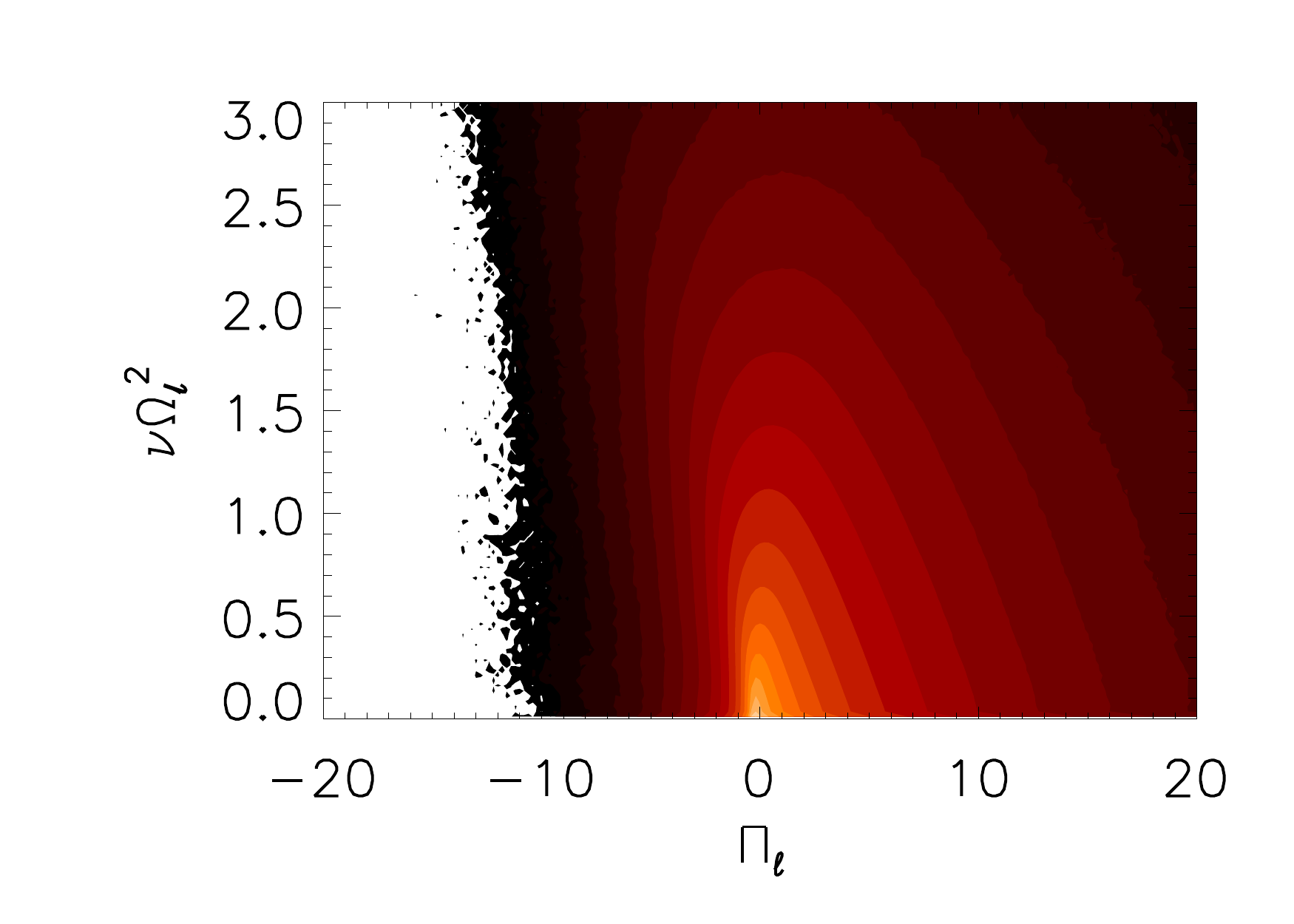}
\includegraphics[width=0.45\textwidth]{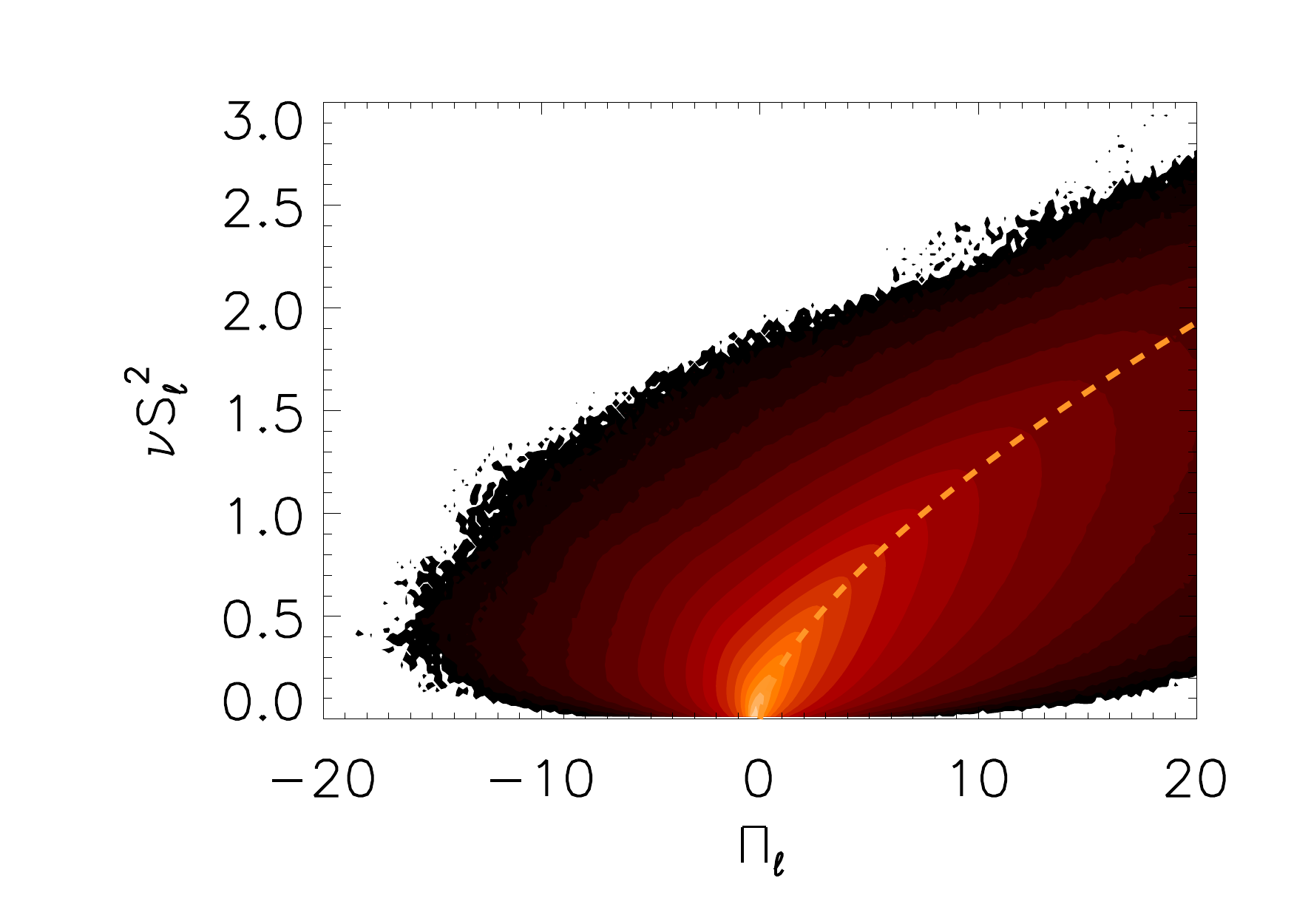}
\caption{ Hydro q=32}
\label{fig6a}
\end{subfigure}\\
%%%%%%%%%%%%%
%%%%%%%%%%%%%%%
\begin{subfigure}[b]{0.32\textwidth}
\centering
\includegraphics[width=\textwidth]{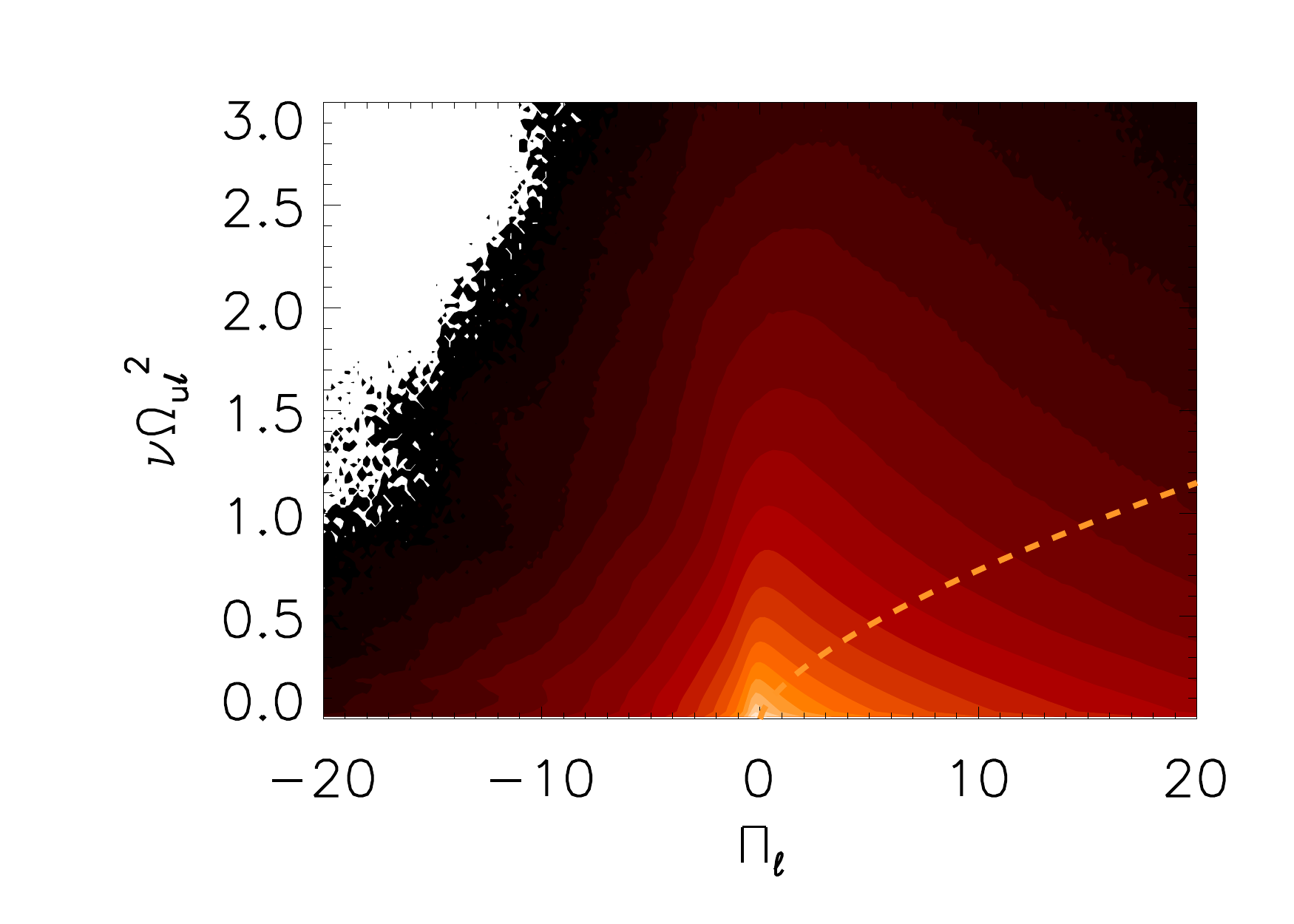}\\
\includegraphics[width=\textwidth]{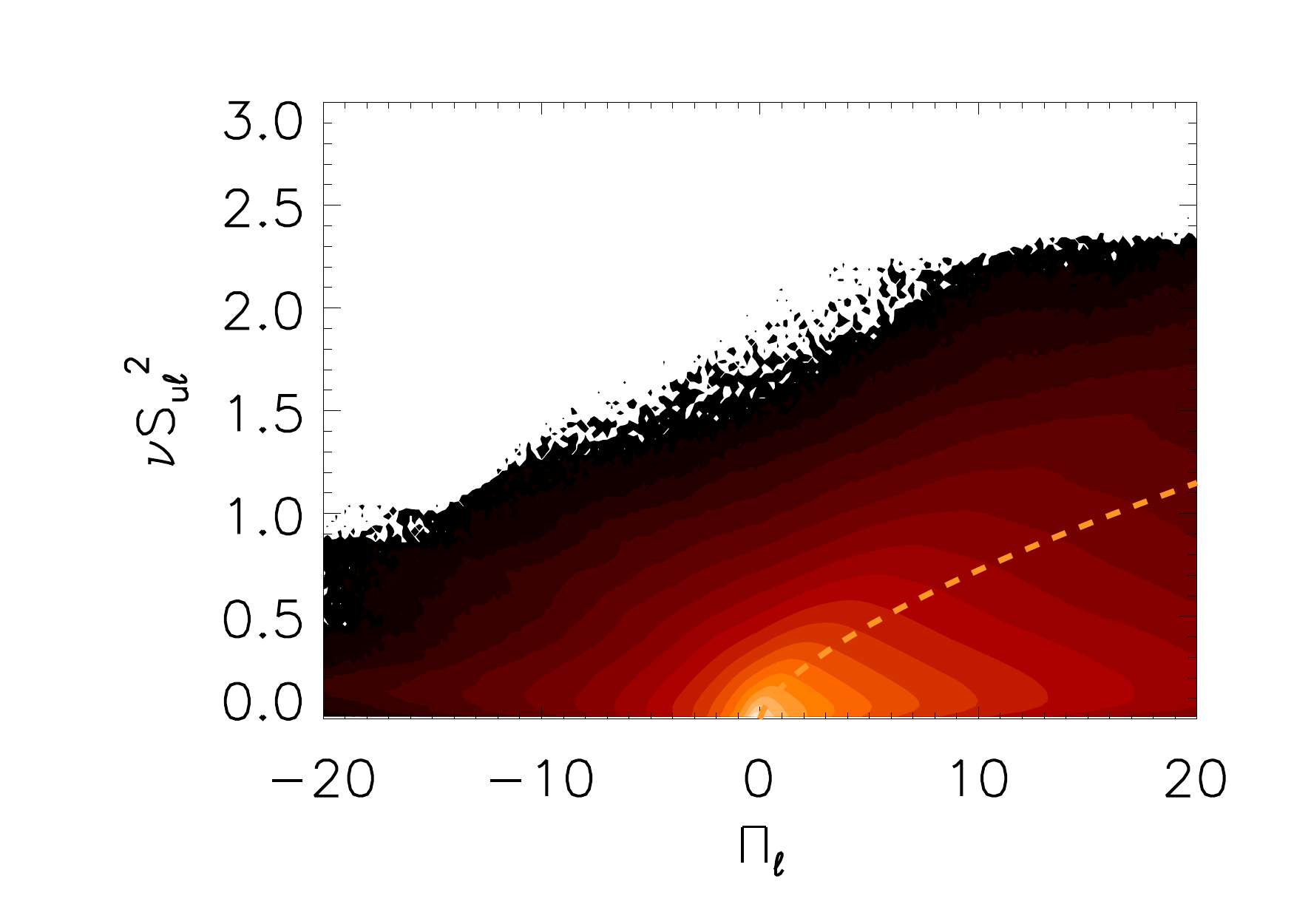}\\
\includegraphics[width=\textwidth]{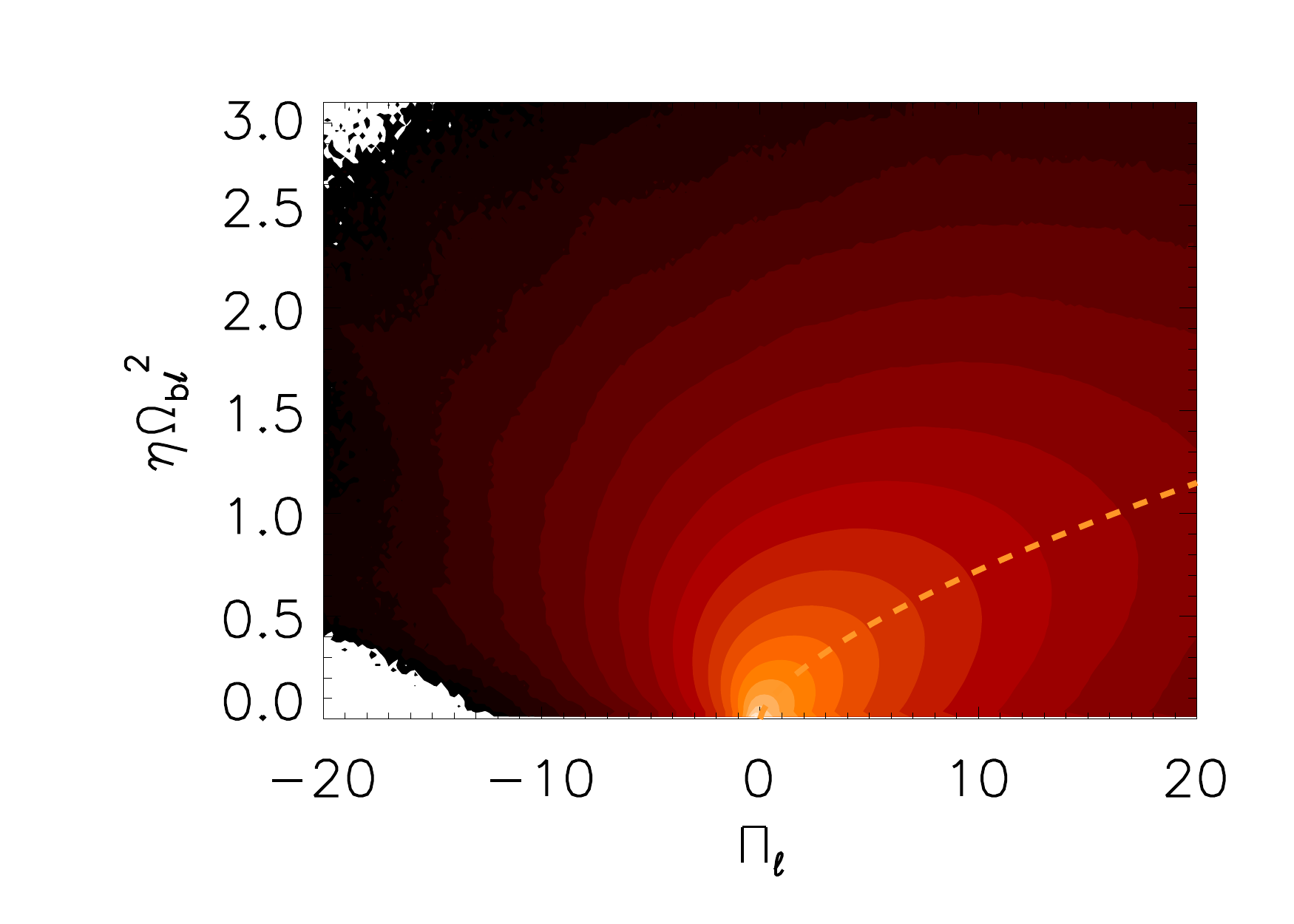}\\
\includegraphics[width=\textwidth]{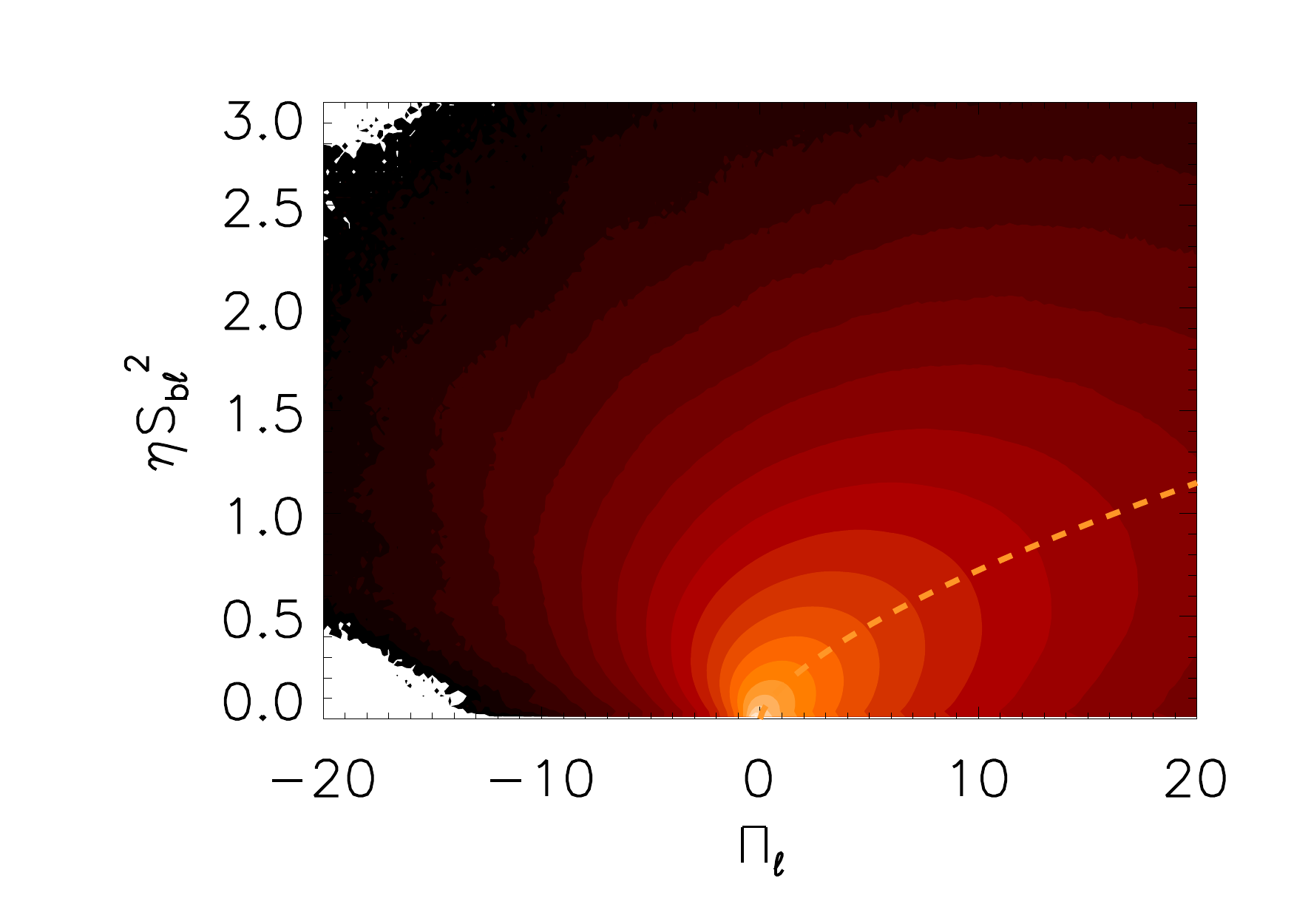}
\caption{ Dynamo q=32}
\label{fig6b}
\end{subfigure}
%%%%%%%%%%%%%
%%%%%%%%%%%%%%%
\begin{subfigure}[b]{0.32\textwidth}
\centering
\includegraphics[width=\textwidth]{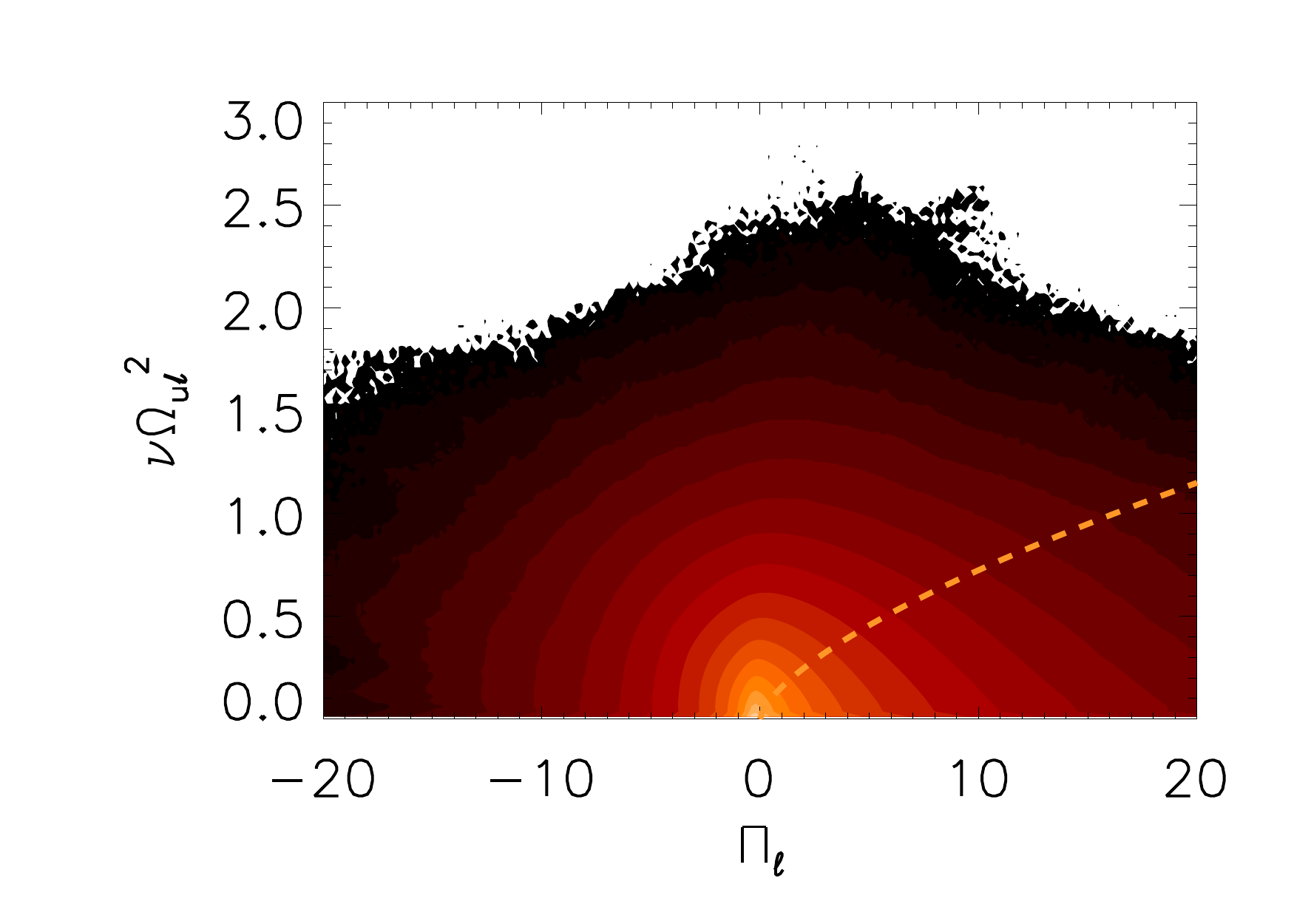}\\
\includegraphics[width=\textwidth]{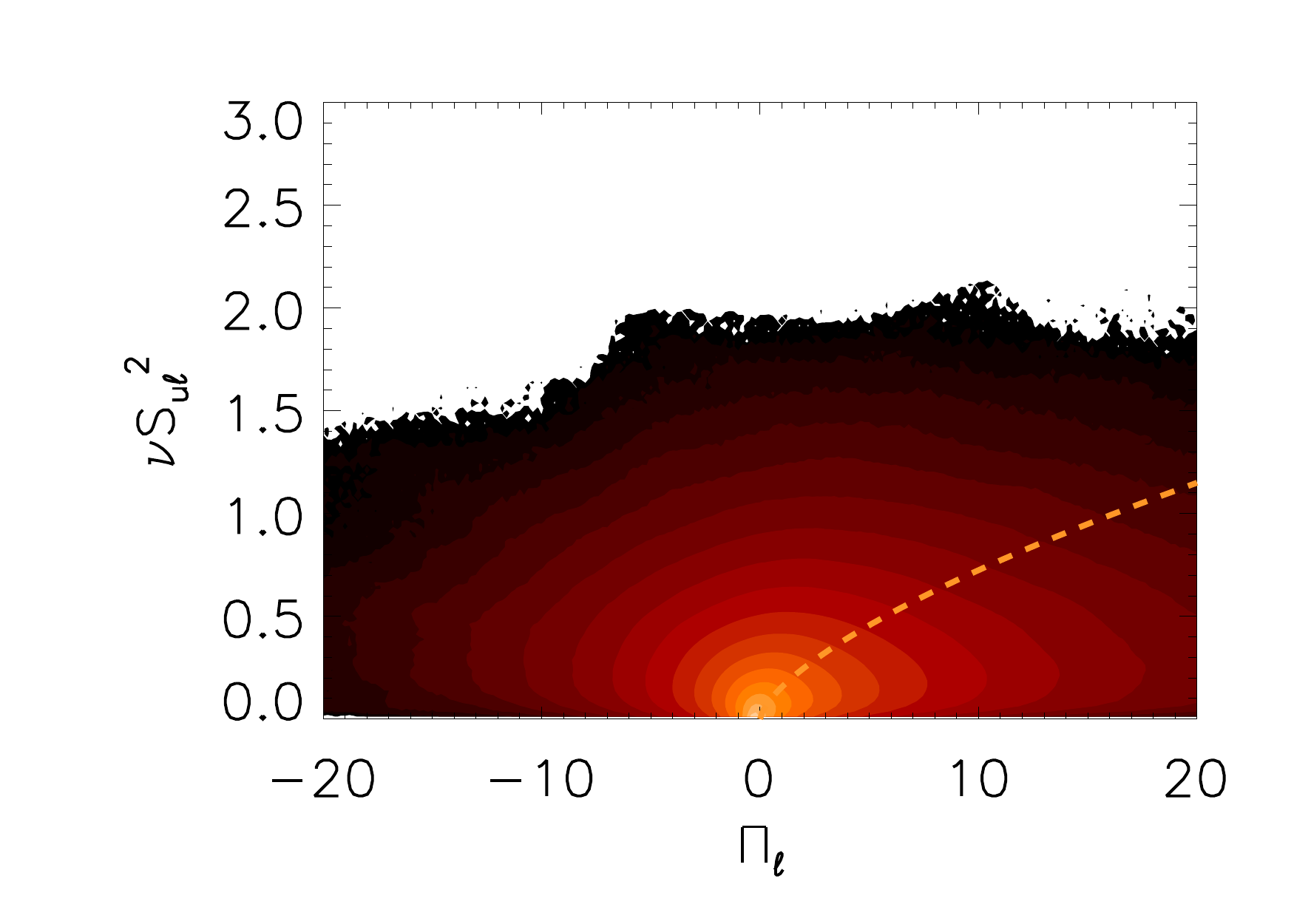}\\
\includegraphics[width=\textwidth]{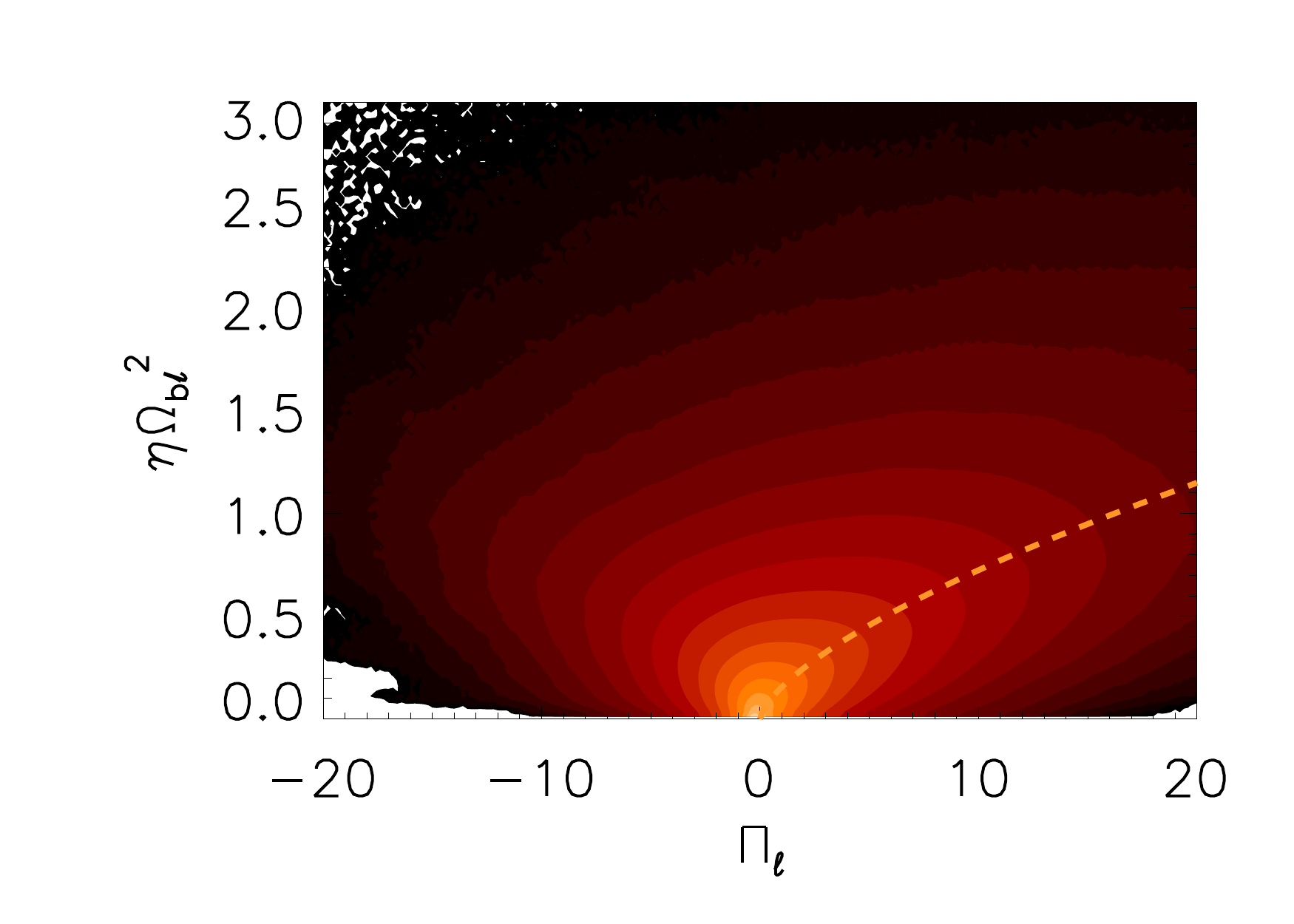}\\
\includegraphics[width=\textwidth]{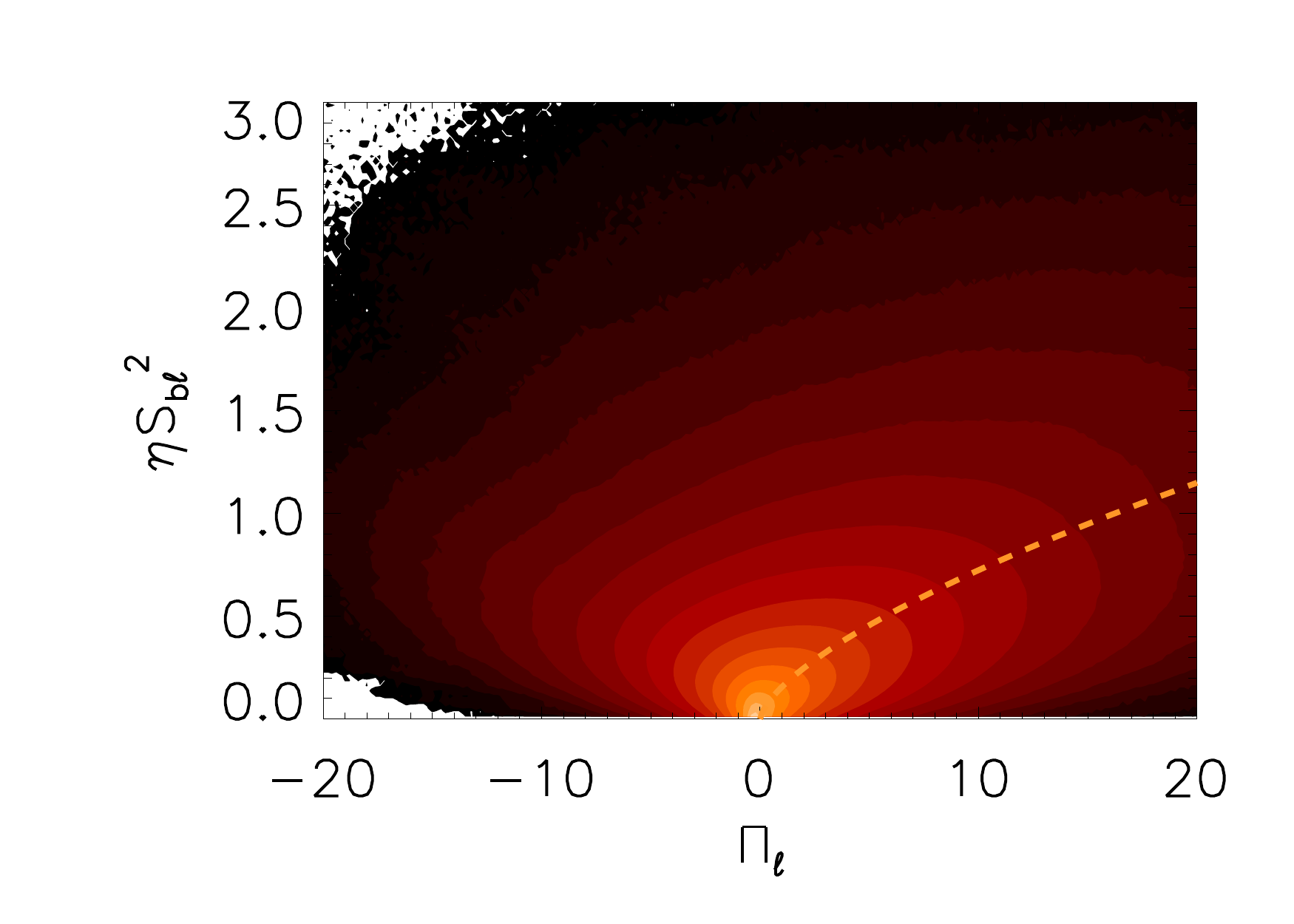}
\caption{ Moderate q=32}
\label{fig6c}
\end{subfigure}
%%%%%%%%%%%%%
%%%%%%%%%%%%%%%
\begin{subfigure}[b]{0.32\textwidth}
\centering
\includegraphics[width=\textwidth]{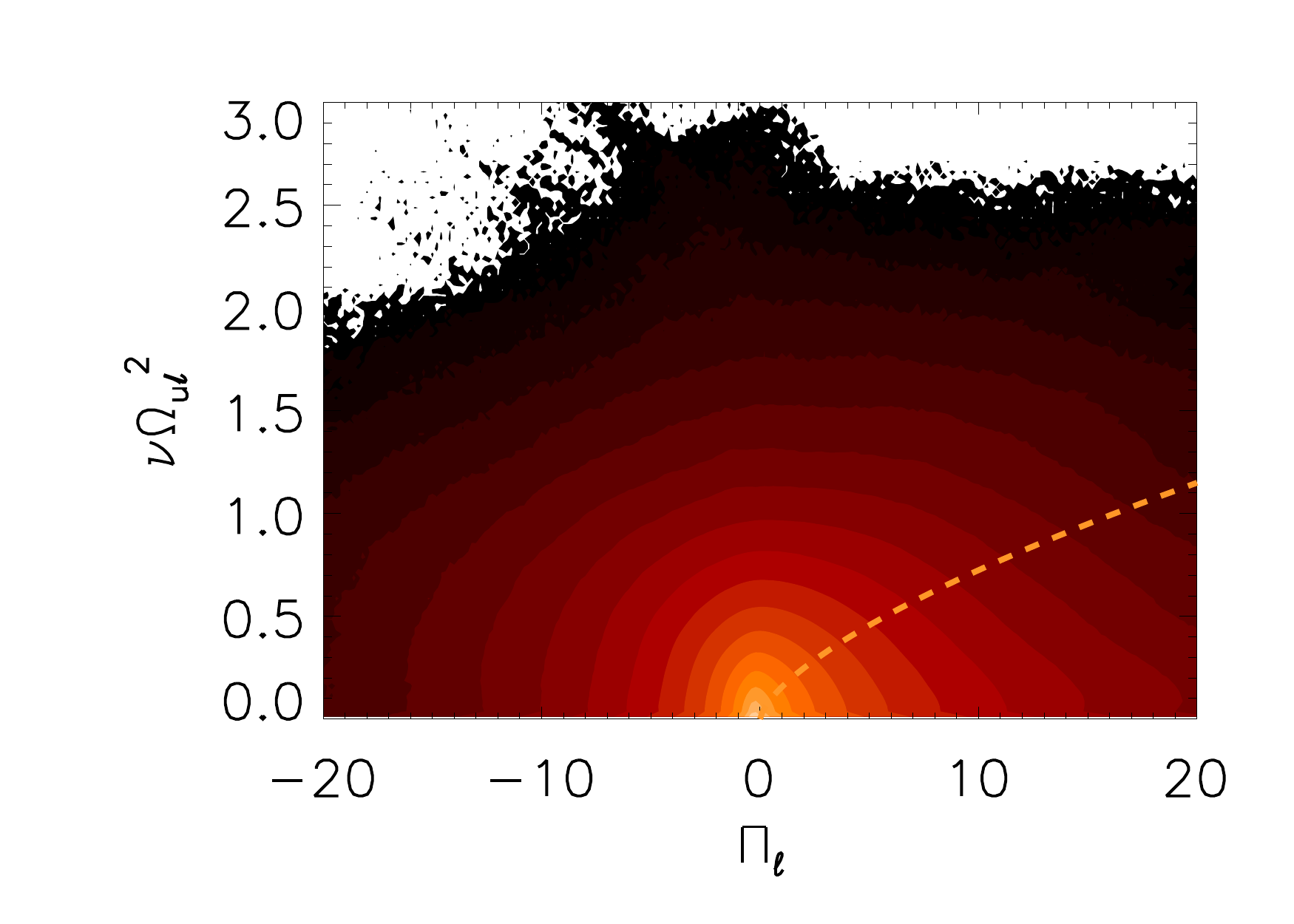}\\
\includegraphics[width=\textwidth]{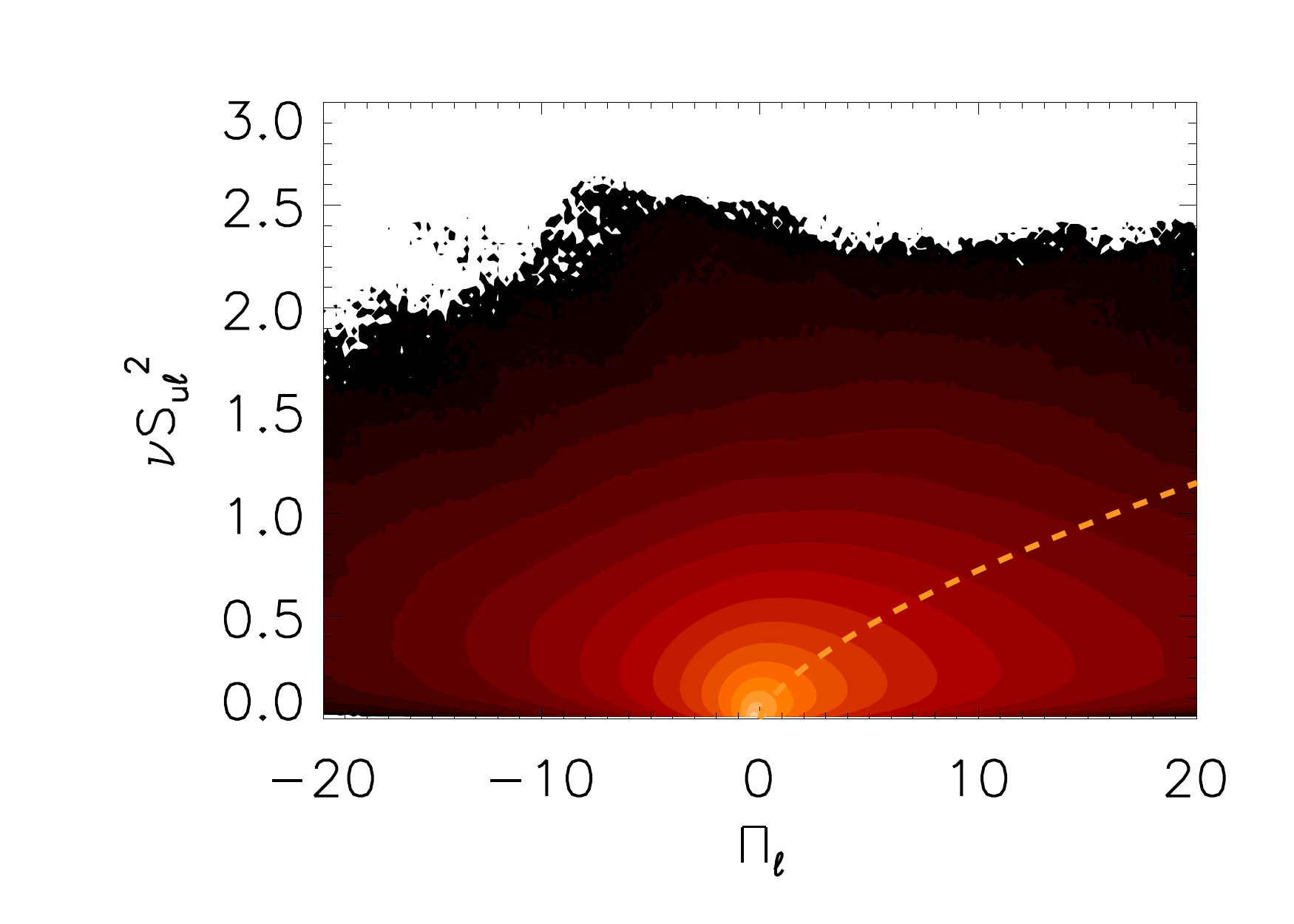}\\
\includegraphics[width=\textwidth]{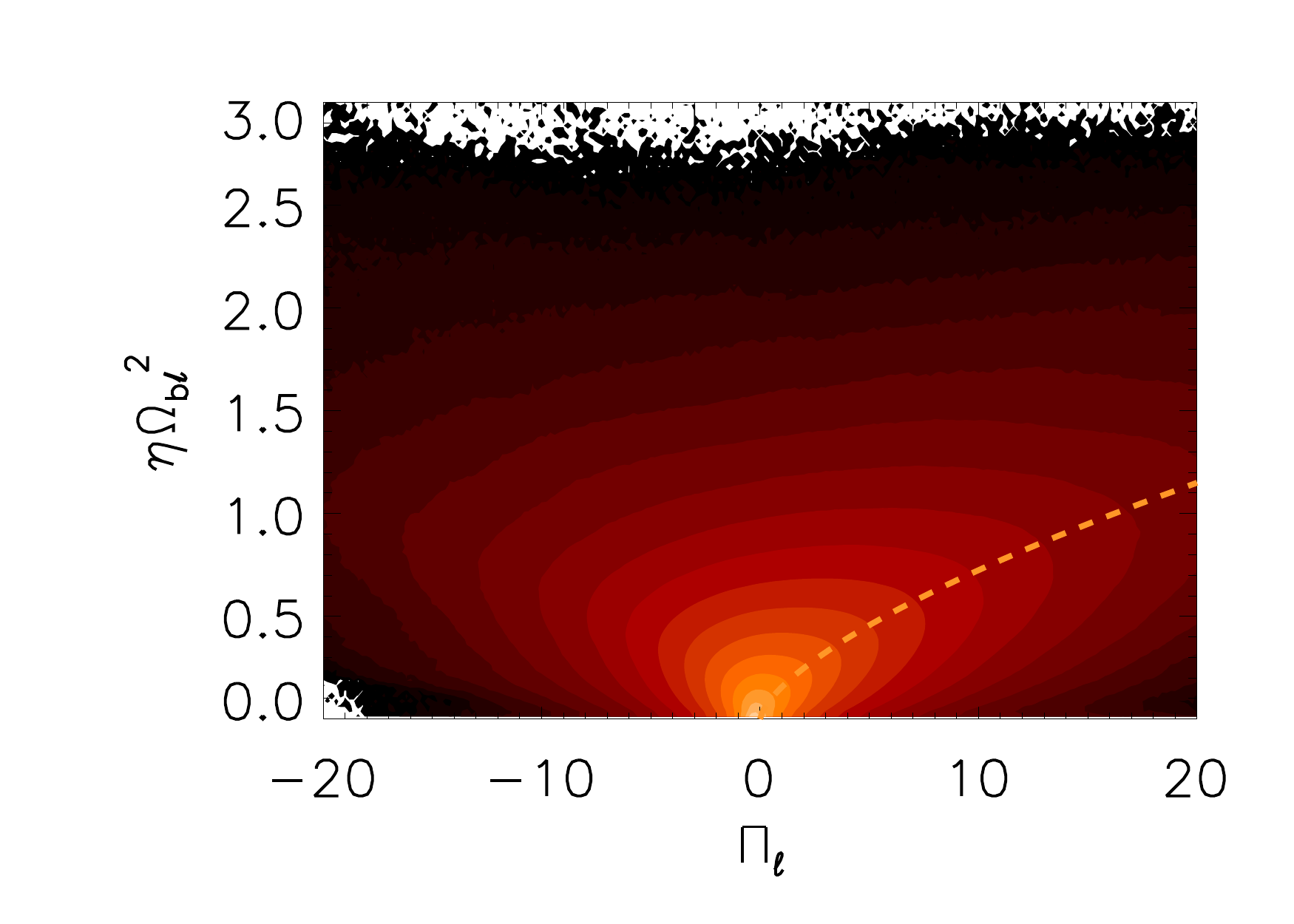}\\
\includegraphics[width=\textwidth]{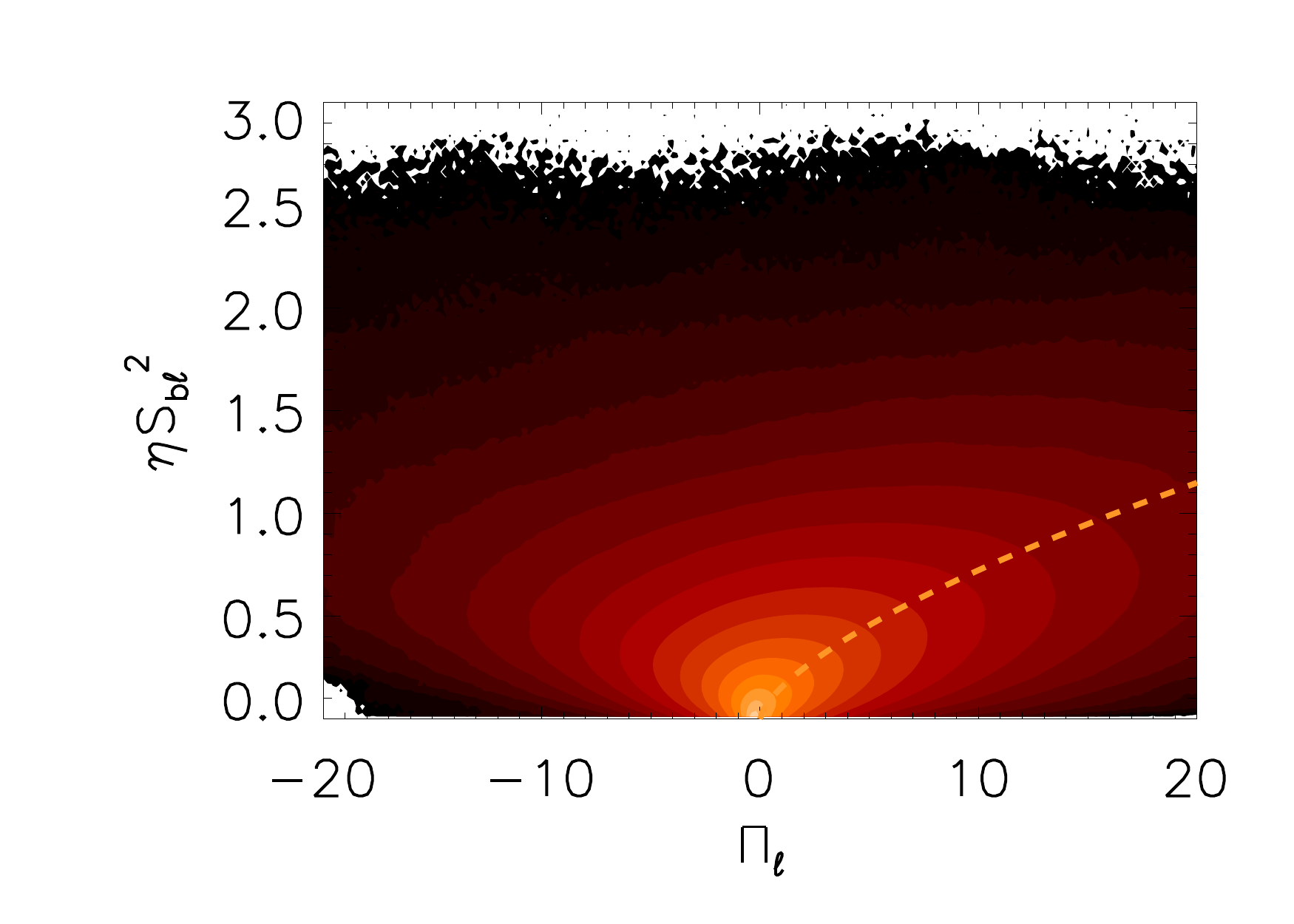}
\caption{ Strong q=32}
\label{fig6d}
\end{subfigure}
\caption{Joint pdfs of $\Pi_\ell$ (for $q=32$) with the modulus of the different strain rates for 
(a) hydrodynamic case
(b) dynamo case
(c) moderate case
and (d) strong case. From left to right the examined strain is 
$\Omega_{u,\ell}^2,S_{u,\ell}^2,\Omega_{b,\ell}^2,S_{b,\ell}^2$.
Bright colors indicate high probability. The yellow dashed line indicates the Smagorinsky scaling
$\Pi_\ell \propto \Omega_\ell^{3/2}$ and $\Pi_\ell \propto S_\ell^{3/2}$.
}
\label{fig6}
\end{center}
\end{figure}

%%%%%%%%%%%%%%%%%%%%%%%%%%%%%%%%%%%%%%%%%%%%%%%%%%%%%%%%%%%%%%%%%%%%%%%
%As in a companion paper \cite{alexakis2020local}, we calculate the joint pdf between $\Pi_\ell$ and $\Omega_\ell^2$ and  between $\Pi_\ell$ and $S_\ell^2$, both for the Kinetic and Magnetic energy fluxes.
In order to help the construction of subgrid scale models that are based on the gradients of the resolved fields we need to reveal what correlation exists between the gradients and the local energy flux.
In figure \ref{fig6} we present the joint pdf between $\Pi_\ell$ and the modulus squared of the symmetric and anti-symmetric stress tensors
$\Omega_{u,\ell}^2,\Omega_{b,\ell}^2,S_{u,\ell}^2,S_{b,\ell}^2$
%
%They are multiplied with the viscosity to be able to compare with the mean value at $\ell=0$ given by 
%\[ \lim_{\ell\to0}\nu \lra{S_\ell^2}      = 
%   \lim_{\ell\to0}\nu \lra{\Omega_\ell^2} =
%   \epsilon=1 .\]
The results are for a given \NEW{wavenumber} in the inertial range $q=32$. 

The results obtained for the pure hydrodynamic configuration are shown for comparison in the top panel.
In this case, the energy flux is essentially uncorrelated with the anti-symmetric part of the strain $\Omega_\ell$
with high probability events for a given $\Omega_\ell$ are concentrated around $\Pi_\ell=0$.  
On the other hand, a visible correlation is observed with the modulus of the symmetric part of the strain tensor $S_{\ell}$, with high probability events for a given $S_\ell$ concentrated around a non zero value of $\Pi_\ell$
that increases with $S_\ell$. The yellow line corresponds to the Smagorinsky scaling 
of eq. (\ref{SmagoRel}). The flux $\Pi_\ell$ is concentrated around this line giving support to the Smagorinsky model.

Such correlations are less clear for the MHD results.
For the dynamo case in the $\Pi_\ell-\Omega_{u\ell}$ diagram 
high-probability events are concentrated around $\Pi_\ell=0$ again.
For the $\Pi_\ell-S_{u\ell}$ diagram a much weaker correlation is observed compared to the hydro case.
The probability density shows long tails with large $\Pi_\ell$ for small $S_{u\ell}$ indicating that
$S_{u\ell}$ is not dominant in driving the cascade. 
A stronger correlation is observed with $\Omega_{b,\ell}^2$ and $S_{b,\ell}^2$.
High probability events are aligned with the Smagorinsky scaling (yellow dashed line)
but the spread is much higher than in the hydro case. The fact that the correlation
with $\Omega_{b,\ell}^2$ and $S_{b,\ell}^2$ look very similar indicates that neither
of the two alone is an optimal proxy to estimate the energy flux to the small scales.

The lack of correlation of $\Pi_\ell$ with the velocity gradients is even more clear for the moderate and strong case. For these cases no correlation is seen neither with $\Omega_{u,\ell}^2$ nor with $S_{u,\ell}^2$.
Although there is some correlation with the magnetic field gradients  $\Omega_{b,\ell}^2$ and $S_{b,\ell}^2$ with the maximum of probability
(for fixed $\Omega_{b,\ell}^2$ or $S_{b,\ell}^2$) following the Smagorinsky scaling (yellow dashed line), the spread is very large with highly probable negative flux events that become more frequent for the strong field case.
This indicates that although there is some correlation with the magnetic field gradients, modeling the subscale stresses with  $\Omega_{b,\ell}^2$ and $S_{b,\ell}^2$
alone is still missing important ingredients.

%The Dynamo case presents a diagram somewhat similar to the Hydro case, whereas the other MHD cases display very different plots, with almost symmetrical distributions.
%Concerning the Strain,  while The two variables appear very strongly correlated in the Hydro case, this is not true when a magnetic field is present.
%Possibly a residual correlation is present in the Dynamo case, yet in the other two MHD cases little difference can be made between strain and entrophy, and both resulting uncorrelated with the flux.
%
%In the MHD framework, we can perform the same analysis for the magnetic energy flux, analysing the joint pdf with the magnetic entrophy and strain. Quite interestingly, the two joint pdfs turn out to be basically indistinguishable.
%
%As shown in appendix, the behaviour is not much changed looking at different scales.
%%%%%%%%%%%%%%%%%%%%%%%%%%%%%%%%%%%%%%%%%%%%%%%%%%%%%%%%%%%%%%%%%%%%%%%
\subsection{Local magnetic field}                         %%%%%%%%%%%%%
\label{fieldb0}                                           %%%%%%%%%%%%%
%%%%%%%%%%%%%%%%%%%%%%%%%%%%%%%%%%%%%%%%%%%%%%%%%%%%%%%%%%%%%%%%%%%%%%%

\begin{figure}
\centering
\includegraphics[width=0.32\textwidth]{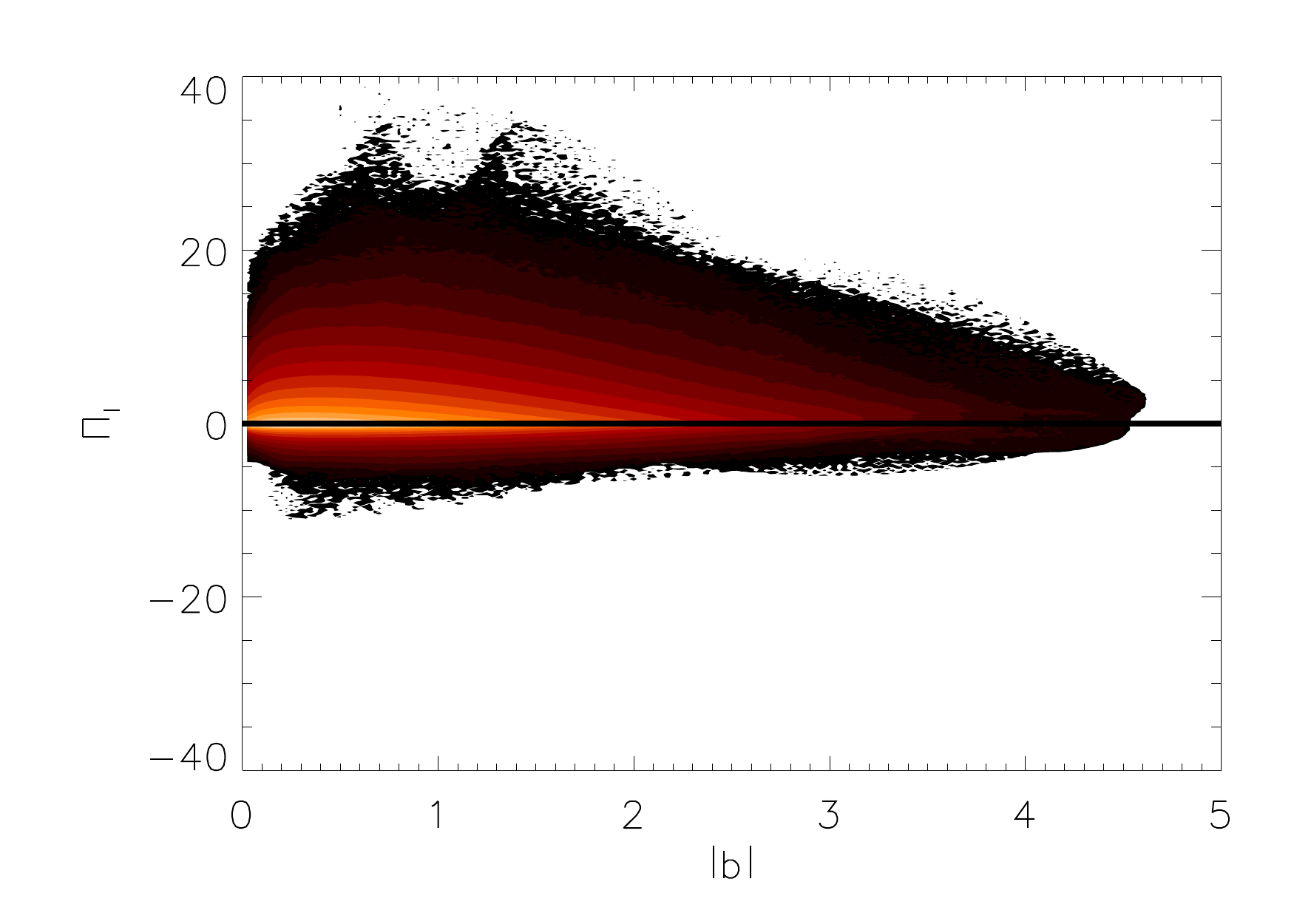}
\includegraphics[width=0.32\textwidth]{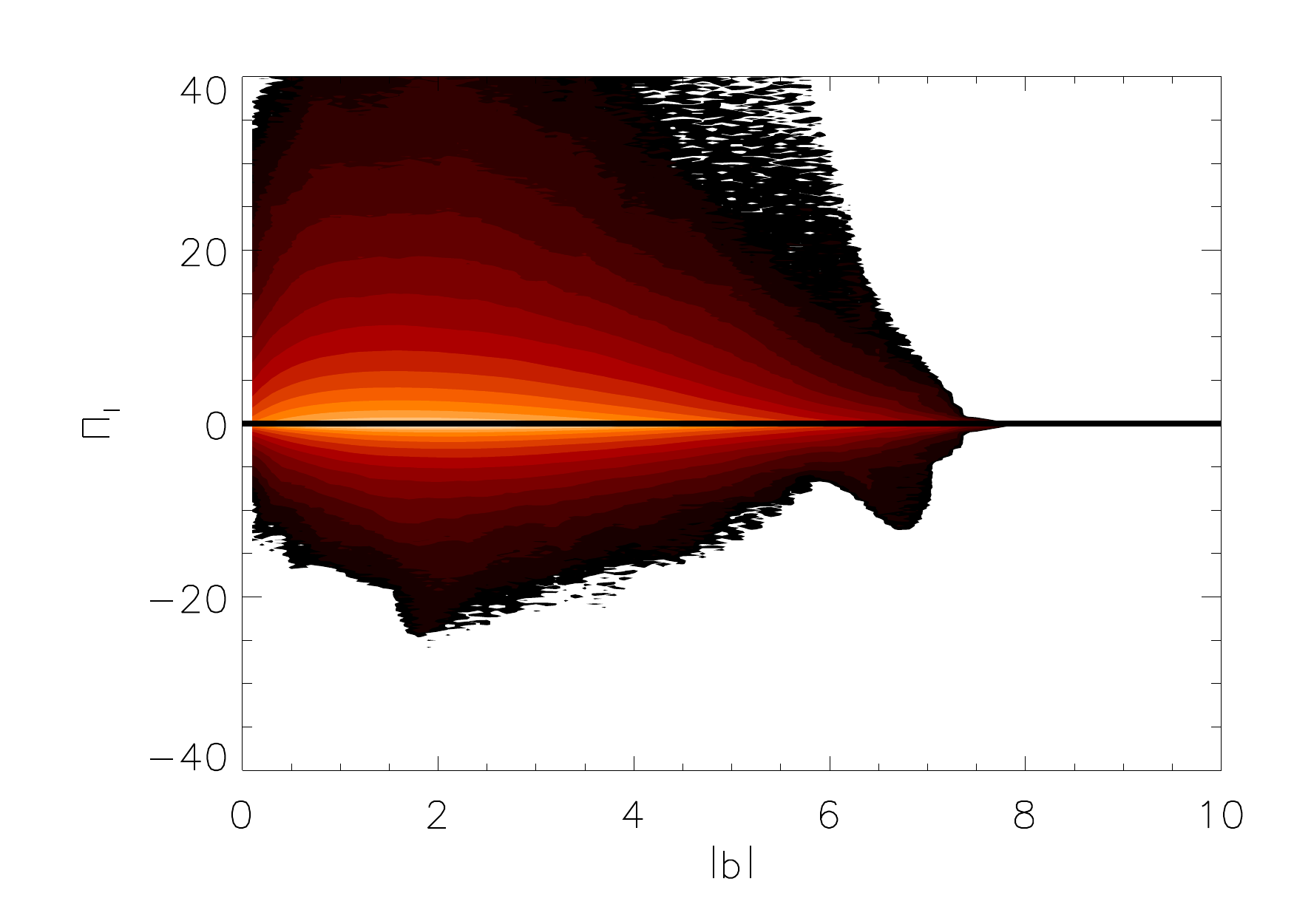}
\includegraphics[width=0.32\textwidth]{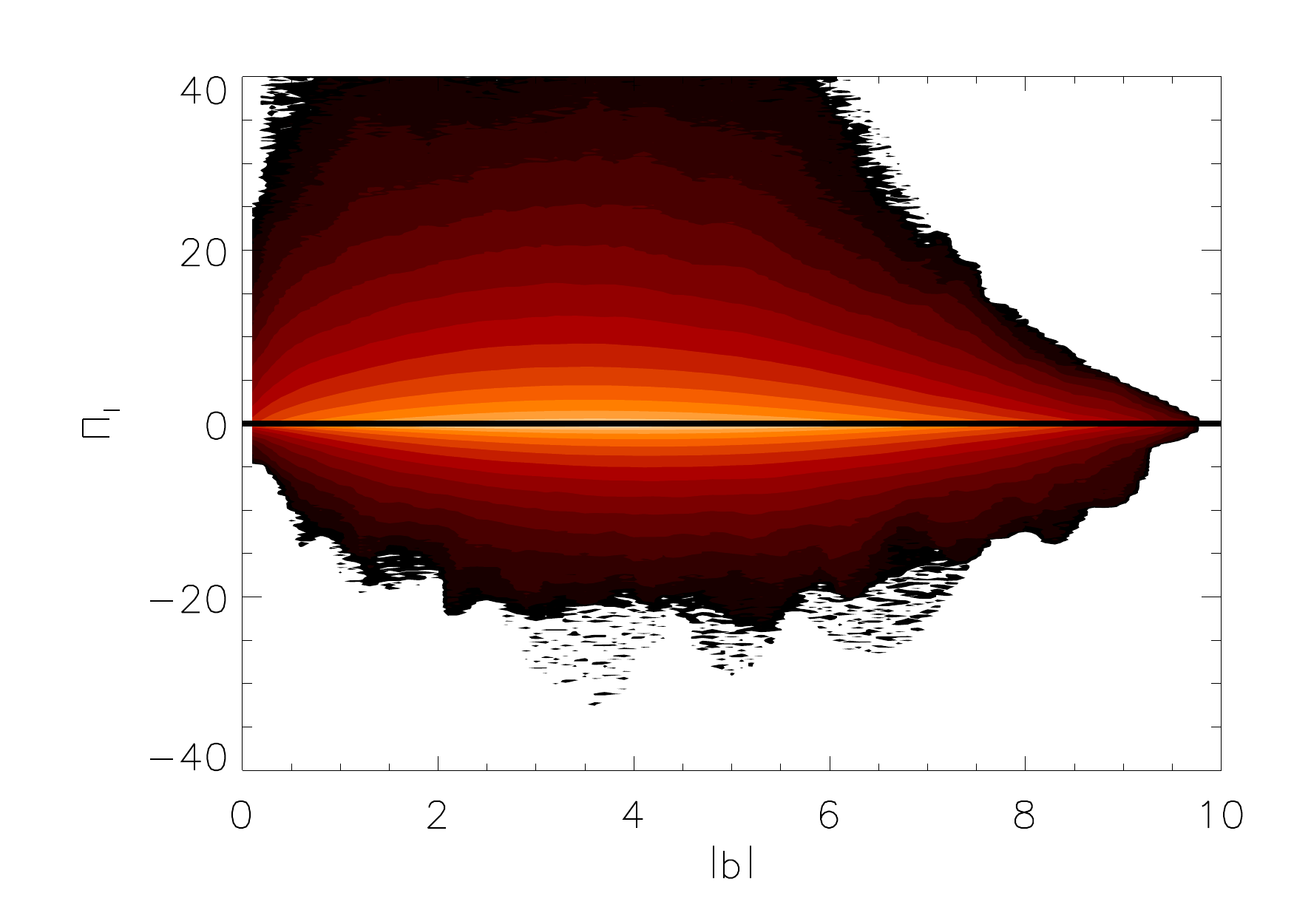}
\caption{ Joint pdf of the local flux $\Pi$ and the local amplitude of the magnetic field $b$ for the three cases; dynamo (left), moderate (center) and strong (right).}
\label{fig7}
\end{figure}

We pursue the analysis of the statistical phenomenology of the cascade process looking at correlations between the flux and the magnetic field. 
Figure \ref{fig7} shows the joint PDF of $\Pi_\ell$ and $\fb$ for the different cases examined. 
%As usual, the Dynamo case is peculiar since the pdf is peaked with very little negative flux.
%Nonetheless, 
The three cases display different distributions, with yet similar characteristics.
Generally speaking, maximum probability of the flux as well as most extreme values of it 
are obtained in the region around the average value of the magnetic field.
Furthermore, the larger the average magnetic field, the wider the distribution.
In this concern, it is interesting to note that the isolines are almost flat over a large range, indicating that in such large region the flux distribution, and therefore the cascade process is basically independent of the value of the magnetic field. That is particularly true for the strong case.

% FIG 7
% FIG 8
\begin{figure}
\centering
\includegraphics[width=0.48\textwidth]{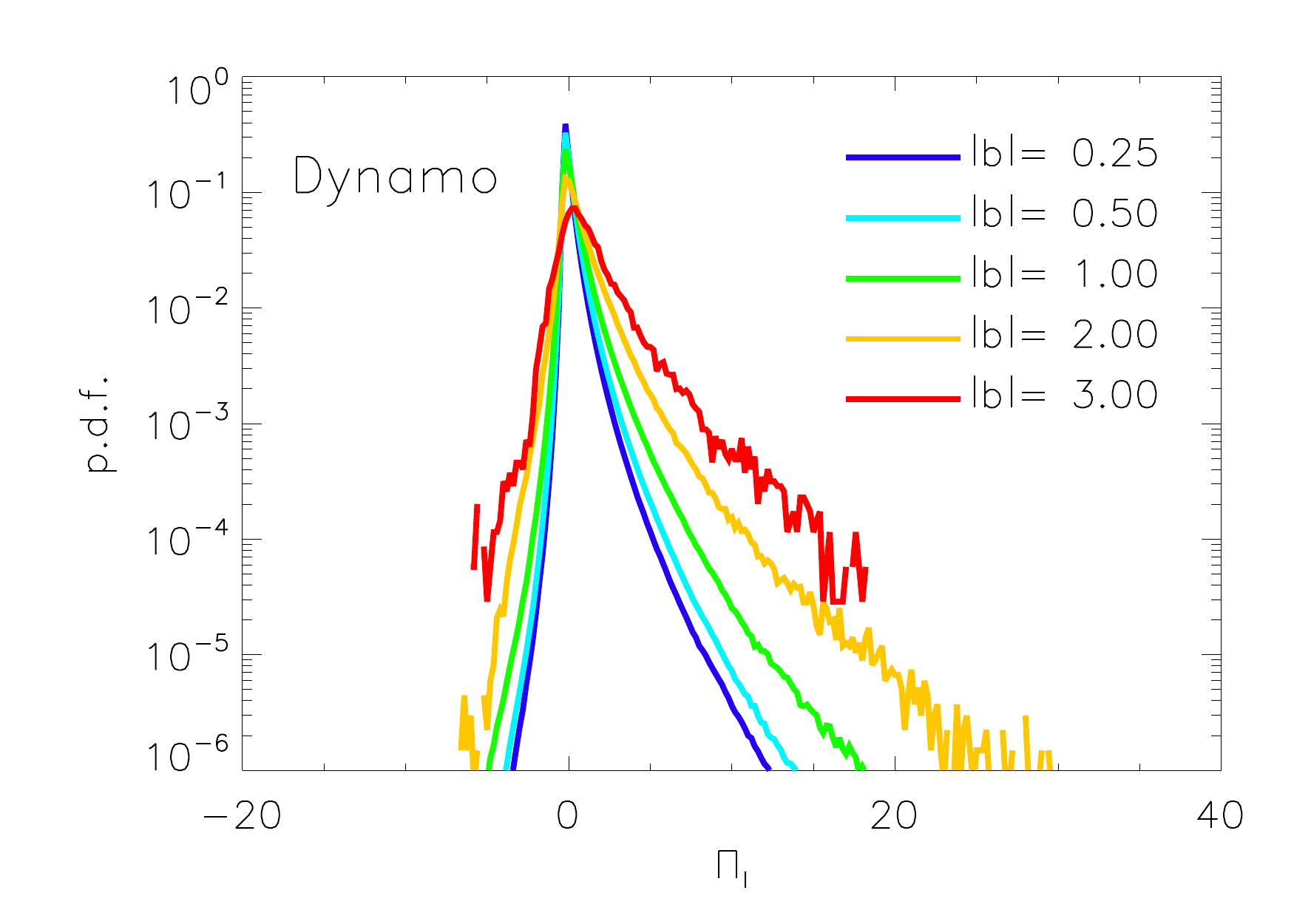}
\includegraphics[width=0.48\textwidth]{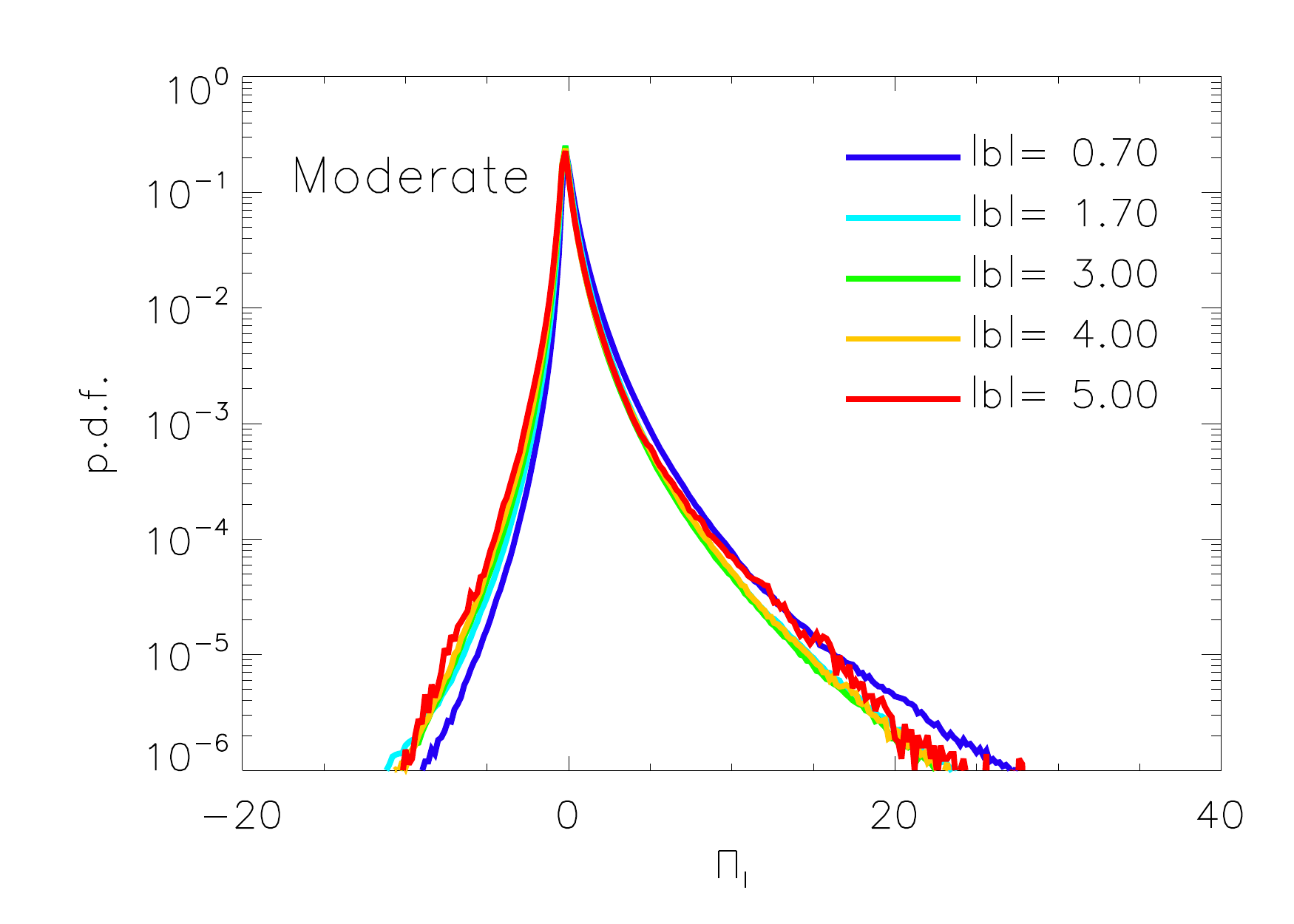}
\includegraphics[width=0.48\textwidth]{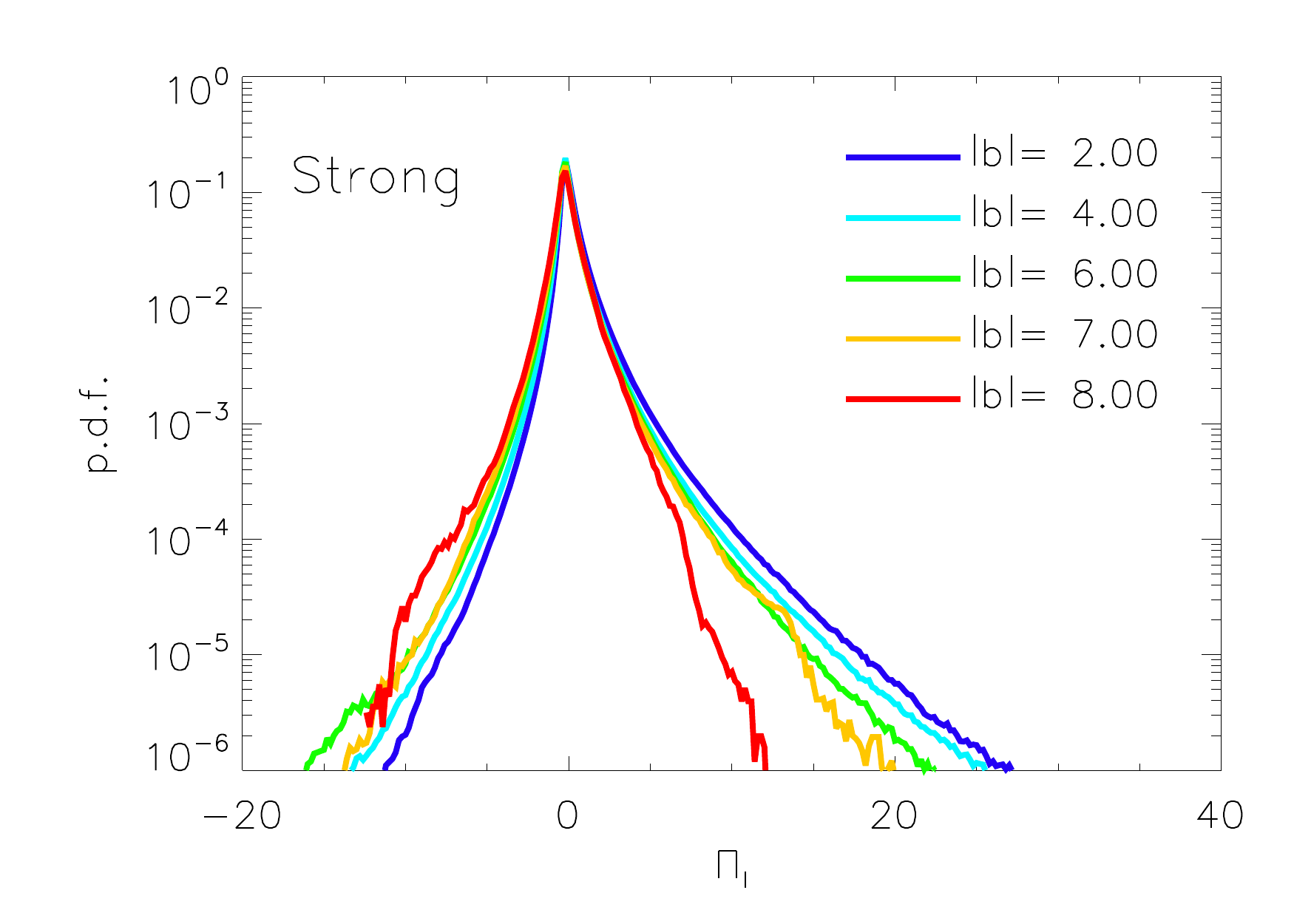}
\caption{Conditioned pdf {\it ie} Probability of having flux $\Pi_\ell$ 
given the magnetic field amplitude $|{\bf b}|$. In dynamo large $|{\bf b}|$ enhances 
large fluctuations. }
\label{fig8}
\end{figure}
%As highlighted by previous results, the impact of magnetic field on the turbulent cascade is considerable, and yet non-trivial.

More insight is gained by looking at 
%We investigate here to conclude the analysis 
the pdf of the energy flux conditioned on the knowledge of the magnetic field amplitude. 
This corresponds at looking at ``slices" of Figure \ref{fig7} for fixed values of $\fb$ and allows
to have a more clear look at the extreme events.
The results are displayed in Fig. \ref{fig8}. Seen this way some differences can be noted between the three different cases.
For the Dynamo there is a monotonic increase of the tails of the distribution with respect to increasing the magnetic field. 
Both negative and positive tails are wider when related to a larger magnetic field amplitude.
In this case large values of $\fb$ indicate more extreme events.
%In this case, however, only moderate values of the magnetic field can be obtained.

In the moderate case, the probability distribution seems to be almost insensitive to changes in the magnetic field amplitude. The %fluxes are overall more important than in the dynamo case, but the 
distribution would point to an energy cascade somewhat independent of the magnetic field amplitude. 
Some small differences are actually present in the distributions, but they are not much more important than  statistical errors.

For the strong case, the negative tails increase monotonically with the amplitude of the magnetic field. The positive tails, on the other hand are monotonically decreased by increasing the magnetic field.
Therefore, in strongly magnetized flows, large $|{\bf b}|$
\NEW{reduces extreme direct cascade fluctuations,
while increases inverse cascade fluctuations. }
%suppresses large direct cascade fluctuations, triggering large inverse cascade fluctuations. 
In this sense, the skewness of the distribution seems to be affected by the amplitude of the magnetic field. It appears that in regions of very large magnetic field the distribution is made symmetric.
%, suppressing in average forward flux events. 
Possibly this behavior occurs because
%Perhaps this should have been anticipated 
locally in regions of large $\fb$ turbulence transitions to a wave turbulence regime that has weaker forward energy flux  or by rendering the flow \NEW{locally }quasi-2D.  
\NEW{This effect however would require further investigation to be fully understood, and to assess possible finite-size effects.}
%and for the largest values the skewness changes sign.

%Consistently with previous results, the local characteristics of the cascade appear scale-dependent.
%When a large-scale helical magnetic field is present, at small scales Alf\'enic effects may become important, 
%even though globally these effects are negligible, as emphasised by the energy spectrum.

%Quasi-2D structures receive energy and local anisotropy may be important, with the trigger of a local inverse cascade. 
%\NOTE{Ok, I am speculating...Do we have also some conditional pdf of the magnetic cascade? I do not know it it is relevant, just for symmetry.}

\section{Conclusions}
In this work we have analysed the statistics of the local energy flux for three different magneto-hydrodynamic flows. A scale-by-scale analysis in physical space has been carried out that allows us to access relevant information on the energy cascade process that could help the construction of sub-grid scale models.

%In this work, we have focused on the total energy flux.

First, we have found that a Gaussian filter should be preferred to a spectral one \NEW{with regard to the representation of the Energy fluxes},  a result previously obtained for fluid turbulence. Then, the results of the total energy flux showed some similarities with the hydrodynamic case with $\Pi_\ell$ showing a skewed distribution with large tails. 
However, the dynamo case displays clear differences with respect to the other two cases. This is actually a general feature detected for all observables. 
On the other hand, moderate and strong cases display overall similar trends and profiles, as exposed by the fluxes computed at different scales.

The analysis of the different components of the energy flux has highlighted that the magnetic field plays a dominant role in the forward cascade. Specifically, the term related to the work made against the sub-scale Reynolds stress is negligible. This is a key information for modelling.  The negative part of the flux, that is the back-scatter flux, is instead basically the same for all components. Moreover, the study of the correlation between the energy flux and the different strain and rotation tensors has indicated that: (i) while results give some support to Smagorinski-like closure, at variance with pure hydrodynamic turbulence MHD energy flux cannot be globally recovered with a simple formula and several eddy viscosities should be used; (ii) as pointed out in previous works, the vorticity tensor has little correlation and can be omitted from modelling, at least as a first approximation; (iii) contrarily to what is commonly used, we have found that the magnetic strain tensor $S_{b,\ell}$ contribution is not negligible with respect to the anti-symmetric rotation $\Omega_{b,\ell}$ one, with regard to the entire flux, so both $S_{b,\ell}$ and $\Omega_{b,\ell}$ need to be used to model subscale stresses. 
%(iv) to hope to represent decently some of the features due to fluctuations, stochastic modelling should be chosen. 
Finally (iv) we have found that the presence of a large magnetic field has an impact on the local shape of the energy flux, and the flux in regions of very high magnetic field \NEW{has reduced} forward fluctuations.

%Present results indicate new directions for optimizing subgrid scale models in MHD turbulent flows.  Future research
%Present results indicate an interesting way for future works based upon scale-by-scale analysis of energy fluxes with central aim to help the construction and improvement of subgrid scale models. 
Present results indicate new directions for optimizing subgrid scale models in MHD turbulent flows. Many open questions still remain that future research should address. In particular 
the correlation between the local energy flux $\Pi_\ell$ and physically motivated combinations of the tensors $\Omega_{u,\ell},\Omega_{b,\ell},S_{u,\ell},S_{b,\ell} $ would be a next step. 
Theoretical work in the same spirit as  in the  Clark model  \citep{meneveau2000scale} can also help in this direction. We plan to address these issues in our future work.

%%%
\acknowledgments 
%%%

%\begin{acknowledgments}

This work was granted access to the HPC resources of MesoPSL financed by the R\'egion 
Ile de France and the project Equip@Meso (reference ANR-10-EQPX-29-01) of the programme Investissements d'Avenir supervised by the Agence Nationale pour la Recherche and the HPC resources of TGCC \& CINES (allocations No. A0090506421 A0110506421 \& No. A0062B10759) attributed by GENCI (Grand Equipement National de Calcul Intensif) where the present numerical simulations were performed. This work was also supported by the Agence nationale de la recherche (ANR DYSTURB project No. ANR-17-CE30-0004)

%\end{acknowledgments}

%%%%%%%%%%%%%%%%%%%%%%%%%%%%%%%%%%%%%%%%%%%%%%%%%%%%%%%%%%%%%%%%%%%%%%%%%%%%%%%%%%%%%%%%%%%%%%%%%%%%%%%%%%%%%%%%%%%%
%%%%%%%%%%%%%%%%%%%%%%%%%%%%%%%%%%%%%%%%%%%%%%%%%%%%%%%%%%%%%%%%%%%%%%%%%%%%%%%%%%%%%%%%%%%%%%%%%%%%%%%%%%%%%%%%%%%%
%%%%%%%%%%%%%%%%%%%%%%%%%%%%%%%%%%%%%%%%%%%%%%%%%%%%%%%%%%%%%%%%%%%%%%%%%%%%%%%%%%%%%%%%%%%%%%%%%%%%%%%%%%%%%%%%%%%%
%%%%%%%%%%%%%%%%%%%%%%%%%%%%%%%%%%%%%%%%%%%%%%%%%%%%%%%%%%%%%%%%%%%%%%%%%%%%%%%%%%%%%%%%%%%%%%%%%%%%%%%%%%%%%%%%%%%%
\appendix %%%%%%%%%%%%%%%%%%%%%%%%%%%%%%%%%%%%%%%%%%%%%%%%%%%%%%%%%%%%%%%%%%%%%%%%%%%%%%%%%%%%%%%%%%%%%%%%%%%%%%%%%%
%%%%%%%%%%%%%%%%%%%%%%%%%%%%%%%%%%%%%%%%%%%%%%%%%%%%%%%%%%%%%%%%%%%%%%%%%%%%%%%%%%%%%%%%%%%%%%%%%%%%%%%%%%%%%%%%%%%%
%%%%%%%%%%%%%%%%%%%%%%%%%%%%%%%%%%%%%%%%%%%%%%%%%%%%%%%%%%%%%%%%%%%%%%%%%%%%%%%%%%%%%%%%%%%%%%%%%%%%%%%%%%%%%%%%%%%%
%%%%%%%%%%%%%%%%%%%%%%%%%%%%%%%%%%%%%%%%%%%%%%%%%%%%%%%%%%%%%%%%%%%%%%%%%%%%%%%%%%%%%%%%%%%%%%%%%%%%%%%%%%%%%%%%%%%%

\bibliographystyle{jpp}
%\bibliography{jfm-instructions}
\bibliography{biblio}

\begin{thebibliography}{93}
\expandafter\ifx\csname natexlab\endcsname\relax\def\natexlab#1{#1}\fi
\def\au#1{#1} \def\ed#1{#1} \def\yr#1{#1}\def\at#1{#1}\def\jt#1{\textit{#1}}
  \def\bt#1{#1}\def\bvol#1{\textbf{#1}} \def\vol#1{#1} \def\pg#1{#1}
  \def\publ#1{#1}\def\arxiv#1{#1}\def\org#1{#1}\def\st#1{\textit{#1}}

\bibitem[Agullo {\em et~al.\/}(2001)Agullo, M{\"u}ller, Knaepen \&
  Carati]{agullo2001large}
{\sc \au{Agullo, Olivier}, \au{M{\"u}ller, W-C}, \au{Knaepen, Bernard} \&
  \au{Carati, Daniele}} \yr{2001}  \at{Large eddy simulation of decaying
  magnetohydrodynamic turbulence with dynamic subgrid-modeling}.  \jt{Physics
  of Plasmas}  \bvol{8}~(7),  \pg{3502--3505}.

\bibitem[Alexakis(2013)]{alexakis2013large}
{\sc \au{Alexakis, Alexandros}} \yr{2013}  \at{Large-scale magnetic fields in
  magnetohydrodynamic turbulence}.  \jt{Physical Review Letters}
  \bvol{110}~(8),  \pg{084502}.

\bibitem[Alexakis \& Biferale(2018)]{alexakis2018cascades}
{\sc \au{Alexakis, Alexandros} \& \au{Biferale, Luca}} \yr{2018}  \at{Cascades
  and transitions in turbulent flows}.  \jt{Physics Reports}  \bvol{767},
  \pg{1--101}.

\bibitem[Alexakis \& Chibbaro(2020)]{alexakis2020local}
{\sc \au{Alexakis, Alexandros} \& \au{Chibbaro, Sergio}} \yr{2020}  \at{Local
  energy flux of turbulent flows}.  \jt{Physical Review Fluids}  \bvol{5}~(9),
  \pg{094604}.

\bibitem[Alexakis {\em et~al.\/}(2005)Alexakis, Mininni \&
  Pouquet]{alexakis2005shell}
{\sc \au{Alexakis, Alexandros}, \au{Mininni, Pablo~D} \& \au{Pouquet, Annick}}
  \yr{2005}  \at{Shell-to-shell energy transfer in magnetohydrodynamics. i.
  steady state turbulence}.  \jt{Physical Review E}  \bvol{72}~(4),
  \pg{046301}.

\bibitem[Alexakis {\em et~al.\/}(2007)Alexakis, Mininni \&
  Pouquet]{alexakis2007turbulent}
{\sc \au{Alexakis, A}, \au{Mininni, Pablo~Daniel} \& \au{Pouquet, A}} \yr{2007}
   \at{Turbulent cascades, transfer, and scale interactions in
  magnetohydrodynamics}.  \jt{New Journal of Physics}  \bvol{9}~(8),  \pg{298}.

\bibitem[Aluie(2017)]{aluie2017coarse}
{\sc \au{Aluie, Hussein}} \yr{2017}  \at{Coarse-grained incompressible
  magnetohydrodynamics: analyzing the turbulent cascades}.  \jt{New Journal of
  Physics}  \bvol{19}~(2),  \pg{025008}.

\bibitem[Aluie \& Eyink(2009)]{aluie2009localness}
{\sc \au{Aluie, Hussein} \& \au{Eyink, Gregory~L}} \yr{2009}  \at{Localness of
  energy cascade in hydrodynamic turbulence. ii. sharp spectral filter}.
  \jt{Physics of Fluids}  \bvol{21}~(11),  \pg{115108}.

\bibitem[Aluie \& Eyink(2010)]{aluie2010scale}
{\sc \au{Aluie, Hussein} \& \au{Eyink, Gregory~L}} \yr{2010}  \at{Scale
  locality of magnetohydrodynamic turbulence}.  \jt{Physical review letters}
  \bvol{104}~(8),  \pg{081101}.

\bibitem[Battaner(1996)]{battaner1996astrophysical}
{\sc \au{Battaner, Eduardo}} \yr{1996} {\em Astrophysical fluid dynamics\/}.
  \publ{Cambridge University Press}.

\bibitem[Beresnyak(2011)]{beresnyak2011spectral}
{\sc \au{Beresnyak, Andrey}} \yr{2011}  \at{Spectral slope and kolmogorov
  constant of mhd turbulence}.  \jt{Physical Review Letters}  \bvol{106}~(7),
  \pg{075001}.

\bibitem[Bian \& Aluie(2019)]{bian2019decoupled}
{\sc \au{Bian, Xin} \& \au{Aluie, Hussein}} \yr{2019}  \at{Decoupled cascades
  of kinetic and magnetic energy in magnetohydrodynamic turbulence}.
  \jt{Physical review letters}  \bvol{122}~(13),  \pg{135101}.

\bibitem[Bian {\em et~al.\/}(2021)Bian, Shang, Blackman, Collins \&
  Aluie]{bian2021scaling}
{\sc \au{Bian, Xin}, \au{Shang, Jessica~K}, \au{Blackman, Eric~G}, \au{Collins,
  Gilbert~W} \& \au{Aluie, Hussein}} \yr{2021}  \at{Scaling of turbulent
  viscosity and resistivity: Extracting a scale-dependent turbulent magnetic
  prandtl number}.  \jt{The Astrophysical Journal Letters}  \bvol{917}~(1),
  \pg{L3}.

\bibitem[Biferale {\em et~al.\/}(2019)Biferale, Bonaccorso, Buzzicotti \&
  Iyer]{biferale2019self}
{\sc \au{Biferale, Luca}, \au{Bonaccorso, Fabio}, \au{Buzzicotti, Michele} \&
  \au{Iyer, Kartik~P}} \yr{2019}  \at{Self-similar subgrid-scale models for
  inertial range turbulence and accurate measurements of intermittency}.
  \jt{Physical review letters}  \bvol{123}~(1),  \pg{014503}.

\bibitem[Biskamp(2003)]{biskamp2003magnetohydrodynamic}
{\sc \au{Biskamp, Dieter}} \yr{2003} {\em Magnetohydrodynamic turbulence\/}.
  \publ{Cambridge University Press}.

\bibitem[Boldyrev(2006)]{boldyrev2006spectrum}
{\sc \au{Boldyrev, Stanislav}} \yr{2006}  \at{Spectrum of magnetohydrodynamic
  turbulence}.  \jt{Phys. Rev. Lett.}  \bvol{96},  \pg{115002}.

\bibitem[Borue \& Orszag(1998)]{borue1998local}
{\sc \au{Borue, Vadim} \& \au{Orszag, Steven~A}} \yr{1998}  \at{Local energy
  flux and subgrid-scale statistics in three-dimensional turbulence}.
  \jt{Journal of Fluid Mechanics}  \bvol{366},  \pg{1--31}.

\bibitem[Brandenburg {\em et~al.\/}(2012)Brandenburg, Sokoloff \&
  Subramanian]{brandenburg2012current}
{\sc \au{Brandenburg, Axel}, \au{Sokoloff, Dmitry} \& \au{Subramanian,
  Kandaswamy}} \yr{2012}  \at{Current status of turbulent dynamo theory}.
  \jt{Space Science Reviews}  \bvol{169}~(1),  \pg{123--157}.

\bibitem[Brandenburg \& Subramanian(2005)]{brandenburg2005astrophysical}
{\sc \au{Brandenburg, Axel} \& \au{Subramanian, Kandaswamy}} \yr{2005}
  \at{Astrophysical magnetic fields and nonlinear dynamo theory}.  \jt{Physics
  Reports}  \bvol{417}~(1-4),  \pg{1--209}.

\bibitem[Bruno \& Carbone(2013)]{bruno2013solar}
{\sc \au{Bruno, Roberto} \& \au{Carbone, Vincenzo}} \yr{2013}  \at{The solar
  wind as a turbulence laboratory}.  \jt{Living Reviews in Solar Physics}
  \bvol{10}~(1),  \pg{1--208}.

\bibitem[Buzzicotti {\em et~al.\/}(2018)Buzzicotti, Linkmann, Aluie, Biferale,
  Brasseur \& Meneveau]{buzzicotti2018effect}
{\sc \au{Buzzicotti, M}, \au{Linkmann, M}, \au{Aluie, H}, \au{Biferale, L},
  \au{Brasseur, J} \& \au{Meneveau, C}} \yr{2018}  \at{Effect of filter type on
  the statistics of energy transfer between resolved and subfilter scales from
  a-priori analysis of direct numerical simulations of isotropic turbulence}.
  \jt{Journal of Turbulence}  \bvol{19}~(2),  \pg{167--197}.

\bibitem[Camporeale {\em et~al.\/}(2018)Camporeale, Sorriso-Valvo, Califano \&
  Retin{\`o}]{camporeale2018coherent}
{\sc \au{Camporeale, Enrico}, \au{Sorriso-Valvo, Luca}, \au{Califano,
  Francesco} \& \au{Retin{\`o}, Alessandro}} \yr{2018}  \at{Coherent structures
  and spectral energy transfer in turbulent plasma: a space-filter approach}.
  \jt{Physical review letters}  \bvol{120}~(12),  \pg{125101}.

\bibitem[Carrasco {\em et~al.\/}(2020)Carrasco, Vigan{\`o} \&
  Palenzuela]{carrasco2020gradient}
{\sc \au{Carrasco, Federico}, \au{Vigan{\`o}, Daniele} \& \au{Palenzuela,
  Carlos}} \yr{2020}  \at{Gradient subgrid-scale model for relativistic mhd
  large-eddy simulations}.  \jt{Physical Review D}  \bvol{101}~(6),
  \pg{063003}.

\bibitem[Casciola {\em et~al.\/}(2003)Casciola, Gualtieri, Benzi \&
  Piva]{casciola2003scale}
{\sc \au{Casciola, CM}, \au{Gualtieri, P}, \au{Benzi, R} \& \au{Piva, R}}
  \yr{2003}  \at{Scale-by-scale budget and similarity laws for shear
  turbulence}.  \jt{Journal of Fluid Mechanics}  \bvol{476},  \pg{105--114}.

\bibitem[Cerri \& Camporeale(2020)]{cerri2020space}
{\sc \au{Cerri, SS} \& \au{Camporeale, E}} \yr{2020}  \at{Space-filter
  techniques for quasi-neutral hybrid-kinetic models}.  \jt{Physics of Plasmas}
   \bvol{27}~(8),  \pg{082102}.

\bibitem[Chen {\em et~al.\/}(2003)Chen, Chen \& Eyink]{chen2003joint}
{\sc \au{Chen, Qiaoning}, \au{Chen, Shiyi} \& \au{Eyink, Gregory~L}} \yr{2003}
  \at{The joint cascade of energy and helicity in three-dimensional
  turbulence}.  \jt{Physics of Fluids}  \bvol{15}~(2),  \pg{361--374}.

\bibitem[Chen {\em et~al.\/}(2006)Chen, Eyink, Wan \& Xiao]{chen2006kelvin}
{\sc \au{Chen, Shiyi}, \au{Eyink, Gregory~L}, \au{Wan, Minping} \& \au{Xiao,
  Zuoli}} \yr{2006}  \at{Is the kelvin theorem valid for high reynolds number
  turbulence?}  \jt{Physical review letters}  \bvol{97}~(14),  \pg{144505}.

\bibitem[Chernyshov {\em et~al.\/}(2014)Chernyshov, Karelsky \&
  Petrosyan]{chernyshov2014subgrid}
{\sc \au{Chernyshov, Alexander~Aleksandrovich}, \au{Karelsky,
  Kirill~Vladimirovich} \& \au{Petrosyan, Arakel~Sarkisovich}} \yr{2014}
  \at{Subgrid-scale modeling for the study of compressible magnetohydrodynamic
  turbulence in space plasmas}.  \jt{Physics-Uspekhi}  \bvol{57}~(5),
  \pg{421}.

\bibitem[Cimarelli {\em et~al.\/}(2013)Cimarelli, De~Angelis \&
  Casciola]{cimarelli2013paths}
{\sc \au{Cimarelli, A}, \au{De~Angelis, E} \& \au{Casciola, CM}} \yr{2013}
  \at{Paths of energy in turbulent channel flows}.  \jt{Journal of Fluid
  Mechanics}  \bvol{715},  \pg{436--451}.

\bibitem[Danaila {\em et~al.\/}(2001)Danaila, Anselmet, Zhou \&
  Antonia]{danaila2001turbulent}
{\sc \au{Danaila, L}, \au{Anselmet, F}, \au{Zhou, Tongming} \& \au{Antonia,
  RA}} \yr{2001}  \at{Turbulent energy scale budget equations in a fully
  developed channel flow}.  \jt{Journal of Fluid Mechanics}  \bvol{430},
  \pg{87--109}.

\bibitem[Dar {\em et~al.\/}(2001)Dar, Verma \& Eswaran]{dar2001energy}
{\sc \au{Dar, Gaurav}, \au{Verma, Mahendra~K} \& \au{Eswaran, V}} \yr{2001}
  \at{Energy transfer in two-dimensional magnetohydrodynamic turbulence:
  formalism and numerical results}.  \jt{Physica D: Nonlinear Phenomena}
  \bvol{157}~(3),  \pg{207--225}.

\bibitem[Davidson(2002)]{davidson2002introduction}
{\sc \au{Davidson, Peter~Alan}} \yr{2002} {\em An introduction to
  magnetohydrodynamics\/}.  \publ{Cambridge University Press}.

\bibitem[Domaradzki {\em et~al.\/}(2009)Domaradzki, Teaca \&
  Carati]{domaradzki2009locality}
{\sc \au{Domaradzki, J~Andrzej}, \au{Teaca, Bogdan} \& \au{Carati, Daniele}}
  \yr{2009}  \at{Locality properties of the energy flux in turbulence}.
  \jt{Physics of fluids}  \bvol{21}~(2),  \pg{025106}.

\bibitem[Dubrulle(2019)]{dubrulle2019beyond}
{\sc \au{Dubrulle, B{\'e}reng{\`e}re}} \yr{2019}  \at{Beyond kolmogorov
  cascades}.  \jt{Journal of Fluid Mechanics}  \bvol{867}.

\bibitem[Eyink \& Sreenivasan(2006)]{Eyink:2006p1379}
{\sc \au{Eyink, G} \& \au{Sreenivasan, K}} \yr{2006}  \at{Onsager and the
  theory of hydrodynamic turbulence}.  \jt{Reviews of Modern Physics} .

\bibitem[Eyink {\em et~al.\/}(2013)Eyink, Vishniac, Lalescu, Aluie, Kanov,
  B{\"u}rger, Burns, Meneveau \& Szalay]{eyink2013flux}
{\sc \au{Eyink, Gregory}, \au{Vishniac, Ethan}, \au{Lalescu, Cristian},
  \au{Aluie, Hussein}, \au{Kanov, Kalin}, \au{B{\"u}rger, Kai}, \au{Burns,
  Randal}, \au{Meneveau, Charles} \& \au{Szalay, Alexander}} \yr{2013}
  \at{Flux-freezing breakdown in high-conductivity magnetohydrodynamic
  turbulence}.  \jt{Nature}  \bvol{497}~(7450),  \pg{466--469}.

\bibitem[Eyink(2005)]{eyink2005locality}
{\sc \au{Eyink, Gregory~L}} \yr{2005}  \at{Locality of turbulent cascades}.
  \jt{Physica D: Nonlinear Phenomena}  \bvol{207}~(1-2),  \pg{91--116}.

\bibitem[Eyink(2018)]{eyink2018cascades}
{\sc \au{Eyink, Gregory~L}} \yr{2018}  \at{Cascades and dissipative anomalies
  in nearly collisionless plasma turbulence}.  \jt{Physical Review X}
  \bvol{8}~(4),  \pg{041020}.

\bibitem[Eyink \& Aluie(2009)]{eyink2009localness}
{\sc \au{Eyink, Gregory~L} \& \au{Aluie, Hussein}} \yr{2009}  \at{Localness of
  energy cascade in hydrodynamic turbulence. i. smooth coarse graining}.
  \jt{Physics of Fluids}  \bvol{21}~(11),  \pg{115107}.

\bibitem[Frisch(1995)]{Fri_95}
{\sc \au{Frisch, U.}} \yr{1995} {\em Turbulence. {T}he legacy of {A.N}
  {K}olmogorov\/}.  \publ{Cambridge, University press}.

\bibitem[Galtier(2009)]{galtier2009exact}
{\sc \au{Galtier, S}} \yr{2009}  \at{Exact vectorial law for axisymmetric
  magnetohydrodynamics turbulence}.  \jt{The Astrophysical Journal}
  \bvol{704}~(2),  \pg{1371}.

\bibitem[Galtier(2016)]{galtier2016introduction}
{\sc \au{Galtier, S{\'e}bastien}} \yr{2016} {\em Introduction to modern
  magnetohydrodynamics\/}.  \publ{Cambridge University Press}.

\bibitem[Galtier(2018)]{galtier2018origin}
{\sc \au{Galtier, S{\'e}bastien}} \yr{2018}  \at{On the origin of the energy
  dissipation anomaly in (hall) magnetohydrodynamics}.  \jt{Journal of Physics
  A: Mathematical and Theoretical}  \bvol{51}~(20),  \pg{205501}.

\bibitem[Germano(1992)]{germano1992turbulence}
{\sc \au{Germano, Massimo}} \yr{1992}  \at{Turbulence: the filtering approach}.
   \jt{Journal of Fluid Mechanics}  \bvol{238},  \pg{325--336}.

\bibitem[Germano {\em et~al.\/}(1991)Germano, Piomelli, Moin \&
  Cabot]{germano1991dynamic}
{\sc \au{Germano, Massimo}, \au{Piomelli, Ugo}, \au{Moin, Parviz} \& \au{Cabot,
  William~H}} \yr{1991}  \at{A dynamic subgrid-scale eddy viscosity model}.
  \jt{Physics of Fluids A: Fluid Dynamics}  \bvol{3}~(7),  \pg{1760--1765}.

\bibitem[Goldreich \& Sridhar(1995)]{goldreich1995toward}
{\sc \au{Goldreich, P} \& \au{Sridhar, S}} \yr{1995}  \at{Toward a theory of
  interstellar turbulence. 2: Strong alfvenic turbulence}.  \jt{The
  Astrophysical Journal}  \bvol{438},  \pg{763--775}.

\bibitem[Goldstein {\em et~al.\/}(1995)Goldstein, Roberts \&
  Matthaeus]{goldstein1995magnetohydrodynamic}
{\sc \au{Goldstein, Melvyn~L}, \au{Roberts, D\_~A} \& \au{Matthaeus, WH}}
  \yr{1995}  \at{Magnetohydrodynamic turbulence in the solar wind}.  \jt{Annual
  review of astronomy and astrophysics}  \bvol{33}~(1),  \pg{283--325}.

\bibitem[Grete {\em et~al.\/}(2017)Grete, Vlaykov, Schmidt \&
  Schleicher]{grete2017comparative}
{\sc \au{Grete, Philipp}, \au{Vlaykov, Dimitar~G}, \au{Schmidt, Wolfram} \&
  \au{Schleicher, Dominik~RG}} \yr{2017}  \at{Comparative statistics of
  selected subgrid-scale models in large-eddy simulations of decaying,
  supersonic magnetohydrodynamic turbulence}.  \jt{Physical Review E}
  \bvol{95}~(3),  \pg{033206}.

\bibitem[Innocenti {\em et~al.\/}(2021)Innocenti, Jaccod, Popinet \&
  Chibbaro]{innocenti2021direct}
{\sc \au{Innocenti, Alessio}, \au{Jaccod, Alice}, \au{Popinet, St{\'e}phane} \&
  \au{Chibbaro, Sergio}} \yr{2021}  \at{Direct numerical simulation of
  bubble-induced turbulence}.  \jt{Journal of Fluid Mechanics}  \bvol{918}.

\bibitem[Iroshnikov(1964)]{iroshnikov1964turbulence}
{\sc \au{Iroshnikov, PS}} \yr{1964}  \at{Turbulence of a conducting fluid in a
  strong magnetic field}.  \jt{Soviet Astronomy}  \bvol{7},  \pg{566}.

\bibitem[Kessar {\em et~al.\/}(2016)Kessar, Balarac \&
  Plunian]{kessar2016effect}
{\sc \au{Kessar, Mouloud}, \au{Balarac, Guillaume} \& \au{Plunian, Franck}}
  \yr{2016}  \at{The effect of subgrid-scale models on grid-scale/subgrid-scale
  energy transfers in large-eddy simulation of incompressible
  magnetohydrodynamic turbulence}.  \jt{Physics of Plasmas}  \bvol{23}~(10),
  \pg{102305}.

\bibitem[Kolmogorov(1941)]{kolmogorov1941local}
{\sc \au{Kolmogorov, Andrey~Nikolaevich}} \yr{1941}  \at{The local structure of
  turbulence in incompressible viscous fluid for very large reynolds numbers}.
  \jt{Cr Acad. Sci. URSS}  \bvol{30},  \pg{301--305}.

\bibitem[Kraichnan(1965)]{kraichnan1965inertial}
{\sc \au{Kraichnan, Robert~H}} \yr{1965}  \at{Inertial-range spectrum of
  hydromagnetic turbulence}.  \jt{The Physics of Fluids}  \bvol{8}~(7),
  \pg{1385--1387}.

\bibitem[Kraichnan(1971)]{kraichnan1971inertial}
{\sc \au{Kraichnan, Robert~H}} \yr{1971}  \at{Inertial-range transfer in
  two-and three-dimensional turbulence}.  \jt{Journal of Fluid Mechanics}
  \bvol{47}~(3),  \pg{525--535}.

\bibitem[Landau {\em et~al.\/}(2013)Landau, Bell, Kearsley, Pitaevskii,
  Lifshitz \& Sykes]{landau2013electrodynamics}
{\sc \au{Landau, Lev~Davidovich}, \au{Bell, JS}, \au{Kearsley, MJ},
  \au{Pitaevskii, LP}, \au{Lifshitz, EM} \& \au{Sykes, JB}} \yr{2013} {\em
  Electrodynamics of continuous media\/}, ,  \vol{vol.~8}.  \publ{elsevier}.

\bibitem[Leonard(1975)]{leonard1975energy}
{\sc \au{Leonard, Athony}} \yr{1975}  \at{Energy cascade in large-eddy
  simulations of turbulent fluid flows}.  \bt{In {\em Advances in
  geophysics\/}}, ,  \vol{vol.~18},  \pg{pp. 237--248}.  \publ{Elsevier}.

\bibitem[Lesieur {\em et~al.\/}(2005)Lesieur, M{\'e}tais, Comte {\em
  et~al.\/}]{lesieur2005large}
{\sc \au{Lesieur, Marcel}, \au{M{\'e}tais, Olivier}, \au{Comte, Pierre} \&
  \au{others}} \yr{2005} {\em Large-eddy simulations of turbulence\/}.
  \publ{Cambridge university press}.

\bibitem[Linkmann {\em et~al.\/}(2018)Linkmann, Buzzicotti \&
  Biferale]{linkmann2018multi}
{\sc \au{Linkmann, Moritz}, \au{Buzzicotti, Michele} \& \au{Biferale, Luca}}
  \yr{2018}  \at{Multi-scale properties of large eddy simulations: correlations
  between resolved-scale velocity-field increments and subgrid-scale
  quantities}.  \jt{Journal of Turbulence}  \bvol{19}~(6),  \pg{493--527}.

\bibitem[Liu {\em et~al.\/}(1994)Liu, Meneveau \& Katz]{liu1994properties}
{\sc \au{Liu, Shewen}, \au{Meneveau, Charles} \& \au{Katz, Joseph}} \yr{1994}
  \at{On the properties of similarity subgrid-scale models as deduced from
  measurements in a turbulent jet}.  \jt{Journal of Fluid Mechanics}
  \bvol{275},  \pg{83--119}.

\bibitem[McKee \& Ostriker(2007)]{mckee2007theory}
{\sc \au{McKee, Christopher~F} \& \au{Ostriker, Eve~C}} \yr{2007}  \at{Theory
  of star formation}.  \jt{Annu. Rev. Astron. Astrophys.}  \bvol{45},
  \pg{565--687}.

\bibitem[Meneveau(1994)]{meneveau1994statistics}
{\sc \au{Meneveau, Charles}} \yr{1994}  \at{Statistics of turbulence
  subgrid-scale stresses: Necessary conditions and experimental tests}.
  \jt{Physics of Fluids}  \bvol{6}~(2),  \pg{815--833}.

\bibitem[Meneveau \& Katz(2000)]{meneveau2000scale}
{\sc \au{Meneveau, Charles} \& \au{Katz, Joseph}} \yr{2000}
  \at{Scale-invariance and turbulence models for large-eddy simulation}.
  \jt{Annual Review of Fluid Mechanics}  \bvol{32}~(1),  \pg{1--32}.

\bibitem[Miesch {\em et~al.\/}(2015)Miesch, Matthaeus, Brandenburg, Petrosyan,
  Pouquet, Cambon, Jenko, Uzdensky, Stone, Tobias {\em
  et~al.\/}]{miesch2015large}
{\sc \au{Miesch, Mark}, \au{Matthaeus, William}, \au{Brandenburg, Axel},
  \au{Petrosyan, Arakel}, \au{Pouquet, Annick}, \au{Cambon, Claude}, \au{Jenko,
  Frank}, \au{Uzdensky, Dmitri}, \au{Stone, James}, \au{Tobias, Steve} \&
  \au{others}} \yr{2015}  \at{Large-eddy simulations of magnetohydrodynamic
  turbulence in heliophysics and astrophysics}.  \jt{Space Science Reviews}
  \bvol{194}~(1),  \pg{97--137}.

\bibitem[Mininni {\em et~al.\/}(2005)Mininni, Alexakis \&
  Pouquet]{mininni2005shell}
{\sc \au{Mininni, Pablo}, \au{Alexakis, Alexandros} \& \au{Pouquet, Annick}}
  \yr{2005}  \at{Shell-to-shell energy transfer in magnetohydrodynamics. ii.
  kinematic dynamo}.  \jt{Physical Review E}  \bvol{72}~(4),  \pg{046302}.

\bibitem[Mininni {\em et~al.\/}(2011)Mininni, Rosenberg, Reddy \&
  Pouquet]{mininni2011hybrid}
{\sc \au{Mininni, Pablo~D}, \au{Rosenberg, Duane}, \au{Reddy, Raghu} \&
  \au{Pouquet, Annick}} \yr{2011}  \at{A hybrid mpi--openmp scheme for scalable
  parallel pseudospectral computations for fluid turbulence}.  \jt{Parallel
  computing}  \bvol{37}~(6-7),  \pg{316--326}.

\bibitem[Misra \& Pullin(1997)]{misra1997vortex}
{\sc \au{Misra, Ashish} \& \au{Pullin, Dale~I}} \yr{1997}  \at{A vortex-based
  subgrid stress model for large-eddy simulation}.  \jt{Physics of Fluids}
  \bvol{9}~(8),  \pg{2443--2454}.

\bibitem[Moll {\em et~al.\/}(2011)Moll, Graham, Pratt, Cameron, M{\"u}ller \&
  Sch{\"u}ssler]{moll2011universality}
{\sc \au{Moll, Rainer}, \au{Graham, J~Pietarila}, \au{Pratt, J}, \au{Cameron,
  RH}, \au{M{\"u}ller, W-C} \& \au{Sch{\"u}ssler, M}} \yr{2011}
  \at{Universality of the small-scale dynamo mechanism}.  \jt{The Astrophysical
  Journal}  \bvol{736}~(1),  \pg{36}.

\bibitem[Monin \& Yaglom(1975)]{Mon_75}
{\sc \au{Monin, A.~S.} \& \au{Yaglom, A.~M.}} \yr{1975} {\em Statistical
  {F}luid {M}echanics\/}.  \publ{MIT Press, Cambridge, Mass}.

\bibitem[M{\"u}ller \& Carati(2002)]{muller2002dynamic}
{\sc \au{M{\"u}ller, Wolf-Christian} \& \au{Carati, Daniele}} \yr{2002}
  \at{Dynamic gradient-diffusion subgrid models for incompressible
  magnetohydrodynamic turbulence}.  \jt{Physics of plasmas}  \bvol{9}~(3),
  \pg{824--834}.

\bibitem[Piomelli {\em et~al.\/}(1991)Piomelli, Cabot, Moin \&
  Lee]{piomelli1991subgrid}
{\sc \au{Piomelli, Ugo}, \au{Cabot, William~H}, \au{Moin, Parviz} \& \au{Lee,
  Sangsan}} \yr{1991}  \at{Subgrid-scale backscatter in turbulent and
  transitional flows}.  \jt{Physics of Fluids A: Fluid Dynamics}  \bvol{3}~(7),
   \pg{1766--1771}.

\bibitem[Politano \& Pouquet(1998)]{politano1998karman}
{\sc \au{Politano, H} \& \au{Pouquet, A}} \yr{1998}  \at{von
  k{\'a}rm{\'a}n--howarth equation for magnetohydrodynamics and its
  consequences on third-order longitudinal structure and correlation
  functions}.  \jt{Physical Review E}  \bvol{57}~(1),  \pg{R21}.

\bibitem[Ponty {\em et~al.\/}(2008)Ponty, Mininni, Laval, Alexakis, Baerenzung,
  Daviaud, Dubrulle, Pinton, Politano \& Pouquet]{ponty2008linear}
{\sc \au{Ponty, Yannick}, \au{Mininni, Pablo~D}, \au{Laval, Jean-Philipe},
  \au{Alexakis, Alexandros}, \au{Baerenzung, Julien}, \au{Daviaud,
  Fran{\c{c}}ois}, \au{Dubrulle, B{\'e}reng{\`e}re}, \au{Pinton,
  Jean-Fran{\c{c}}ois}, \au{Politano, H{\'e}l{\'e}ne} \& \au{Pouquet, Annick}}
  \yr{2008}  \at{Linear and non-linear features of the taylor--green dynamo}.
  \jt{Comptes Rendus Physique}  \bvol{9}~(7),  \pg{749--756}.

\bibitem[Pope(2000)]{Pope_turbulent}
{\sc \au{Pope, S.~B.}} \yr{2000} {\em Turbulent Flows\/}.  \publ{Cambridge
  University Press}.

\bibitem[Sagaut(2001)]{sagaut2001large}
{\sc \au{Sagaut, P.}} \yr{2001} {\em Large eddy simulation for incompressible
  flows\/}, ,  \vol{vol.~20}.  \publ{Springer}.

\bibitem[Schekochihin {\em et~al.\/}(2004)Schekochihin, Cowley, Taylor, Maron
  \& McWilliams]{schekochihin2004simulations}
{\sc \au{Schekochihin, Alexander~A}, \au{Cowley, Steven~C}, \au{Taylor,
  Samuel~F}, \au{Maron, Jason~L} \& \au{McWilliams, James~C}} \yr{2004}
  \at{Simulations of the small-scale turbulent dynamo}.  \jt{The Astrophysical
  Journal}  \bvol{612}~(1),  \pg{276}.

\bibitem[Smagorinsky(1963)]{smagorinsky1963general}
{\sc \au{Smagorinsky, Joseph}} \yr{1963}  \at{General circulation experiments
  with the primitive equations: I. the basic experiment}.  \jt{Monthly weather
  review}  \bvol{91}~(3),  \pg{99--164}.

\bibitem[Sorriso-Valvo {\em et~al.\/}(2007)Sorriso-Valvo, Marino, Carbone,
  Noullez, Lepreti, Veltri, Bruno, Bavassano \&
  Pietropaolo]{sorriso2007observation}
{\sc \au{Sorriso-Valvo, Luca}, \au{Marino, Raffaele}, \au{Carbone, Vincenzo},
  \au{Noullez, A}, \au{Lepreti, F}, \au{Veltri, P}, \au{Bruno, Roberto},
  \au{Bavassano, Bruno} \& \au{Pietropaolo, Ermanno}} \yr{2007}
  \at{Observation of inertial energy cascade in interplanetary space plasma}.
  \jt{Physical review letters}  \bvol{99}~(11),  \pg{115001}.

\bibitem[Speziale(1985)]{speziale1985galilean}
{\sc \au{Speziale, Charles~G}} \yr{1985}  \at{Galilean invariance of
  subgrid-scale stress models in the large-eddy simulation of turbulence}.
  \jt{Journal of fluid mechanics}  \bvol{156},  \pg{55--62}.

\bibitem[Teaca {\em et~al.\/}(2011)Teaca, Carati \&
  Andrzej~Domaradzki]{teaca2011locality}
{\sc \au{Teaca, Bogdan}, \au{Carati, Daniele} \& \au{Andrzej~Domaradzki, J}}
  \yr{2011}  \at{On the locality of magnetohydrodynamic turbulence scale
  fluxes}.  \jt{Physics of Plasmas}  \bvol{18}~(11),  \pg{112307}.

\bibitem[Teaca {\em et~al.\/}(2021)Teaca, Gorbunov, Told, Navarro \&
  Jenko]{teaca2021sub}
{\sc \au{Teaca, Bogdan}, \au{Gorbunov, Evgeny~A}, \au{Told, Daniel},
  \au{Navarro, Alejandro~Ba{\~n}{\'o}n} \& \au{Jenko, Frank}} \yr{2021}
  \at{Sub-grid-scale effects in magnetised plasma turbulence}.  \jt{Journal of
  Plasma Physics}  \bvol{87}~(2).

\bibitem[Teaca {\em et~al.\/}(2009)Teaca, Verma, Knaepen \&
  Carati]{teaca2009energy}
{\sc \au{Teaca, Bogdan}, \au{Verma, MK}, \au{Knaepen, Bernard} \& \au{Carati,
  Daniele}} \yr{2009}  \at{Energy transfer in anisotropic magnetohydrodynamic
  turbulence}.  \jt{Physical Review E}  \bvol{79}~(4),  \pg{046312}.

\bibitem[Tennekes \& Lumley(1990)]{Ten_90}
{\sc \au{Tennekes, H.} \& \au{Lumley, J.~L.}} \yr{1990} {\em A First Course in
  Turbulence\/}.  \publ{The MIT Press, Cambridge, Massachusetts}.

\bibitem[Theobald {\em et~al.\/}(1994)Theobald, Fox \&
  Sofia]{theobald1994subgrid}
{\sc \au{Theobald, Michael~L}, \au{Fox, Peter~A} \& \au{Sofia, Sabatino}}
  \yr{1994}  \at{A subgrid-scale resistivity for magnetohydrodynamics}.
  \jt{Physics of plasmas}  \bvol{1}~(9),  \pg{3016--3032}.

\bibitem[Valori {\em et~al.\/}(2020)Valori, Innocenti, Dubrulle \&
  Chibbaro]{valori2020weak}
{\sc \au{Valori, Valentina}, \au{Innocenti, Alessio}, \au{Dubrulle,
  B{\'e}reng{\`e}re} \& \au{Chibbaro, Sergio}} \yr{2020}  \at{Weak formulation
  and scaling properties of energy fluxes in three-dimensional numerical
  turbulent rayleigh--b{\'e}nard convection}.  \jt{Journal of Fluid Mechanics}
  \bvol{885}.

\bibitem[Verma(2004)]{verma2004statistical}
{\sc \au{Verma, Mahendra~K}} \yr{2004}  \at{Statistical theory of
  magnetohydrodynamic turbulence: recent results}.  \jt{Physics Reports}
  \bvol{401}~(5-6),  \pg{229--380}.

\bibitem[Verma(2019)]{verma2019energy}
{\sc \au{Verma, Mahendra~K}} \yr{2019} {\em Energy transfers in fluid flows:
  multiscale and spectral perspectives\/}.  \publ{Cambridge University Press}.

\bibitem[Verma(2021)]{verma2021variable}
{\sc \au{Verma, Mahendra~K}} \yr{2021}  \at{Variable energy flux in
  turbulence}.  \jt{Journal of Physics A: Mathematical and Theoretical} .

\bibitem[Verma \& Kumar(2004)]{verma2004large}
{\sc \au{Verma, Mahendra~K} \& \au{Kumar, Shishir}} \yr{2004}  \at{Large-eddy
  simulations of fluid and magnetohydrodynamic turbulence using renormalized
  parameters}.  \jt{Pramana}  \bvol{63}~(3),  \pg{553--561}.

\bibitem[Vigan{\`o} {\em et~al.\/}(2019)Vigan{\`o}, Aguilera-Miret \&
  Palenzuela]{vigano2019extension}
{\sc \au{Vigan{\`o}, Daniele}, \au{Aguilera-Miret, Ricard} \& \au{Palenzuela,
  Carlos}} \yr{2019}  \at{Extension of the subgrid-scale gradient model for
  compressible magnetohydrodynamics turbulent instabilities}.  \jt{Physics of
  Fluids}  \bvol{31}~(10),  \pg{105102}.

\bibitem[Vlaykov {\em et~al.\/}(2016)Vlaykov, Grete, Schmidt \&
  Schleicher]{vlaykov2016nonlinear}
{\sc \au{Vlaykov, Dimitar~G}, \au{Grete, Philipp}, \au{Schmidt, Wolfram} \&
  \au{Schleicher, Dominik~RG}} \yr{2016}  \at{A nonlinear structural
  subgrid-scale closure for compressible mhd. i. derivation and energy
  dissipation properties}.  \jt{Physics of Plasmas}  \bvol{23}~(6),
  \pg{062316}.

\bibitem[Vreman {\em et~al.\/}(1994)Vreman, Geurts \&
  Kuerten]{vreman1994realizability}
{\sc \au{Vreman, Bert}, \au{Geurts, Bernard} \& \au{Kuerten, Hans}} \yr{1994}
  \at{Realizability conditions for the turbulent stress tensor in large-eddy
  simulation}.  \jt{Journal of Fluid Mechanics}  \bvol{278},  \pg{351--362}.

\bibitem[Vreman {\em et~al.\/}(1997)Vreman, Geurts \& Kuerten]{vreman1997large}
{\sc \au{Vreman, Bert}, \au{Geurts, Bernard} \& \au{Kuerten, Hans}} \yr{1997}
  \at{Large-eddy simulation of the turbulent mixing layer}.  \jt{Journal of
  fluid mechanics}  \bvol{339},  \pg{357--390}.

\bibitem[Yang {\em et~al.\/}(2022)Yang, Matthaeus, Roy, Roytershteyn, Parashar,
  Bandyopadhyay \& Wan]{yang2022pressure}
{\sc \au{Yang, Yan}, \au{Matthaeus, William~H}, \au{Roy, Sohom},
  \au{Roytershteyn, Vadim}, \au{Parashar, Tulasi~N}, \au{Bandyopadhyay, Riddhi}
  \& \au{Wan, Minping}} \yr{2022}  \at{Pressure--strain interaction as the
  energy dissipation estimate in collisionless plasma}.  \jt{The Astrophysical
  Journal}  \bvol{929}~(2),  \pg{142}.

\end{thebibliography}

%%%
%%%

%%%
%%%

\end{document}